\documentclass[a4paper,11pt]{article}
\pdfoutput=1 

\usepackage{jheppub}
\usepackage[T1]{fontenc} 

\usepackage[dvipsnames]{xcolor}
\usepackage{tikz}
\usetikzlibrary{patterns}
\usetikzlibrary{snakes}

\newcommand{\gen}[1]{\mathbf{#1}}

\newcommand{\phase}{\sigma}
\newcommand{\BES}{\sigma_{\text{\tiny BES}}}

\newcommand{\curvearrowurl}{\curvearrowleft}

\newcommand{\curvearrowdlr}{\rotatebox[origin=c]{180}{$\curvearrowleft$}}

\newcommand{\inturl}{\;{\int \negthickspace \negthickspace \negthickspace\negthickspace \negthinspace \curvearrowurl}\mbox{ }\,} 
 
\newcommand{\intdlr}{\;{\int \negthickspace \negthickspace \negthickspace \negthinspace \curvearrowdlr}\mbox{ }\,}

\newcommand{\sigmabb}{\varsigma^{\bullet\bullet}}
\newcommand{\sigmabbt}{\tilde{\varsigma}^{\bullet\bullet}}
\newcommand{\sigmabc}{\varsigma^{\bullet\circ}}
\newcommand{\sigmacb}{\varsigma^{\circ\bullet}}
\newcommand{\sigmacc}{\varsigma^{\circ\circ}}
\newcommand{\sigmacct}{\tilde{\varsigma}^{\circ\circ}}
\newcommand{\sigmap}{\varsigma^{\boldsymbol{+}}}
\newcommand{\sigmam}{\varsigma^{\boldsymbol{-}}}

\newcommand{\PhiSG}{\varPhi}
\newcommand{\PhiMON}{\widehat{\varPhi}}
\newcommand{\phiSG}{\varphi}
\newcommand{\phiMON}{\widehat{\varphi}}
\newcommand{\RSG}{R}
\newcommand{\RMON}{\widehat{R}}

\newcommand{\arcsinh}{\text{arcsinh}}

\newcommand{\bal}{\begin{equation}\begin{aligned}}
\newcommand{\eal}{\end{aligned}\end{equation}}

\newcommand{\ov}{\over}
\newcommand{\g}{\gamma}

\newcommand{\sn}{\text{sn}}
\newcommand{\cn}{\text{cn}}
\newcommand{\dn}{\text{dn}}
\newcommand{\am}{\text{am}}

\definecolor{grey}{rgb}{0.4,0.4,0.5}
\definecolor{darkgreen}{rgb}{0,0.5,0}
\definecolor{darkred}{rgb}{0.6,0.0,0}
\definecolor{lightbrown}{rgb}{1,0.9,0.8}
\definecolor{brown}{rgb}{0.6,0.3,0.3}
\definecolor{darkblue}{rgb}{0,0,0.5}
\definecolor{darkmagenta}{rgb}{0.5,0,0.5}

\title{\boldmath New Dressing Factors for AdS3/CFT2}

\author[a,1]{Sergey Frolov,%
\note{Correspondent fellow at Steklov Mathematical Institute, Moscow.}}
\author[b,c,2]{Alessandro Sfondrini%
\note{IBM Einstein Fellow.}}

\affiliation[a]{School of Mathematics and Hamilton Mathematics Institute,\\
Trinity College, Dublin 2, Ireland}
\affiliation[b]{Institute for Advanced Study,\\
Einstein Drive, Princeton, New Jersey, 08540 USA}
\affiliation[c]{Dipartimento di Fisica e Astronomia, Universit\`a degli Studi di Padova,\\
\& Istituto Nazionale di Fisica Nucleare, Sezione di Padova,\\
via Marzolo 8, 35131 Padova, Italy}

\emailAdd{frolovs@maths.tcd.ie}
\emailAdd{alessandro@ias.edu}

\abstract{The worldsheet S matrix of strings on the $AdS_3\times S^3\times T^4$ background is almost entirely fixed by symmetries, up to five functions --- the dressing factors. These must satisfy several consistency conditions, in particular a set of crossing equations. We find that the existing proposal for the dressing factors, while crossing invariant, violates some of these consistency conditions. We put forward a new set of dressing factors and discuss in detail their analytic properties in the string and mirror region, as well as under bound-state fusion.}

\begin{document} 
\maketitle
\flushbottom

\section{Introduction}
\label{sec:introduction}
The ability to compute observables at arbitrary values of  coupling constant(s) is crucial to understand and test a model. In the context of string theory and AdS/CFT this is very hard for backgrounds that are supported by Ramond-Ramond  (RR) fluxes. In some cases, however, it is possible to relate the spectral problem of a given (RR) background to the determination of the finite-volume spectrum of  an integrable two-dimensional quantum field theory (IQFT). This approach has led to spectacular results for the study of type IIB strings on $AdS_5\times S^5$ and their holographic dual, $\mathcal{N}=4$ supersymmetric Yang-Mills theory (SYM) in the planar limit~\cite{'tHooft:1973alw}. The two-dimensional IQFT here  arises on the string worldsheet in light-cone gauge, and it is a rather intricate non-relativistic theory, see~\cite{Arutyunov:2009ga,Beisert:2010jr} for reviews.
More recently, a similar program has been undertaken for backgrounds of the $AdS_3$ type, see~\cite{Sfondrini:2014via} for a review of early efforts in this direction. The first step to determine the finite-volume quantum spectrum of an IQFT is to have a firm handle on its S~matrix in infinite volume. For a string-theory in lightcone gauge the S~matrix should feature 8 Bosons and 8 Fermions, plus possibly bound states thereof. Hence, we are dealing at least wtth a $256\times 256$ matrix. The study of this object can be broken down in two parts: first, we may use the symmetries of the theory to constrain as many entries as possible. This was done for the $AdS_3\times S^3\times S^3\times S^1$ background in~\cite{Borsato:2012ud,Borsato:2015mma} and for $AdS_3\times S^3\times T^4$ in~\cite{Borsato:2013qpa, Borsato:2014exa}. The S~matrix of the latter background turns out to be simpler than the one of former, and it will be the focus of this paper. In $AdS_3\times S^3\times T^4$ it was found~\cite{Borsato:2014exa} that symmetries determine the S~matrix up to ten dressing factors --- compared to the single dressing factor of $AdS_5\times S^5$. Moreover, ``braiding'' unitarity (a symmetry expected in IQFTs, see \textit{e.g.}~\cite{Arutyunov:2009ga} for a review) and parity can be used to further reduce the number of independent dressing factors to five~\cite{Borsato:2014hja,upcoming:massless}. These dressing factors are not arbitrary. Rather, they have to satisfy some constraints due to (``physical'' or ``generalised'') unitarity as well as to a non-relativistic analogue of crossing symmetry~\cite{Janik:2006dc}. These constraints have been derived in~\cite{Borsato:2014hja,upcoming:massless} for the $AdS_3\times S^3\times T^4$ S~matrix.

The crossing equations typically force the dressing factors to have a very non-trivial structure, see \textit{e.g.}~\cite{Bombardelli:2016rwb}. For instance in relativistic theories it is convenient to introduce a rapidity variable~$\theta$ satisfying
\begin{equation}
    p_i = M\,\sinh\theta_i\,,\qquad E_i =M\,\cosh\theta_i\,.
\end{equation}
As long as we disregard the dressing factor, the entries of the S~matrix are $2\pi i$-periodic functions of~$(\theta_1-\theta_2)$. However, the dressing factors are typically non-periodic functions on the~$(\theta_1-\theta_2)$ plane. Equivalently, they encode the information on all sheets of the Mandelstam plane. Still, the crossing and unitarity conditions do not have a unique solution; there are non-trivial solutions of the \textit{homogeneous} associated equations, the so-called Castillejo-Dalitz-Dyson (CDD) factors~\cite{Castillejo:1955ed},  by which we can multiply   any particular candidate dressing factor to obtain a new one. In relativistic theories, however, it is usually possible to fix a solution based on its expected asymptotic behaviour and poles in $(\theta_1-\theta_2)$. Unfortunately, things are substantially harder for non-relativistic models like the one emerging from the worldsheet of $AdS$ strings.

Here, the S~matrix depends on two distinct rapidities, one for each particle, and it is not a meromorphic function of such variables. Rather, it has infinitely many pairs of branch points. Quite remarkably, the solution to the $AdS_5\times S^5$ crossing equation was found by Beisert, Eden and Staudacher (BES)~\cite{Beisert:2006ez}, who also relied on intuition coming from perturbative computations in the dual $\mathcal{N}=4$ SYM. That dressing factor turns out to have remarkable properties~\cite{Dorey:2007xn,Arutyunov:2009kf}, which are essential to define a ``mirror'' theory through analytic continuation --- a necessary step to eventually describe the finite-volume spectrum of the theory~\cite{Ambjorn:2005wa,Arutyunov:2007tc}.
For $AdS_3\times S^3\times T^4$ things are more complicated, at least in general.%
\footnote{There is one exception: the $AdS_3\times S^3\times T^4$ background can be realised without any RR flux. That case can be understood as a Wess-Zumino-Witten model~\cite{Maldacena:2000hw} and its S~matrix, dressing factor and spectrum can also be easily obtained by integrability~\cite{Baggio:2018gct, Dei:2018mfl}.}
In fact, the analytic structure of the worldsheet theory is further complicated by the presence of \textit{massless} excitations~\cite{Borsato:2014exa,Borsato:2014hja,upcoming:massless}, whose non-relativistic dispersion relation has no mass gap. Besides, very little is known about the dual theory.%
\footnote{%
Once again, with the exception of backgrounds that can be realised as WZW models, see~\cite{Giribet:2018ada,Eberhardt:2018ouy,Eberhardt:2021vsx}.} 
Notwithstanding these difficulties, the dressing factors for $AdS_3\times S^3\times T^4$ were proposed in the case where the background is supported by RR flux only (this is the case most similar to $AdS_5\times S^5$) in~\cite{Borsato:2013hoa,Borsato:2016xns}.
While that proposal does solve the crossing equations and unitarity conditions, some of its analytic properties are troublesome, especially for what concerns the scattering processes involving massless modes. Qualitatively, when studying the proposal of~\cite{Borsato:2013hoa,Borsato:2016xns}, we encounter the following stumbling blocks:
\begin{enumerate}
    \item The dressing factors scattering massive particles violate the parity invariance of the model. This fact was not appreciated in the existing literature, but can be easily checked as we do in appendix~\ref{app:monodromy}.
    \item The dressing factors involving massless particles contained the Arutyunov-Frolov-Staudacher (AFS) dressing factor~\cite{Arutyunov:2004vx}, which is the leading-order term of the asymptotic expansion of the BES phase. However, unlike the BES factor, the AFS one has a rather complicated analytic structure that makes it hard to analytically continue it.
    \item The massless-massive dressing factor has additional apparent square-root branch points whose position depends on the relative value of the momenta of the two scattered particles.
    \item The form of the dressing factors makes it hard to ``fuse'' them, \textit{i.e.}\ to use them as building blocks of the bound-state S matrices.
    \item It is not clear how to analytically continue the dressing factors to the ``mirror'' region.
    \item The dressing factors are not compatible with some of the perturbative computations in the literature (though this may be due to subtleties in the perturbative computations related to infrared divergences).
\end{enumerate}

In our effort to resolve these issues, we have found a different solution of the crossing equations, which resolves problems 1--4. To do this, we have reformulated the crossing equations in terms of rapidities following~\cite{Beisert:2006ib,Fontanella:2019baq} so that part of the solution is of difference-type. Moreover, a careful analysis of the analytic properties of massless particles suggests that the path used for analytic continuation in the crossing transformation in~\cite{Borsato:2016xns} is not correct. As a consequence, the proposal of~\cite{Borsato:2016xns} could not be correct. Our proposal passes several nontrivial consistency checks, including having good properties in the ``mirror'' region and under ``fusion''. As for point 5 of the list above, we will still encounter discrepancies in the one-loop expansion of the dressing factors, which may indeed be explained by infrared issues in the perturbative computations.

This article is structured as  follows. We start in section~\ref{sec:smatrix} by fixing our convention for the normalisation of the S~matrix and listing the constraints on the dressing factors; in section~\ref{sec:smatrix:BESnormalisation} we rewrite those conditions when a BES factor is stripped out of each dressing factor, which will facilitate our subsequent analysis.
In section~\ref{sec:rapidities} we discuss the analytic properties of massive and massless particles in the string region, mirror region, and anti-string (or crossed) region; in particular, we introduce rapidity variables by means of which we can write the crossing equations in difference form, which we do in section~\ref{sec:rapidities:crossing}. In section~\ref{sec:buildingblocks} we introduce the building blocks that we will need to construct our solutions, starting from the BES one in section~\ref{sec:buildingblocks:BES}, which we extend to the massive-massless and massless-massless kinematics; we then introduce three additional functions which we will use to solve the difference-form part of the crossing equations. Based on that, in section~\ref{sec:proposal} we present our proposal for the dressing factors, and discuss the analytic properties and perturbative expansion of the various functions. We also compare our proposal with the existing one, in section~\ref{sec:proposal:relations},  and discuss how it can be adapted to backgrounds with RR and NSNS fluxes too in section~\ref{sec:proposal:mixedflux}. Finally, in section~\ref{sec:BYE} we write the appropriately-normalised Bethe-Yang equations, which serve as a summary of our proposal as well as the starting point for the study of the spectrum of the theory, and we present our conclusions in section~\ref{sec:conclusions}.
The appendices contain a number slightly more technical discussions and derivations. The properties of the BES factor, especially in the mirror region, as well as its perturbative expansion, are discussed in appendix~\ref{app:BES} and~\ref{app:BESexpansion}, respectively.
In appendix~\ref{app:Fourier} we derive one of the new functions which we need for our solution.
In appendix~\ref{app:lcgauge} we discuss how to normalise the S-matrix elements in order to compare with perturbative results. In appendix~\ref{app:Zhukovsky} and~\ref{app:BMNexpansion} we discuss the perturbative expansion of rapidities and dressing factors. In appendix~\ref{app:monodromy} we show that the existing proposal for the massive dressing factors violates parity invariance of the~model.

\section{The S matrix and crossing equations}
\label{sec:smatrix}
The S~matrix for fundamental particles in $AdS_3\times S^3\times T^4$ is composed of several blocks, corresponding to different irreducible representations of the light-cone symmetry algebra. In terms of the Cartan elements of $\mathfrak{psu}(1,1|2)^{\oplus2}$ we consider in particular
\begin{equation}
    \gen{E}=\gen{L}_0-\gen{J}^3 +\tilde{\gen{L}}_0-\tilde{\gen{J}}^3\,,\qquad
    \gen{M}=\gen{L}_0-\gen{J}^3 -\tilde{\gen{L}}_0+\tilde{\gen{J}}^3\,.
\end{equation}
While $\gen{E}$ is the light-cone Hamiltonian, $\gen{M}$ is a combination of spin, which is quantised for the pure-RR backgrounds.%
\footnote{When NSNS fluxes are present, the quantisation of $\gen{M}$ only holds for states that satisfy the level-matching condition~\cite{Hoare:2013lja,Lloyd:2014bsa}.}
In fact, for fundamental particles the eigenvalue $M$ of $\gen{M}$ is
\begin{equation}
    M = \begin{cases}
    +1& \text{``left'':}\quad (Y,\psi^{\alpha},Z)\\
    -1& \text{``right'':}\quad (\bar{Z},\bar{\psi}^{\alpha},\bar{Y})\\
    0& \text{``massless'':}\quad (\chi^{\dot{\alpha}},T^{\alpha\dot{\alpha}},\tilde{\chi}^{\dot{\alpha}})
    \end{cases}
\end{equation}
It is possible also to define bound states, which have $M\in\mathbb{Z}$~\cite{Borsato:2013hoa}. 
Particles transform in short representations, and the shortening condition allows to write down a dispersion relation
\begin{equation}
    E(p) =\sqrt{M^2+4^2\sin^2\Big(\frac{p}{2}\Big)}\,,
\end{equation}
where $h$ acts as a coupling constant.  The dispersion is $2\pi$-periodic, and real particles have momentum $-\pi< p<\pi$. Note that for $M=0$ the dispersion is not gapped, and it is non-analytic. To properly account for this fact, it is necessary to split massless excitations between anti-chiral ($-\pi< p <0$) and chiral ($0<p<\pi$), and distinguish the limits $p\to0^\pm$~\cite{upcoming:massless}.
The worldsheet theory is weakly coupled when $h\gg1$ and the momentum is rescaled as $p/h$~\cite{Berenstein:2002jq}. In that case, the dispersion becomes relativistic with massive and massless particles.

By virtue of a discrete symmetry under flipping the sign of $M$ for all particles involved in a given process, we have only five distinct blocks in the S~matrix of fundamental particles:
\begin{enumerate}
    \item left-left scattering, or equivalently right-right scattering,
    \item left-right scattering, or equivalently right-left scattering.
    \item left-massless scattering, or equivalently right-massless scattering.
    \item massless-left scattering, or equivalently massless-right scattering.
    \item massless-massless scattering.
\end{enumerate}
However, every time we encounter a massless block we have two options: the particle can be chiral (positive velocity) or anti-chiral (negative velocity). Effectively, this doubles the number of blocks for points 3.~and 4., and it quadruples it for point 5., at least in principle. However, as we will review, there are a number of discrete symmetries that relates several of these blocks to each other~\cite{upcoming:massless}.

In order to describe the S~matrix for the various blocks it is convenient to introduce the Zhukovsky variables, defined as
\begin{equation}
\label{eq:Zhukovsky}
    x^\pm(p,|M|)=e^{\pm i p/2}\frac{|M|+\sqrt{M^2+4h^2\sin^2(p/2)}}{2h\sin(p/2)}\,.
\end{equation}
Note that for $M=0$ we have 
\begin{equation}
    x^+(p,0)\,x^-(p,0)=1\,.
\end{equation}
As a result, we introduce two distinct notations for $|M|=1$ and $M=0$, namely
\begin{equation}
x_p^\pm = x^\pm(p,1)\,,\qquad x_p = x^+(p,0)=\frac{1}{x^-(p,0)}\,.
\end{equation}
In terms of these expressions we have, for massive particles,
\begin{equation}
    e^{ip}=\frac{x^+_p}{x^-_p}\,,\qquad
    E=\frac{h}{2i}\left(
    x^+_p-\frac{1}{x^+_p}-x^-_p+\frac{1}{x^-_p}\right)\,,
\end{equation}
which for massless particles takes the slightly simpler form
\begin{equation}
    e^{ip}=(x_p)^2\,,\qquad
    E=\frac{h}{i}\left(
    x_p-\frac{1}{x_p}\right)\,.
\end{equation}

\subsection{S-matrix normalisation}
\label{sec:smatrix:normalisation}
Symmetries fix the scattering matrix up to a dressing factor for each block. In order to define in an unambiguous way our choice of normalisation, we fix the following scattering processes among highest-weight states in each representation, in the notation of~\cite{Borsato:2014hja}). For the massive states,
\begin{equation}
\label{eq:massivenorm}
    \begin{aligned}
    \mathbf{S}\,\big|Y_{p_1}Y_{p_2}\big\rangle&=&
    e^{+i p_1}e^{-i p_2}
    \frac{x^-_1-x^+_2}{x^+_1-x^-_2}
    \frac{1-\frac{1}{x^-_1x^+_2}}{1-\frac{1}{x^+_1x^-_2}}\big(\phase^{\bullet\bullet}_{12}\big)^{-2}\,
    \big|Y_{p_1}Y_{p_2}\big\rangle,\\
    \mathbf{S}\,\big|Y_{p_1}\bar{Z}_{p_2}\big\rangle&=&
    e^{-i p_2}
    \frac{1-\frac{1}{x^-_1x^-_2}}{1-\frac{1}{x^+_1x^+_2}}
    \frac{1-\frac{1}{x^-_1x^+_2}}{1-\frac{1}{x^+_1x^-_2}}\big(\tilde{\phase}^{\bullet\bullet}_{12}\big)^{-2}\,
    \big|Y_{p_1}\bar{Z}_{p_2}\big\rangle,\\
    \mathbf{S}\,\big|\bar{Z}_{p_1}Y_{p_2}\big\rangle&=&
    e^{+ip_1}
    \frac{1-\frac{1}{x^+_1x^+_2}}{1-\frac{1}{x^-_1x^-_2}}
    \frac{1-\frac{1}{x^-_1x^+_2}}{1-\frac{1}{x^+_1x^-_2}}\big(\widetilde{\phase}^{\bullet\bullet}_{12}\big)^{-2}\,
    \big|\bar{Z}_{p_1}Y_{p_2}\big\rangle,\\
    \mathbf{S}\,\big|\bar{Z}_{p_1}\bar{Z}_{p_2}\big\rangle&=&
    \frac{x^+_1-x^-_2}{x^-_1-x^+_2}
    \frac{1-\frac{1}{x^-_1x^+_2}}{1-\frac{1}{x^+_1x^-_2}}\big(\phase^{\bullet\bullet}_{12}\big)^{-2}\,
    \big|\bar{Z}_{p_1}\bar{Z}_{p_2}\big\rangle,
    \end{aligned}
\end{equation}
where the latter two equations are related to the former two by left-right symmetry~\cite{Borsato:2012ud} (the middle two equations are also related to each other by braiding unitarity).

In the mixed-mass sector we have to pick a chirality for the massless particle. We will assume that the first particle has always positive velocity and the second one has negative velocity. Hence we have
\begin{equation}
\label{eq:mixednormalisation}
    \begin{aligned}
    \mathbf{S}\,\big|Y_{p_1}\chi^{\dot{\alpha}}_{p_2}\big\rangle&=&
    e^{+\frac{i}{2} p_1}e^{-i p_2}
    \frac{x^-_1-x_2}{1-x^+_1x_2}\big(\phase^{\bullet-}_{12}\big)^{-2}\,
    \big|Y_{p_1}\chi^{\dot{\alpha}}_{p_2}\big\rangle,\\
    \mathbf{S}\,\big|\bar{Z}_{p_1}\chi^{\dot{\alpha}}_{p_2}\big\rangle&=&
    e^{-\frac{i}{2} p_1}e^{-i p_2}
    \frac{1-x^+_1x_2}{x^-_1-x_2}\big(\phase^{\bullet-}_{12}\big)^{-2}\,
    \big|\bar{Z}_{p_1}\chi^{\dot{\alpha}}_{p_2}\big\rangle,\\
    \mathbf{S}\,\big|\chi^{\dot{\alpha}}_{p_1}Y_{p_2}\big\rangle&=&
    e^{+i p_1}e^{-\frac{i}{2} p_2}
    \frac{1-x_1x^+_2}{x_1-x^-_2}\big(\phase^{+\bullet}_{12}\big)^{-2}\,
    \big|\chi^{\dot{\alpha}}_{p_1}Y_{p_2}\big\rangle,\\
    \mathbf{S}\,\big|\chi^{\dot{\alpha}}_{p_1}\bar{Z}_{p_2}\big\rangle&=&
    e^{+i p_1}e^{+\frac{i}{2} p_2}
    \frac{x_1-x^-_2}{1-x_1x^+_2}\big(\phase^{+\bullet}_{12}\big)^{-2}\,
    \big|\chi^{\dot{\alpha}}_{p_1}\bar{Z}_{p_2}\big\rangle,
    \end{aligned}
\end{equation}
where the first and second lines, as well as the third and fourth, are related by left-right symmetry, while the former two lines and the latter two are related by a combination of braiding unitarity and parity. As we will see the two remaining choices for the chiralities of the massive particle, which involve $\phase^{-\bullet}_{12}$ and $\phase^{\bullet+}_{12}$, are related to the ones above by braiding unitarity. Effectively, we will be dealing with a single dressing factor in the mixed-mass sector.

Finally, for massless modes in the kinematics regime where the first particle is chiral and the second particle is antichiral we have simply,
\begin{equation}
\label{eq:masslessnorm}
    \mathbf{S}\,\big|\chi^{\dot{\alpha}}_{p_1}\chi^{\dot{\beta}}_{p_2}\big\rangle=
    \big(\phase^{+-}_{12}\big)^{-2}\,
    \big|\chi^{\dot{\alpha}}_{p_1}\chi^{\dot{\beta}}_{p_2}\big\rangle\,.
\end{equation}
There are three more choices for the velocities of the massless particles. One, which involves $\phase^{-+}_{12}$ is related to the above by braiding unitarity as well as by parity. The remaining two involve $\phase^{\pm\pm}_{12}$. They are related to each other by parity but are in principle independent from $\phase^{\pm\mp}_{12}$. Hence for completeness let us re-write the normalisation for the same-chirality scattering, which is
\begin{equation}
    \mathbf{S}\,\big|\chi^{\dot{\alpha}}_{p_1}\chi^{\dot{\beta}}_{p_2}\big\rangle=
    \big(\phase^{++}_{12}\big)^{-2}\,
    \big|\chi^{\dot{\alpha}}_{p_1}\chi^{\dot{\beta}}_{p_2}\big\rangle\,.
\end{equation}

Of course there is quite some freedom in the choice of the above normalisations. To justify them somewhat, let us briefly look at the behaviour which we expect from these dressing factors in the BMN limit~\cite{Berenstein:2002jq}, where the coupling is large, $h\gg1$ and momenta are small, scaling like $1/h$. We will later more systematically consider a near-BMN expansion, but for now let us just take $h\to\infty$ and keep only the leading term. From the definition of the Zhukovsky variables~\eqref{eq:Zhukovsky}, see also appendix~\ref{app:Zhukovsky} we see that in this limit $p\to0$, $x^\pm\to\infty$ and $x\to\pm1$, where the sign depends on whether the momentum goes to $0^\pm$, respectively. The theory is meant to become free, so that all of our diagonal S-matrix elements introduced above should go to one. As a consequence, we see that by our definition all of the dressing factors introduced above should also go to one when the theory becomes free.

It is worth remarking that they differ from what appeared in early literature for the normalisation of the mixed-mass dressing factor, see~\cite{Borsato:2014hja,Borsato:2016xns}. Those expressions involved a term of the form
\begin{equation}
    \sqrt{\frac{x_1^--\frac{1}{x_2}}{x_1^+-x_2}\,\frac{x_1^--x_2}{x_1^+-\frac{1}{x_2}}}\,,
\end{equation}
which manifestly features branch points as $x^\pm(p_1)$ approaches $x(p_2)$ or $1/x(p_2)$. One of our  initial motivations was investigating whether it is possible to self-consistently define the S~matrix so that such branch points are manifestly absent.

Given the normalisations above, the rest of the S~matrix elements are fixed by theory's symmetries in light-cone gauge~\cite{Borsato:2012ud,Borsato:2014hja,Borsato:2016xns}. Note that throughout the cited literature, unfortunately quite a variety of conventions have been used. The complete expression for the S-matrix elements has been recently collected in~\cite{Eden:2021xhe}.
Finally, the functions denoted by $\sigma$'s are the various dressing factors, whose discussion is the purpose of this work.

\subsection{Symmetries and constraints on the dressing factors}
\label{sec:smatrix:symmetries}
There are several discrete symmetries that may be used to relate the dressing factors to each other or to themselves. As discussed elsewhere~\cite{upcoming:massless} the most important such symmetries are parity, braiding unitarity, and physical unitarity, as well as crossing symmetry.

\subsubsection{Parity}
The parity transformation allows us to get constraints on the dressing factors under $(p_1,p_2)\to(-p_1,-p_2)$. In $AdS_3\times S^3\times T^4$ this is particularly interesting because it allows us to relate the chiral and anti-chiral kinematics for massless particles. 
Let us start from the massive sectors, where we get the simple conditions on the dressing factors%
\footnote{Notice that all the constraints will hold, strictly speaking, for the square of the dressing factors, as that is what appears in the S-matrix elements.}
\begin{equation}
    \phase^{\bullet\bullet}(-p_1,-p_2)^2\,\phase^{\bullet\bullet}(p_1,p_2)^2=1\,,\qquad
    \widetilde{\phase}^{\bullet\bullet}(-p_1,-p_2)^2\,\widetilde{\phase}^{\bullet\bullet}(p_1,p_2)^2=1\,.
\end{equation}

For mixed-mass scattering we can write an expression valid for arbitrary chirality,
\begin{equation}
    \begin{aligned}
    \mathbf{S}\,\big|Y_{p_1}\chi^{\dot{\alpha}}_{p_2}\big\rangle&=&
    e^{+\frac{i}{2} p_1}e^{-i p_2}
    \frac{x^-_1-x_2}{1-x^+_1x_2}\big(\phase^{\bullet\circ}_{12}\big)^{-2}\,
    \big|Y_{p_1}\chi^{\dot{\alpha}}_{p_2}\big\rangle,\\
    \mathbf{S}\,\big|\bar{Z}_{p_1}\chi^{\dot{\alpha}}_{p_2}\big\rangle&=&
    e^{-\frac{i}{2} p_1}e^{-i p_2}
    \frac{1-x^+_1x_2}{x^-_1-x_2}\big(\phase^{\bullet\circ}_{12}\big)^{-2}\,
    \big|\bar{Z}_{p_1}\chi^{\dot{\alpha}}_{p_2}\big\rangle,\\
    \mathbf{S}\,\big|\chi^{\dot{\alpha}}_{p_1}Y_{p_2}\big\rangle&=&
    e^{+i p_1}e^{-\frac{i}{2} p_2}
    \frac{1-x_1x^+_2}{x^-_2-x_1}\big(\phase^{\circ\bullet}_{12}\big)^{-2}\,
    \big|\chi^{\dot{\alpha}}_{p_1}Y_{p_2}\big\rangle,\\
    \mathbf{S}\,\big|\chi^{\dot{\alpha}}_{p_1}\bar{Z}_{p_2}\big\rangle&=&
    e^{+i p_1}e^{+\frac{i}{2} p_2}
    \frac{x^-_2-x_1}{1-x_1x^+_2}\big(\phase^{\circ\bullet}_{12}\big)^{-2}\,
    \big|\chi^{\dot{\alpha}}_{p_1}\bar{Z}_{p_2}\big\rangle,
    \end{aligned}
\end{equation}
which reduces to the ones given in~\eqref{eq:mixednormalisation} by choosing the correct kinematics. Parity poses a constraint on the dressing factor that we just introduced, namely
\begin{equation}
    \phase^{\circ\bullet}(p_1,p_2)^2\,\phase^{\circ\bullet}(-p_1,-p_2)^2=-1\,,\qquad
    \phase^{\bullet\circ}(p_1,p_2)^2\,\phase^{\bullet\circ}(-p_1,-p_2)^2=-1\,.
\end{equation}
It is also worth looking at the behaviour which we expect from these dressing factors in the BMN limit~\cite{Berenstein:2002jq}. When $h\to\infty$ the theory must become free, which forces for $p_2<0$
\begin{equation}
    \big(\phase^{\bullet\circ}_{12}\big)^{-2}=\big(\phase^{\bullet-}_{12}\big)^{-2}\approx +1\,,
\end{equation}
while for $p_1>0$
\begin{equation}
    \big(\phase^{\circ\bullet}_{12}\big)^{-2}=-\big(\phase^{+\bullet}_{12}\big)^{-2}\approx-1\,,
\end{equation}
where we need to be careful to specify physical regions for the momenta, see~\cite{upcoming:massless}.

Let us come now to massless--massless scattering. Here we get
\begin{equation}
    \phase^{+-}(p_1,p_2)^2\,\phase^{-+}(-p_1,-p_2)^2=1\,,\qquad
    \phase^{++}(p_1,p_2)^2\,\phase^{--}(-p_1,-p_2)^2=1\,.
\end{equation}
We can think of this as a way of defining $\phase^{-+}(p_1,p_2)$ and $\phase^{--}(p_1,p_2)$ in terms of $\phase^{+-}(p_1,p_2)$ and $\phase^{++}(p_1,p_2)$, respectively. As a consequence, we can replace the four massless phases with \textit{two phases}, one for opposite-chirality and one for same-chirality. We set
\begin{equation}
    \phase^{\circ\circ}(p_1,p_2)=\phase^{++}(p_1,p_2)\,,\qquad
    \widetilde{\phase}^{\circ\circ}(p_1,p_2)=\phase^{+-}(p_1,p_2)\,.
\end{equation}
We summarise the independent dressing factors in table~\ref{table:dressing}.
Despite the fact that this is not required by any physical symmetry, we will eventually be looking for a solution which may be expressed in terms of \textit{a single function} for all massless dressing factors,  $\widetilde{\phase}^{\circ\circ}=\phase^{\circ\circ}$. We will return to this point in section~\ref{sec:proposal:massless}.

\begin{table}[t]
\centering
\begin{tabular}{|l | l|} 
 \hline
 Dressing factor & Particles scattered \\
 \hline
 $\phase^{\bullet\bullet}(p_1,p_2)$ & Two massive particles with $M_1=M_2=\pm1$.\\
 $\widetilde{\phase}^{\bullet\bullet}(p_1,p_2)$ & Two massive particles with $M_1=-M_2=\pm1$.\\
 $\phase^{\circ\bullet}(p_1,p_2)$ & One massive and one massless particle, $M_1=0$, $|M_2|=1$.\\
 $\phase^{\bullet\circ}(p_1,p_2)$ & One massless and one massive particle, $|M_1|=1$, $M_2=0$.\\
 $\phase^{\circ\circ}(p_1,p_2)$ & Two massless particles, $M_1=M_2=0$, same chirality.\\
 $\widetilde{\phase}^{\circ\circ}(p_1,p_2)$ & Two massless particles, $M_1=M_2=0$, opposite chirality.\\
 \hline
\end{tabular}
\caption{A summary of the independent dressing factors once parity has been imposed.}
\label{table:dressing}
\end{table}

\subsubsection{Braiding unitarity}
Braiding unitarity is a consistency condition of the Zamolodchikov-Faddeev algebra (see \textit{e.g.}~\cite{Arutyunov:2009ga} for a review). On the massive dressing factors it imposes
\begin{equation}
    \phase^{\bullet\bullet}(p_1,p_2)^2\,\phase^{\bullet\bullet}(p_2,p_1)^2=1\,,\qquad
    \widetilde{\phase}^{\bullet\bullet}(p_1,p_2)^2\,\widetilde{\phase}^{\bullet\bullet}(p_2,p_1)^2=1\,.
\end{equation}
For the mixed-mass factors we get
\begin{equation}\label{eq:braidunit}
    \phase^{\bullet\circ}(p_1,p_2)^2\,\phase^{\circ\bullet}(p_2,p_1)^2=1\,,
\end{equation}
which we may think of as the definition of $\phase^{\circ\bullet}(p_2,p_1)$ in terms of $\phase^{\bullet\circ}(p_1,p_2)$. Finally, for massless-massless dressing factors, we distinguish the case of opposite-chirality scattering and the case of same-chirality scattering. In the former we have
\begin{equation}
    \phase^{+-}(p_1,p_2)^2\,\phase^{-+}(p_2,p_1)^2=1\,,
\end{equation}
which, in terms of the opposite-chirality phase $\widetilde{\phase}^{\circ\circ}$, gives the constraint
\begin{equation}
    \widetilde{\phase}^{\circ\circ}(p_1,p_2)^2=\widetilde{\phase}^{\circ\circ}(-p_2,-p_1)^2\,.
\end{equation}
Instead, for the same-chirality scattering we get two constraints
\begin{equation}
    \phase^{++}(p_1,p_2)^2\,\phase^{++}(p_2,p_1)^2=1\,,\qquad
    \phase^{--}(p_1,p_2)^2\,\phase^{--}(p_2,p_1)^2=1\,,
\end{equation}
which boil down to a constraint on $\phase^{\circ\circ}$,
\begin{equation}
    \phase^{\circ\circ}(p_1,p_2)^2\,\phase^{\circ\circ}(p_2,p_1)^2=1\,.
\end{equation}
We observe that, while for opposite-chirality scattering we can rescale the phase by a constant $\phase^{\pm\mp}\to e^{\pm i \xi}\phase^{\pm\mp}$ without spoiling braiding unitarity, this is not possible for same-chirality scattering.

\subsubsection{Physical unitarity}
The weakest unitarity requirement that we may impose is that, when we are scattering particles with real energy and we are considering a physical kinematics regime, the S~matrix is a unitary matrix. In formulae,
\begin{equation}
    \gen{S}(p_1,p_2)^\dagger\,\gen{S}(p_1,p_2)=\gen{1}\,,\qquad
    p_1,p_2\in\mathbb{R}\,,\quad v_1>v_2\,.
\end{equation}
The last condition, on the velocity of the particles, ensures that scattering can happen. For the various dressing factors this means simply that , when the velocities are as above,
\begin{equation}
    |\phase^{\bullet\bullet}|=|\widetilde{\phase}^{\bullet\bullet}|=|\phase^{\bullet\circ}|=|\phase^{\circ\bullet}|=|\phase^{\circ\circ}|=|\widetilde{\phase}^{\circ\circ}|=1\qquad\text{(physical scattering)}.
\end{equation}
In general, and in particular for particles that may form bound states, we also want to consider complex values of  momenta.  In that case we can impose a stronger condition, generalised unitarity, which reads
\begin{equation}
\begin{aligned}
    \Big(\phase^{\bullet\bullet}(p_1^*,p_2^*)^2\Big)^*\phase^{\bullet\bullet}(p_1,p_2)^2=
    \Big(\widetilde{\phase}^{\bullet\bullet}(p_1^*,p_2^*)^2\Big)^*\widetilde{\phase}^{\bullet\bullet}(p_1,p_2)^2=
    1\,,\\
    \Big(\phase^{\bullet\circ}(p_1^*,p_2^*)^2\Big)^*\phase^{\bullet\circ}(p_1,p_2)^2=
    \Big(\phase^{\circ\bullet}(p_1^*,p_2^*)^2\Big)^*\phase^{\circ\bullet}(p_1,p_2)^2=
    1\,,\\
    \Big(\phase^{\circ\circ}(p_1^*,p_2^*)^2\Big)^*\phase^{\circ\circ}(p_1,p_2)^2=
    \Big(\widetilde{\phase}^{\circ\circ}(p_1^*,p_2^*)^2\Big)^*\widetilde{\phase}^{\circ\circ}(p_1,p_2)^2=
    1\,.
\end{aligned}
\end{equation}
It is not completely clear, a priori, whether this strong condition should also hold for massless modes.

\subsubsection{Crossing symmetry}
The dressing factors must satisfy another  highly nontrivial constraint --- the  crossing equations, which were given  originally in~\cite{Borsato:2014hja}, see in particular appendix~P there for their normalisation-independent form. The crossing transformation flips the sign of energy and momentum $(E,p)\to(-E, -p)$ and it can be realised by analytically continuing the S~matrix in an appropriate variable. As it turns out, the S~matrix is not a meromorphic function of the Zhukovsky variables (even before considering the dressing factors). Fortunately, there are other parametrisations that may be used to resolve this issue. For massive particles, in analogy with the case of $AdS_{5}\times S^5$, see \textit{e.g.}~\cite{Arutyunov:2009ga}, we may express the equations on the $z$-torus, where it amounts to a shift of the $z$-variable by half of the imaginary period~$\omega_2$. We will return in the next section to a more detailed discussion of the $z$-torus parametrisation. Similarly, for massless particles, it is possible to introduce a rapidity that greatly simplifies the form of the S~matrix~\cite{Fontanella:2019baq}. However, as described in detail in~\cite{upcoming:massless}, the S~matrix splits in different blocks depending on the chirality of the massless particles. As also reviewed above, by using braiding unitarity and parity we just have to distinguish two cases (in the massless-massless kinematics): particles of the same chirality, \textit{versus} particle of opposite chirality.
Postponing for a moment a detailed discussion of the various rapidity parametrisations and the explicit form of the crossing transformation, we introduce the crossed variables
\begin{equation}
\bar{x}^\pm\quad \text{(massive)}\,,\qquad
\bar{x}\quad \text{(massless)}\,.
\end{equation}
We have, for the dressing factors involving massive particles,
\begin{equation}
\begin{aligned}
\left(\sigma^{\bullet\bullet} (x_1^\pm,x_2^\pm)\right)^2\left(\tilde\sigma^{\bullet\bullet} (\bar{x}_1^\pm,x_2^\pm)\right)^2&=
\left(\frac{x_2^-}{x_2^+}\right)^2
\frac{(x_1^- - x_2^+)^2}{(x_1^- - x_2^-)(x_1^+ - x_2^+)}\frac{1-\frac{1}{x_1^-x_2^+}}{1-\frac{1}{x_1^+x_2^-}}\,,\\
\left(\sigma^{\bullet\bullet} (\bar{x}_1^\pm,x_2^\pm)\right)^2\left(\tilde\sigma^{\bullet\bullet} (x_1^\pm,x_2^\pm)\right)^2&=
\left(\frac{x_2^-}{x_2^+}\right)^2
\frac{\left(1-\frac{1}{x^+_1x^+_2}\right)\left(1-\frac{1}{x^-_1x^-_2}\right)}{\left(1-\frac{1}{x^+_1x^-_2}\right)^2}\frac{x_1^--x_2^+}{x_1^+-x_2^-}\,.
\end{aligned}
\end{equation}
For the dressing factors involving one massive and one  massless particle we have%
\footnote{While writing this article, we noticed a misprint in the mixed-mass crossing equations of ref.~\cite{Borsato:2016xns}.}
\begin{equation}
\begin{aligned}
\left(\sigma^{\bullet\circ} (x_1^\pm,x_2)\right)^2\left(\sigma^{\bullet\circ} (\bar{x}_1^\pm,x_2)\right)^2&={
\frac{1}{(x_2)^4}}\frac{f(x_1^+,x_2)}{f(x_1^-,x_2)}\,,\\
\left(\sigma^{\circ\bullet} (x_1,x_2^\pm)\right)^2\left(\sigma^{\circ\bullet} (\bar{x}_1,x_2^\pm)\right)^2&=\frac{f(x_1,x_2^+)}{f(x_1,x_2^-)}\,,
\end{aligned}
\end{equation}
where $\bar{x}$  denotes crossing in the massless variable (which we will discuss later), and
\begin{equation}
\label{eq:ffunction}
    f(x,y)=i\,\frac{1-xy}{x-y}\,.
\end{equation}
Finally, for massless particles we have
\begin{equation}
\begin{aligned}
    &\big(\sigma^{\circ\circ} (x_1,x_2)\big)^2\big(\sigma^{\circ\circ} (\bar{x}_1,x_2)\big)^2&=&-f(x_1,x_2)^2\,,\\
    &\big(\widetilde{\sigma}^{\circ\circ} (x_1,x_2)\big)^2\big(\widetilde{\sigma}^{\circ\circ} (\bar{x}_1,x_2)\big)^2&=&-f(x_1,x_2)^2\,,
\end{aligned}
\end{equation}
for same-chirality scattering and opposite-chirality scattering, respectively.

\subsection{Stripping out a BES factor}
\label{sec:smatrix:BESnormalisation}
We now redefine the dressing factors by stripping away the Beisert-Eden-Staudacher (BES) factor~\cite{Beisert:2006ez}. This is quite natural at least in the massive sector, where the BES structure is crucial to reproduce the analytic features related to the scattering of bound states~\cite{Dorey:2007xn,  Arutyunov:2009kf, OhlssonSax:2019nlj}. It is possible to extend the BES factor to the case where one or both of the particles are in the massless kinematics, and to strip out a BES-like factor from the corresponding dressing factors. This is done simply by analytically continuing the massive BES phase to the massless kinematics.
Postponing a more complete discussion of the BES dressing factor and of its generalisations to section~\ref{sec:buildingblocks:BES} and appendix~\ref{app:BES}, for now we note its crossing equations. For massive particles we have the celebrated crossing equation by Janik~\cite{Janik:2006dc,Beisert:2006ez, Arutyunov:2009kf}
\begin{equation}
\BES (x_1^\pm,x_2^\pm)\BES (\bar{x}_1^\pm,x_2^\pm)
=\frac{x_2^-}{x_2^+}g(x_1^\pm,x_2^\pm)\,,
\end{equation}
where
\begin{equation}
\label{eq:gfunction}
g(x_1^\pm,x_2^\pm)
=\frac{x_1^--x_2^+}{x_1^--x_2^-}\frac{1-\frac{1}{x_1^+x_2^+}}{1-\frac{1}{x_1^+x_2^-}}=\frac{x_1^--x_2^+}{x_1^+-x_2^+}\frac{1-\frac{1}{x_1^-x_2^-}}{1-\frac{1}{x_1^+x_2^-}}\,.
\end{equation}
Denoting the massless variable by $x$ rather than $x^\pm$ we can also write
\begin{equation}
    \BES(\bar{x}_1,x_2^\pm)\,\BES(x_1,x_2^\pm)
    =1\,,\qquad
    \BES(\bar{x}_1^\pm,x_2)\,\BES(x_1^\pm,x_2)
    =\frac{1}{x_2^2} \frac{f(x_1^+,x_2)}{f(x_1^-,x_2)}\,,
\end{equation}
where $f(x,y)$ is given by eq.~\eqref{eq:ffunction}, and
\begin{equation}
    \BES(\bar{x}_1,x_2)\,\BES(x_1,x_2)=1\,.
\end{equation}

\subsubsection{Massive-massive kinematics}
We want to define new factor by ``stripping out'' the BES factor. Hence we define
\begin{equation}
    \begin{aligned}
    \sigmabb(x_1^\pm,x_2^\pm) = \frac{\phase^{\bullet\bullet}(x_1^\pm,x_2^\pm)}{\BES(x_1^\pm,x_2^\pm)}\,,&&\qquad
    \sigmabbt(x_1^\pm,x_2^\pm) = \frac{\tilde{\phase}^{\bullet\bullet}(x_1^\pm,x_2^\pm)}{\BES(x_1^\pm,x_2^\pm)}\,,\\
    \sigmabc(x_1^\pm,x_2) = \frac{\phase^{\bullet\circ}(x_1^\pm,x_2)}{\BES(x_1^\pm,x_2)}\,,&&\qquad
    \sigmacb(x_1,x_2^\pm) = \frac{\phase^{\circ\bullet}(x_1,x_2^\pm)}{\BES(x_1,x_2^\pm)}\,,\\
    \sigmacc(x_1,x_2) = \frac{\phase^{\circ\circ}(x_1,x_2)}{\BES(x_1,x_2)}\,,&&\qquad
    \sigmacct(x_1,x_2) = \frac{\widetilde{\phase}^{\circ\circ}(x_1,x_2)}{\BES(x_1,x_2)}\,.
    \end{aligned}
\end{equation}
Note that the function $\BES(x_1^\pm,x_2^\pm)$ indicates the BES factor in the massive-massive kinematics, whereas \textit{e.g.}\ $\BES(x_1^\pm,x_2)$ indicates the BES factor in the massive-massless kinematics, see appendix~\ref{app:BES} for details. For the massive-massive kinematics, it is also convenient to define
\begin{equation}
    \sigmap(x_1^\pm,x_2^\pm) = \sigmabb(x_1^\pm,x_2^\pm)\sigmabbt(x_1^\pm,x_2^\pm)\,,\qquad
    \sigmam(x_1^\pm,x_2^\pm) = \frac{\sigmabb(x_1^\pm,x_2^\pm)}{\sigmabbt(x_1^\pm,x_2^\pm)}\,.
\end{equation}

After this redefinition, the massive-massive factors satisfy
\begin{equation}
\begin{aligned}
    \Big(\sigmabb(x_1^\pm,x_2^\pm)\Big)^{-2}\Big(\sigmabbt(\bar{x}_1^\pm,x_2^\pm)\Big)^{-2}&=\frac{1-x_1^+x_2^+}{1-x_1^-x_2^+}\frac{1-x_1^-x_2^-}{1-x_1^+x_2^-}\,,\\
    \Big(\sigmabb(\bar{x}_1^\pm,x_2^\pm)\Big)^{-2}\big(\sigmabbt(x_1^\pm,x_2^\pm)\Big)^{-2}&=\frac{x_1^+ - x_2^-}{x_1^+ - x_2^+}\frac{x_1^- - x_2^+}{x_1^- - x_2^-}\,,
\end{aligned}
\end{equation}
which we may also rewrite as one crossing equation for the product of the dressing factors
\begin{equation}
\Big(\sigmap(\bar{x}_1^\pm,x_2^\pm)\Big)^{-2}
\Big(\sigmap(x_1^\pm,x_2^\pm)\Big)^{-2}
=
\frac{f(x_1^+,x_2^+)\,f(x_1^-,x_2^-)}{f(x_1^+,x_2^-)\,f(x_1^-,x_2^+)}\,,
\end{equation}
and a monodromy equation for the ratio of the factors,
\begin{equation}
    \frac{\Big(\sigmam(\bar{x}_1^\pm,x_2^\pm)\Big)^{-2}}{\Big(\sigmam(x_1^\pm,x_2^\pm)\Big)^{-2}} 
    = \frac{(u_1-u_2+\frac{i}{h})(u_1-u_2-\frac{i}{h})}{(u_1-u_2)^2}\,,
\end{equation}
which we have rewritten in terms of the rapidity $u$ which obeys
\begin{equation}
    u = x^++\frac{1}{x^+}-\frac{i}{h} = x^-+\frac{1}{x^-}+\frac{i}{h}\,.
\end{equation}

\subsubsection{Mixed-mass kinematics}
The crossing equations in the mixed-mass kinematics are
\begin{equation}
\begin{aligned}
    \Big(\sigmabc(\bar{x}_1^\pm,x_2)\Big)^{-2} \Big(\sigmabc(x_1^\pm,x_2)\Big)^{-2}&=
    \frac{f(x_1^+,x_2)}{f(x_1^-,x_2)}\,,
    \\
    \Big(\sigmacb(\bar{x}_1,x_2^\pm)\Big)^{-2} \Big(\sigmacb(x_1,x_2^\pm)\Big)^{-2}&=
    \frac{f(x_1,x_2^-)}{f(x_1,x_2^+)}\,.
\end{aligned}
\end{equation}
As a consequence of braiding unitarity we have
\begin{equation}
\begin{aligned}
    \Big(\sigmacb(x_1,\bar{x}_2^\pm)\Big)^{-2} \Big(\sigmacb(x_1,x_2^\pm)\Big)^{-2}&=
    \frac{f(x_1,x_2^-)}{f(x_1,x_2^+)},\\
    \Big(\sigmabc(x_1^\pm,\bar{x}_2)\Big)^{-2} \Big(\sigmabc(x_1^\pm,x_2)\Big)^{-2}&=
    \frac{f(x_1^+,x_2)}{f(x_1^-,x_2)}\,,
\end{aligned}
\end{equation}
where we recall that $\bar{x}^\pm(z)=x^\pm(z+\omega_2)$. This is unlike how the crossing equation is normally written in the second variable: one would need to shift $z\to z-\omega_2$ to recast the last two equations in a more familiar form.

\subsubsection{Massless-massless kinematics}
For the massless-massless dressing factor we have
\begin{equation}
\begin{aligned}
        \Big(\sigmacc(\bar{x}_1,x_2)\Big)^{-2} \Big(\sigmacc(x_1,x_2)\Big)^{-2}
    =
    -\frac{1}{f(x,y)^2}\,.\\
        \Big(\sigmacct(\bar{x}_1,x_2)\Big)^{-2} \Big(\sigmacct(x_1,x_2)\Big)^{-2}
    =
    -\frac{1}{f(x,y)^2}\,.\\
\end{aligned}
\end{equation}

\section{Rapidity variables}
\label{sec:rapidities}
There are a number of useful ways to parametrise the S~matrix. It is particularly convenient to introduce a set of rapidity variables following Fontanella and Torrielli~\cite{Fontanella:2019baq} (see also Beisert, Hernandez and Lopez~\cite{Beisert:2006ib}). It will turn out that the equations for the ``stripped out'' factors admit a  simple difference-form solution when expressed in terms of these variables.

\subsection{Massive variables}
\label{sec:rapidities:massive}

First of all, let us recall that it is possible to parametrise $x^\pm$ in terms of Jacobi elliptic functions of~$z$ in such a way that the matrix part of the S~matrix is meromorphic as a fuction of $(z_1,z_2)$.
Let us briefly review the construction~\cite{Arutyunov:2007tc}, introducing 
\begin{equation}
    p=2 \am z\,,\qquad E=|M|\,\dn(z,k)\,,\qquad
    \sin\frac{p}{2}=\sn(z,k)\,,
\end{equation}
so that
\begin{equation}
    x^\pm =\frac{|M|}{2h}\left(\frac{\cn(z,k)}{\sn(z,k)}\pm i\right)\big(1+\dn(z,k)\big)\,,
\end{equation}
where the elliptic modulus is~$k$ and is related to the periods of the tours as
\begin{equation}
    k=-\frac{4h^2}{M^2}\,,\qquad
    \omega_1=2K(k)\,,\qquad \omega_2=2iK(1-k)-2K(k)\,,
\end{equation}
in terms of the complete elliptic integral of the first kind~$K(k)$. Clearly these formulae make sense for $|M|>0$, though we will be mostly interested in the case $|M|=1$. For that case, we can introduce a new set of variables
\begin{equation}
\label{eq:gammapm}
x^+= \frac{i - e^{\gamma^+}}{i +  e^{\gamma^+}}\,,\qquad x^-= \frac{i + e^{\gamma^-}}{i -  e^{\gamma^-}}\,,
\end{equation}
which can be thought of as functions of~$z$.
The inverse of this parametrisation requires chosing a branch,
\begin{equation}
    \gamma^+ = \log \left(-i \frac{x^+ - 1}{x^++1}\right)\quad \text{mod}\ 2\pi i\,,
    \qquad
    \gamma^- = \log\left(i\frac{ x^- - 1}{  x^-+1}\right)\quad\text{mod}\ 2\pi i\,,
\end{equation}
and we will discuss below how this should be done in various kinematics regimes.

\subsubsection{Physical region and crossing}
\begin{figure}
    \centering
\begin{tikzpicture}
\fill[pattern=horizontal lines, pattern color=red] (-2cm,-2cm) rectangle (2cm,2cm);
\node[rounded corners, draw=black, fill=white, fill opacity=0.7, text opacity=1] at (1.6cm,1.6cm) {$x^+$};
\draw[thick, fill=white] (0,0) circle (1cm);
\draw[->, thick] (-2cm,0)--(2cm,0);
\draw[->, thick] (0,-2cm)--(0,2cm);
\end{tikzpicture}
\begin{tikzpicture}
\fill[pattern=vertical lines, pattern color=blue] (-2cm,-2cm) rectangle (2cm,2cm);
\node[rounded corners, draw=black, fill=white, fill opacity=0.7, text opacity=1] at (1.6cm,1.6cm) {$x^-$};
\draw[thick, fill=white] (0,0) circle (1cm);
\draw[->, thick] (-2cm,0)--(2cm,0);
\draw[->, thick] (0,-2cm)--(0,2cm);
\end{tikzpicture}\ \ 
\begin{tikzpicture}
\fill[pattern=vertical lines, pattern color=blue] (-3cm,0cm) rectangle (3cm,1cm);
\fill[pattern=horizontal lines, pattern color=red] (-3cm,-1cm) rectangle (3cm,0cm);
\draw[->, thick] (-3cm,0)--(3cm,0);
\draw[->, thick] (0,-2cm)--(0,2cm);
\draw[densely dashed,draw=gray] (-3cm,1cm)--(3cm,1cm);
\draw[densely dashed,draw=gray] (-3cm,-1cm)--(3cm,-1cm);
\node[rounded corners, draw=black, fill=white, fill opacity=0.7, text opacity=1] at (2.6cm,1.6cm) {$\gamma^\pm$};
\node[anchor=south west,rounded corners, fill opacity=0.7, text opacity=1] at (0,1cm) {$\scriptsize{\color{gray}+i\pi}$};
\node[anchor=north west,rounded corners, fill opacity=0.7, text opacity=1] at (0,-1cm) {$\scriptsize{\color{gray}-i\pi}$};
\end{tikzpicture}
    \caption{The physical string region. In the leftmost panel, the physical string region for $x^+$ is $|x^+|>1$; in the middle panel, for $x^-$ is $|x^-|>1$; in the rightmost panel, we draw it for the $\gamma^+$ and $\gamma^-$ panel, where it takes the form of the strips between $-i\pi$ and $0$, and between $0$ and $+i\pi$, respectively.}
    \label{fig:stringregion}
\end{figure}
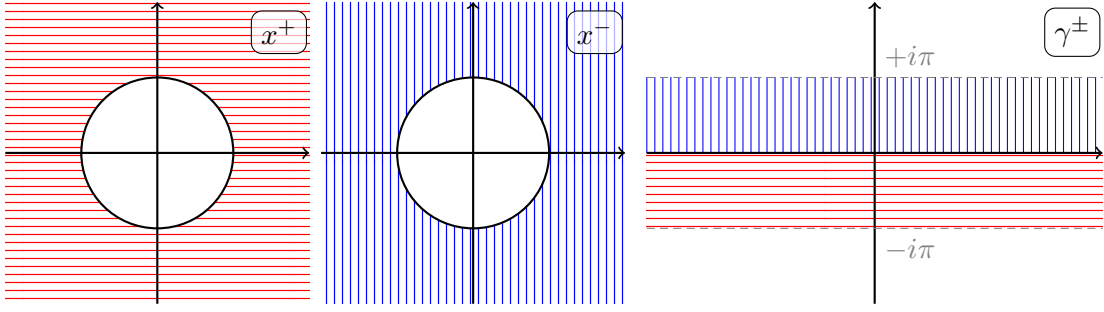

In the string region, physical values of $x^+$ satisfy $|x^+|>1$, which is the case if
\begin{equation}
\label{eq:physicalgammap}
    -\pi+2k\pi<\mathfrak{I}[\gamma^+]<2k\pi\,,\qquad
    k\in\mathbb{Z}\,.
\end{equation}
In what follows, we will choose as physical region the one given by $k=0$. Similarly, for $x^-$ the physical region is $|x^-|>1$, which gives
\begin{equation}
\label{eq:physicalgammam}
    2k\pi<\mathfrak{I}[\gamma^-]<\pi+2k\pi\,,\qquad k\in\mathbb{Z}\,.
\end{equation}
Again, we pick the region with $k=0$, see figure~\ref{fig:stringregion}. Then we have%
\footnote{We choose the branches of the logarithms to avoid any cuts in the path that we will use to define the crossing transformation.}
\begin{equation}
\label{eq:stringgammaofx}
    \gamma^+ = \log \left(-i \frac{x^+ - 1}{x^++1}\right),
    \quad
    \gamma^- = \log\left(i\frac{ x^- - 1}{  x^-+1}\right),\qquad\text{(string region)},
\end{equation}

In the string region, the energy is
\begin{equation}
\label{eq:gammaenergy}
    E= \frac{ih}{2}\left(x^- - \frac{1}{x^-}-x^++\frac{1}{x^+}\right)=h\left(\frac{1}{\cosh\gamma^+}+\frac{1}{\cosh\gamma^-}\right)\,,
\end{equation}
the momentum is given by
\begin{equation}
    e^{i p }=\frac{i - e^{\gamma^+}}{i +  e^{\gamma^+}}\frac{i - e^{\gamma^-}}{i +  e^{\gamma^-}}\,,
\end{equation}
while the shortening condition reads
\begin{equation}
    \tanh\gamma^- - \tanh\gamma^+ = \frac{i}{h}\,.
\end{equation}
It is worth emphasising that the energy is real and positive only when
\begin{equation}
    \mathfrak{I}[x^+]>0,\qquad |x^+|>1\,,\qquad
    x^- = (x^+)^*\,,
\end{equation}
which corresponds to
\begin{equation}
    -\frac{1}{2}\pi<  \mathfrak{I}[\gamma^+]<0\,,\qquad
    \gamma^- = (\gamma^+)^*\,,
\end{equation}
which is the region of \textit{real-momentum} physical particles. More generally, for complex values of the momentum (or of the torus rapidity~$z$) we have
\begin{equation}
    \big(x^+(z)\big)^* = x^-(z^*)\,,\qquad
    \big(\gamma^+(z)\big)^* = \gamma^-(z^*)\,.
\end{equation}
Finally, we note that in the small-momentum limit
\begin{equation}
\label{eq:zeromomentummassive}
    \lim_{p\to0}\gamma^{\pm}(p)=\mp \frac{i\pi}{2}\,.
\end{equation}

\subsubsection{Mirror kinematics}
The mirror theory is defined through a double Wick rotation, as discussed in detail in~\cite{Arutyunov:2007tc}. Starting from a particle of the string-worldsheet model with real momentum $p$ and positive energy $E\geq0$, we get mirror energies and momentum (denoted by tildes)
\begin{equation}
    \tilde{E}=-i\,p\,,\qquad \tilde{p}=-i\,E\,.
\end{equation}
Now for a real mirror particle $\tilde{p}$ is real and $\tilde{E}\geq0$. Indeed by this defintion we have the mirror dispersion relation
\begin{equation}
\label{eq:mirrordispersion}
    \tilde{E}(\tilde{p}) = 2\arcsinh\left(\frac{\sqrt{M^2+\tilde{p}^2}}{2h}\right)\,.
\end{equation}
For massive particles it is understood~\cite{Arutyunov:2007tc} that the mirror region may be reached from the string region by analytic continuation on the rapidity torus.
It is well-known that we may obtain the mirror dispersion by shifting $z$ by half the imaginary period of the torus~$\omega_2$~\cite{Arutyunov:2007tc,Arutyunov:2009ga}, setting
\begin{equation}
    z= \tilde{z}+\frac{\omega_2}{2}\,.
\end{equation}
Using explicit formulae for the energy and momentum in terms of Jacobi elliptic functions it is easy to verify that indeed
\begin{equation}
    \tilde{E}(\tilde{z})=-i\,p(\tilde{z}+\tfrac{1}{2}\omega_2)\,,\qquad
    \tilde{p}(\tilde{z})=-i\,E(\tilde{z}+\tfrac{1}{2}\omega_2)\,,
\end{equation}
are positive-semidefinite and real, respectively, for real rapidity~$\tilde{z}$ satisfying the condition $-\omega_1/2<\tilde z <\omega_1/2$. Moreover, we can explicitly evaluate
\begin{equation}
    \tilde{p}(\tilde{z})=\sqrt{M^2+4h^2}\,\frac{\sn(\tilde{z},k)}{\cn(\tilde{z},k)}\,,
\end{equation}
as well as%
\footnote{%
It is worth noting that  $\tilde{x}^\pm$ does not have a simple form in terms of $x^\pm$. In particular, it is not true that $\tilde{x}^+=1/x^+$ and $\tilde{x}^-=x^-$. This would be the case under the transformation $z\to-z+\tfrac{\omega_2}{2}+\tfrac{\omega_1}{2}$~\cite{Arutyunov:2009ga}.}
\begin{equation}
    \tilde{x}^\pm(\tilde{z})=x^\pm(\tilde{z}+\tfrac{\omega_2}{2})=
    -i\frac{\sqrt{1-k}\mp \dn(\tilde{z},k)}{\sqrt{-k}\dn(\tilde{z},k)}\left(
    1+i\sqrt{1-k}\frac{\sn(\tilde{z},k)}{\cn(\tilde{z},k)}
    \right)\,.
\end{equation}
From this we can find the relation valid in the mirror region
\begin{equation}
    \tilde{x}^\pm(\tilde{p}) ={1\ov 2h} \left(\sqrt{1+\frac{4h^2}{M^2+\tilde{p}^2}}\mp 1\right)\big(\tilde{p}-i |M|\big)\,.
\end{equation}
It is interesting to note that our expression for $\tilde{x}^\pm(\tilde{p})$ happens to be regular as $M\to0$.
From this formula we can find how mirror particles behave under complex conjugation. In fact, even if our discussion was mostly focused around real particles (with real $\tilde{p}$) we can analytically continue the above expression to complex mirror momentum to find that
\begin{equation}
    \big(\tilde{x}^\pm(\tilde{p})\big)^* = \frac{1}{\tilde{x}^{\mp}(\tilde{p}^*)}\,,
\end{equation}
that is a different reality condition from the string model. 
For massive particle it is necessary to consider such complex momenta. Indeed, there exist  mirror bound-states, whose complex momenta live in a ``strip'' of sorts in the $z$-torus~\cite{Arutyunov:2007tc}.
More precisely, the mirror physical region is defined by having $\mathfrak{I}[x^\pm]<0$. This region has therefore overlaps with the string physical region $|x^\pm|>1$.

\begin{figure}
    \centering
\begin{tikzpicture}
\fill[pattern=horizontal lines, pattern color=red] (-2cm,-2cm) rectangle (2cm,0);
\node[rounded corners, draw=black, fill=white, fill opacity=0.7, text opacity=1] at (1.6cm,1.6cm) {$\tilde{x}^+$};
\draw[thick] (0,0) circle (1cm);
\draw[->, thick] (-2cm,0)--(2cm,0);
\draw[->, thick] (0,-2cm)--(0,2cm);
\end{tikzpicture}
\begin{tikzpicture}
\fill[pattern=vertical lines, pattern color=blue] (-2cm,-2cm) rectangle (2cm,0);
\node[rounded corners, draw=black, fill=white, fill opacity=0.7, text opacity=1] at (1.6cm,1.6cm) {$\tilde{x}^-$};
\draw[thick] (0,0) circle (1cm);
\draw[->, thick] (-2cm,0)--(2cm,0);
\draw[->, thick] (0,-2cm)--(0,2cm);
\end{tikzpicture}\ \ 
\begin{tikzpicture}
\fill[pattern=vertical lines, pattern color=blue] (-3cm,-0.5cm) rectangle (3cm,0.5cm);
\fill[pattern=horizontal lines, pattern color=red] (-3cm,-1.5cm) rectangle (3cm,-0.5cm);
\draw[->, thick] (-3cm,0)--(3cm,0);
\draw[->, thick] (0,-2cm)--(0,2cm);
\draw[densely dashed,draw=gray] (-3cm,1cm)--(3cm,1cm);
\draw[densely dashed,draw=gray] (-3cm,-1cm)--(3cm,-1cm);
\node[rounded corners, draw=black, fill=white, fill opacity=0.7, text opacity=1] at (2.6cm,1.6cm) {$\tilde{\gamma}^\pm$};
\node[anchor=south west,rounded corners, fill opacity=0.7, text opacity=1] at (0,1cm) {$\scriptsize{\color{gray}+i\pi}$};
\node[anchor=north west,rounded corners, fill=white, fill opacity=0.6, text opacity=1] at (0,-1cm) {$\scriptsize{\color{gray}-i\pi}$};
\end{tikzpicture}
    \caption{The mirror region. In the leftmost panel, the mirror region for $x^+$ is $\mathfrak{I}[x^+]<0$; in the middle panel, for $x^-$ is $\mathfrak{I}[x^-]<0$; in the rightmost panel, we draw it for the $\gamma^+$ and $\gamma^-$ panel, where it takes the form of the strips between $-\tfrac{3}{2}i\pi$ and $-\tfrac{1}{2}i\pi$, and between $-\tfrac{1}{2}i\pi$ and $+\tfrac{1}{2}i\pi$, respectively.}
    \label{fig:mirrorregion}
\end{figure}
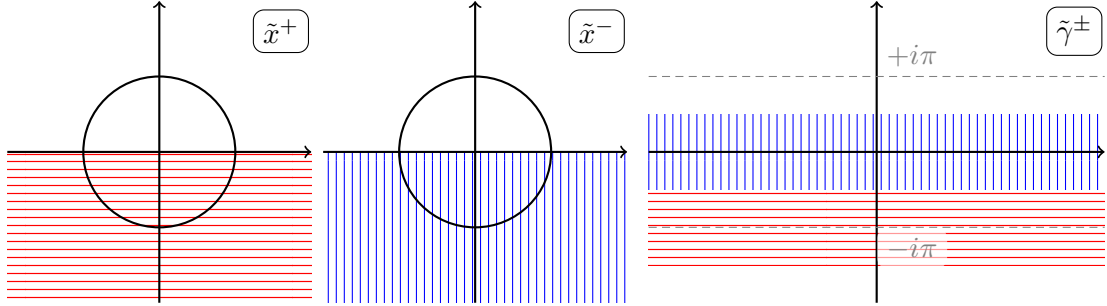

We now want to see how to parametrise the mirror kinematics in terms of the $\gamma^\pm$ variables, or of suitable mirror counterparts~$\tilde{\gamma}^\pm$. We begin by observing that $\mathfrak{I}[x^\pm]<0$ occurs when the rapidities~$\gamma^\pm$ live in the strips
\begin{equation}
    -\frac{3}{2}\pi+2k\pi<\mathfrak{I}[\gamma^+]<-\frac{1}{2}\pi+2k\pi\,,\qquad
    -\frac{1}{2}\pi+2k\pi<\mathfrak{I}[\gamma^-]<\frac{1}{2}\pi+2k\pi\,,\qquad
    k\in\mathbb{Z}\,.
\end{equation}
Once again, we choose the regions given by $k=0$, see figure~\ref{fig:mirrorregion}. We see that the mirror regions are shifted downwards by $-\tfrac{i}{2}\pi$ with respect to the physical region. 

To find $\tilde{\gamma}^\pm$ in the mirror theory let us once again start from real-momentum particles. First we want to start from $\gamma^\pm(z)$ using~\eqref{eq:stringgammaofx}, where the torus rapidity $z$ is on the real-string line, use it to define  $\tilde{\gamma}^\pm(\tilde{z})=\gamma^\pm(\tilde z+\tfrac{1}{2}\omega_2)$. It should be noted that the effect of the $\tfrac{1}{2}\omega_2$-shift on the $\gamma^\pm$ variables is not particularly simple (just like it was not particularly simple for~$x^\pm$). More specifically, it results in a $z$-dependent imaginary shift of $\gamma^\pm$. We find that
\begin{equation}
\begin{aligned}
    &\tilde{\gamma}^+(\tilde{z})=\gamma^+(\tilde z+\tfrac{1}{2}\omega_2)=
    \log \left(i \frac{\tilde{x}^+ - 1}{\tilde{x}^++1}\right)-i \pi,\\
    &\tilde{\gamma}^-(\tilde{z})=
    \gamma^-(\tilde{z}+\tfrac{1}{2}\omega_2) = \log\left(i\frac{ \tilde{x}^- - 1}{  \tilde{x}^-+1}\right),
\end{aligned}
\end{equation}
where $\tilde{x}^\pm=x^\pm(\tilde{z}+\tfrac{1}{2}\omega_2)$.
Using this definition it is also possible to verify that, under complex conjugation
\begin{equation}
\label{eq:massivemirrorconj}
    \left(\tilde{\gamma}^\pm(z)\right)^*
    =\tilde{\gamma}^\mp(z^*)+i\pi\,.
\end{equation}
While below we shall use the definition of the mirror rapidities~$\tilde{\gamma}^\pm$, in the mirror region it can be convenient to redefine $\widetilde{\gamma}^\pm= \tilde{\gamma}^\pm(\tilde{z})+ i\pi /2$, in such a way that~$(\widetilde{\gamma}^\pm)^* = \widetilde{\gamma}^\mp$. This makes the discussion of the mirror dressing factors quite a bit simpler, and in fact we will employ this redefinition when discussing the mirror thermodynamic Bethe ansatz~\cite{upcoming:mirror}.

\subsubsection{Crossing transformation}
In a similar spirit, we can describe the crossing transformation which takes us to the ``anti-string'' region and flips the sign of energy and momentum. We denote this transformation with a bar:
\begin{equation}
    \bar{E} = -E\,,\qquad
    \bar{p}=-p\,.
\end{equation}
As remarked it is not sufficient to describe crossing in terms of $x^\pm$, as the S~matrix has cuts in the $x^\pm$ plane.
Fortunately, the above transformations may  be easily realised on the $z$-torus by shifting $z$ by $\omega_2$ --- consistently with the relativistic intuition that the mirror transformation is ``half-crossing''. Hence we have%
\footnote{%
In analogy with what we have done for the mirror model, we may introduce the variable~$\bar{z}$; we avoid doing so because, unlike the mirror theory, the crossed theory is essentially identical to the original theory, \textit{i.e.}\ it has the same kinematics. Moreover, strictly speaking we may have wanted to define $\bar{E}(z)=-E(z+\omega_2)$ so that it is positive; we prefer to make the energy of the anti-string region negative to emphasise that it is not a physical region.}
\begin{equation}
    \bar{E}(z)=E(z+\omega_2)=-E\,,\qquad
    \bar{p}(z)=p(z+\omega_2)=-p\,,
\end{equation}
as well as
\begin{equation}
    \bar{x}^\pm(z)=x^\pm(z+\omega_2)=\frac{1}{x^\pm(z)}\,.
\end{equation}
The same prescription, in terms of the $\gamma^\pm$ variables, gives
\begin{equation}
    \bar{\gamma}^\pm(z)=\gamma^\pm(z+\omega_2)=\gamma^\pm(z)-i\pi\,.
\end{equation}
Interestingly, while the mirror transformation did not have a simple action on the $\gamma^\pm$, the crossing transformation simply amounts to a shift of $\gamma^\pm$ by $-i\pi$. This fact will play a crucial role in what follows.

We conclude this section by noting some roots of useful expressions involving two Zhukovsky variables, in terms of $\gamma^\pm$ rapidities. They are reported in table~\ref{table:zeros}.

\begin{table}[t]
\centering
\renewcommand{\arraystretch}{1.4}
\begin{tabular}{|l|lr|}
\hline
$x_1^\pm = x_2^\pm\quad$ & $\gamma_1^\pm=\gamma_2^\pm\quad$ &$\text{mod}\ 2\pi i$ \\
$x_1^\pm = \displaystyle\frac{1}{x_2^\pm}$ & $\gamma_1^\pm=\gamma_2^\pm+i\pi\quad$& $\text{mod}\ 2\pi i$ \\
$x_1^+ = x_2^-$ & $\gamma_1^+=\gamma_2^-+i\pi\quad$& $\text{mod}\ 2\pi i$ \\
$x_1^+ = \displaystyle\frac{1}{x_2^-}$ & $\gamma_1^+=\gamma_2^-$& $\text{mod}\ 2\pi i$ \\[0.3cm]
\hline
\end{tabular}
\caption{Some notable zeros of expressions involving $x^\pm_1$ and $x^\pm_2$, and their equivalent expressions in terms of the $\gamma^\pm_1$ and $\gamma^\pm_2$ rapidities. The physical interpretation of such zeros will depend on whether the correspondent rapidities fall in the physical region for string or mirror particles. We will return to this point in section~\ref{sec:proposal}.}
\label{table:zeros}
\end{table}

\subsection{Massless variables}
\label{sec:rapidities:massless}
We firstly observe that for massless variables the dispersion relation is not analytic,
\begin{equation}
    E=2h\,\left|\sin\Big(\frac{p}{2}\Big)\right|\,,
\end{equation}
and similarly the relation between the massless Zhukovsky variable and the momentum is not analytic,
\begin{equation}
    x_p =\text{sgn}\Big(\sin\tfrac{p}{2}\Big)\, e^{i p /2}\,,
\end{equation}
so that strictly speaking we need to treat separately the case of anti-chiral and chiral particles ($-\pi<p<0$ and $0<p<\pi$), see ref.~\cite{upcoming:massless} for a detailed discussion of this point. As we have summarised above, this means that the dressing factors which we will discuss correspond to a definite choice of chirality, and that the remaining ones arise by using parity and braiding unitarity.
Bearing all this in mind, in the massless case we set~\cite{Fontanella:2019baq}
\begin{equation}
\label{eq:gamma}
    x=\frac{i-e^{\gamma}}{i+e^{\gamma}}\,.
\end{equation}
Notice that for real momenta $x$ is on the upper half of the unit circle, which corresponds to $\gamma$ being real.
The relation between $x_p$ and the energy is analytic, so that in terms of $\gamma$ we have simply
\begin{equation}
    E(\gamma) = \frac{2h}{\cosh\gamma}\,,
\end{equation}
which is indeed positive semi-definite on the real line. Notice that the region where the momentum and the energy are small is mapped to (plus and minus) infinity in terms of $\gamma$. The map between the rapidity $\gamma$ and momentum is not analytic (as expected) and it is
\begin{equation}
    \gamma =\log\left(-\cot\frac{p}{4}\right)\,,\quad
    p=-2i\log\left(-\frac{i-e^{\gamma}}{i+e^{\gamma}}\right)\,,\qquad
    \text{for}\quad\gamma_p\geq0\,,\quad -\pi\leq p \leq 0\,,
\end{equation}
for anti-chiral particles, whereas for chiral particles
\begin{equation}
    \gamma =\log\left(\tan\frac{p}{4}\right)\,,\quad
    p=-2i\log\left(+\frac{i-e^{\gamma}}{i+e^{\gamma}}\right)\,,\qquad
    \text{for}\quad \gamma\leq0\,,\quad 0\leq p \leq \pi\,.
\end{equation}

\subsubsection{String region, mirror region, and crossing}
The discussion now is similar in spirit to what we have done for massive particles, with some important difference: massless particles cannot create bound states. As a consequence, it does not make sense to talk about a ``physical region'' for complex momenta. In fact, the physical region is just the line where momentum is real and the energy is positive. This occurs when $\gamma$ is on the real line of the string region. Despite this important difference, it still makes sense to require that we can analytically continue the rapidity to complex values and reach, for instance, the mirror or the anti-string (\textit{i.e.}, crossed) region. In fact, we have implicitly used the existence of such an analytic continuation when writing the crossing equation, while the possibility of analytically continuing the S~matrix to the mirror region is crucial to obtain the thermodynamic Bethe ansatz equations for excited states~\cite{Dorey:1996re}.

Let us now discuss the mirror ``region''. Once again, we expect this region to consist only of one line --- where the mirror energy~$\tilde{E}$ is positive semi-definite and  the mirror momentum~$\tilde{p}$ is real.
In the massless sector we do not have access to the $z$-torus parametrisation. Still, since we can uniformise the dispersion in terms of a single real variable, we can use any such parametrisation (\textit{e.g.}, in terms of $x$ or in terms of~$\gamma$) to define the mirror energy. Working in terms of $\gamma$, we find
\begin{equation}
    \tilde{E}(\tilde{\gamma}) = - i\,p(\tilde{\gamma}+\tfrac{i}{2}\pi)\,,\qquad
    \tilde{p}(\tilde{\gamma}) = - i\,E(\tilde{\gamma}+\tfrac{i}{2}\pi)\,,
\end{equation}
so that, bearing in mind that the two branches of $p$ result in two branches of~$\tilde{E}$,
\begin{equation}
    \tilde{E}(\tilde{\gamma}) = -2\log\Big|\tanh\frac{\tilde{\gamma}}{2} \Big|\,,\qquad \tilde{p}=-\frac{2h}{\sinh\tilde{\gamma}}\,.
\end{equation}
It is also worth working out the velocity of a mirror particle, which is
\begin{equation}
    \tilde{v}=\frac{\partial\tilde{E}}{\partial\tilde{p}} = \frac{\pm 2}{\sqrt{4h^2+\tilde{p}^2}}\,,
\end{equation}
where the sign is negative for anti-chiral particles, and positive for chiral ones. This is compatible with the $M=0$ expression in~\eqref{eq:mirrordispersion}.

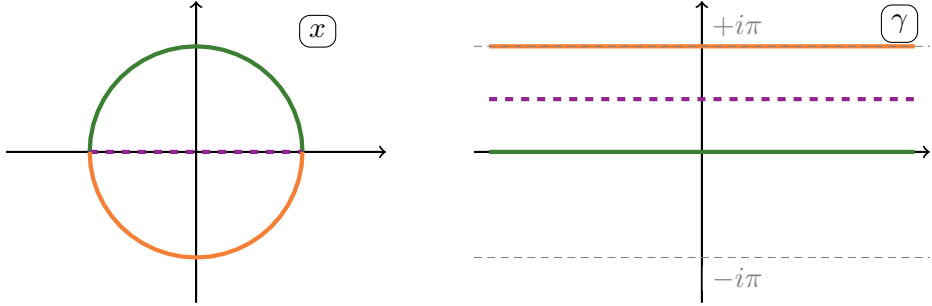
\begin{figure}
    \centering
\begin{tikzpicture}
\node[rounded corners, draw=black, fill=white, fill opacity=0.7, text opacity=1] at (1.6cm,1.6cm) {$x$};
\draw[->, thick] (-2.5cm,0)--(2.5cm,0);
\draw[->, thick] (0,-2cm)--(0,2cm);
\draw[ultra thick,OliveGreen,domain=0:180] plot ({1.4*cos(\x)}, {1.4*sin(\x)});
\draw[ultra thick, dashed,Plum,domain=-1:1] plot ({1.4*\x}, {0});
\draw[ultra thick,Orange,domain=180:360] plot ({1.4*cos(\x)}, {1.4*sin(\x)});
\end{tikzpicture}\hspace{1cm}
\begin{tikzpicture}
\draw[->, thick] (-3cm,0)--(3cm,0);
\draw[->, thick] (0,-2cm)--(0,2cm);
\draw[ultra thick,OliveGreen] (-2.8cm,0)--(2.8cm,0);
\draw[ultra thick,dashed,Plum] (-2.8cm,0.7cm)--(2.8cm,0.7cm);
\draw[ultra thick,Orange] (-2.8cm,1.4cm)--(2.8cm,1.4cm);
\draw[densely dashed,draw=gray] (-3cm,1.4cm)--(3cm,1.4cm);
\draw[densely dashed,draw=gray] (-3cm,-1.4cm)--(3cm,-1.4cm);
\node[rounded corners, draw=black, fill=white, fill opacity=0.7, text opacity=1] at (2.6cm,1.7cm) {$\gamma$};
\node[anchor=south west,rounded corners, fill opacity=0.7, text opacity=1] at (0,1.4cm) {$\scriptsize{\color{gray}+i\pi}$};
\node[anchor=north west,rounded corners, fill=white, fill opacity=0.6, text opacity=1] at (0,-1.4cm) {$\scriptsize{\color{gray}-i\pi}$};
\end{tikzpicture}
    \caption{The string, mirror and anti-string region in the massless kinematics. In all three cases, the ``region'' is actually a line, corresponding to real momentum particles. We denote the string region by a solid green line (upper-half-circle in the $x$-plane), the mirror region by a dashed purple line (real segment in the $x$-plane), and the anti-string region by a solid orange line (lower-half-circle in the $x$-plane).}
    \label{fig:masslessgammax}
\end{figure}

Some remarks are in order. In the mirror region, the momentum is no longer periodic; rather, it takes any real value. This is also what happens for massive particles. We see that large (positive or negative) values of the mirror momentum are mapped to (negative or positive) values of~$\tilde{\gamma}$ in the vicinity of zero. Conversely, when the mirror rapidity is large in modulus, the momentum of the particles is close to zero; here, the mirror energy vanishes, while the velocity is in modulus as large as it can be, $\tilde{v}=\pm 1/h$. Finally, we remark once again that the transformation that takes us from the real to the mirror line should be defined separately  for chiral and anti-chiral particles.

Let us now comment on the  qualitative behaviour of the Zhukovsky map under crossing. We have
\begin{equation}
    \tilde{x}(\tilde{\gamma})=x(\tilde{\gamma}+\tfrac{i}{2}\pi)=-\tanh{\frac{\tilde{\gamma}}{2}}\,.
\end{equation}
For real~$\tilde{\gamma}$, this takes values on the real line with~$-1<\tilde{x}<1$. In particular, the points $\tilde{x}=\pm1$ correspond to zero mirror momentum and mirror energy. In terms of the $x$ plane, we may think of taking the Zhukovski variable from the upper-half circle straight down to the mirror line, through a suitable path inside the unit disk.

In a similar way as we have discussed crossing, we may now define the crossing transformation to the anti-string region,
\begin{equation}
    \bar{E}(\gamma) = E(\gamma+i\pi)= - E(\gamma)\,,\qquad
    \bar{p}(\gamma) = p(\gamma+i\pi)=-p(\gamma)\,,
\end{equation}
under which we flip the sign of energy and momentum. In terms of the Zhukovsky variable $x$ this corresponds to taking
\begin{equation}
    \bar{x} = \frac{1}{x}\,,
\end{equation}
where now real value of the momentum corresponds to the lower-half-circle. The transformation that takes us from the string region to the anti-string (crossed) region therefore takes us inside the unit disk, through the real line between $(-1,1)$, and down to the lower-half-circle.%
\footnote{%
Note that it is different from the transformation used in ref.~\cite{Borsato:2016xns}, which went outside the unit disk. That choice was taken in analogy with the path of $x^+(z)$ in the massive case, but upon closer inspection we see that such a path does not go through the mirror region in the $M\to0$ limit.
} 
We represent the $x$- and $\gamma$-plane in figure~\ref{fig:masslessgammax}.

\begin{figure}
    \centering
\begin{tikzpicture}
\draw[->, thick] (-3cm,0)--(3cm,0);
\draw[->, thick] (0,-2cm)--(0,2cm);
\node[rounded corners, draw=black, fill=white, fill opacity=0.7, text opacity=1] at (2.8cm,1.8cm) {$\gamma^\pm(u)$};
\node[anchor=south west,rounded corners, fill opacity=0.7, text opacity=1] at (0,1cm) {$\scriptsize{\color{gray}+\tfrac{i}{h}}$};
\node[anchor=north west,rounded corners, fill=white, fill opacity=0.6, text opacity=1] at (0,-1cm) {$\scriptsize{\color{gray}-\tfrac{i}{h}}$};
\draw[loosely dashed,draw=gray] (-3cm,1cm)--(3cm,1cm);
\draw[loosely dashed,draw=gray] (-3cm,-1cm)--(3cm,-1cm);
\draw[blue, snake, ultra thick] (-1.5cm,1cm) -- (1.5cm,1cm);
\draw[red, snake, ultra thick] (-1.5cm,-1cm) -- (1.5cm,-1cm);
\fill[fill=black] (-1.5cm,1cm) circle (0.05cm);
\fill[fill=black] (+1.5cm,1cm) circle (0.05cm);
\fill[fill=black] (-1.5cm,-1cm) circle (0.05cm);
\fill[fill=black] (+1.5cm,-1cm) circle (0.05cm);
\end{tikzpicture} \hspace{1.5cm}
\begin{tikzpicture}
\draw[->, thick] (-3cm,0)--(3cm,0);
\draw[->, thick] (0,-2cm)--(0,2cm);
\node[rounded corners, draw=black, fill=white, fill opacity=0.7, text opacity=1] at (2.8cm,1.8cm) {$\tilde{\gamma}^\pm(u)$};
\node[anchor=south west,rounded corners, fill opacity=0.7, text opacity=1] at (0,1cm) {$\scriptsize{\color{gray}+\tfrac{i}{h}}$};
\node[anchor=north west,rounded corners, fill=white, fill opacity=0.6, text opacity=1] at (0,-1cm) {$\scriptsize{\color{gray}-\tfrac{i}{h}}$};
\draw[loosely dashed,draw=gray] (-3cm,1cm)--(3cm,1cm);
\draw[loosely dashed,draw=gray] (-3cm,-1cm)--(3cm,-1cm);
\draw[blue, snake, ultra thick] (-1.5cm,1cm) -- (-3cm,1cm);
\draw[red, snake, ultra thick] (-1.5cm,-1cm) -- (-3cm,-1cm);
\draw[blue, snake, ultra thick] (1.5cm,1cm) -- (3cm,1cm);
\draw[red, snake, ultra thick] (1.5cm,-1cm) -- (3cm,-1cm);
\fill[fill=black] (-1.5cm,1cm) circle (0.05cm);
\fill[fill=black] (+1.5cm,1cm) circle (0.05cm);
\fill[fill=black] (-1.5cm,-1cm) circle (0.05cm);
\fill[fill=black] (+1.5cm,-1cm) circle (0.05cm);
\end{tikzpicture}
    \caption{The $u$ plane. In the left panel, the function $\gamma^+(u)$ has branch points at $u=\pm 2-\tfrac{i}{h}$ and the branch cut runs on the red wavy line; the function $\gamma^-(u)$ has branch points at $u=\pm 2+\tfrac{i}{h}$ and the branch cut runs on the blue wavy line. In the right panel, we make a similar drawing for $\tilde{\gamma}^\pm(u)$; this time the branch cuts are ``long'', \textit{i.e.}\ they go to infinity.}
    \label{fig:uplane}
\end{figure}
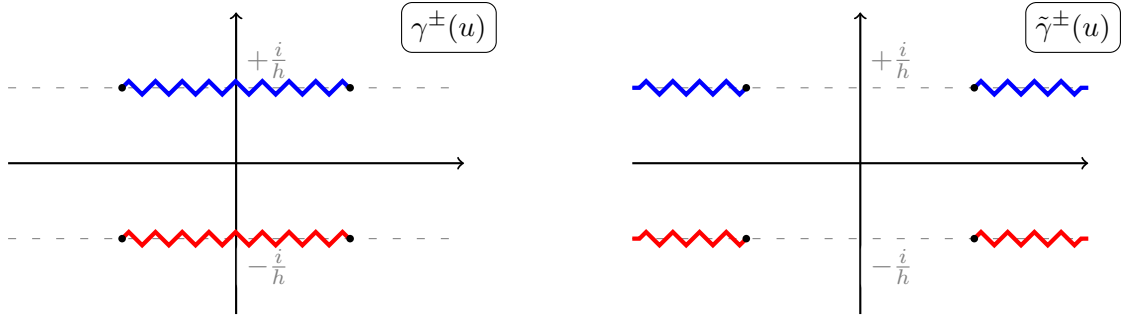

\subsection{The $u$-plane}
\label{sec:rapidities:uplane}
It can be useful to introduce the rapidity $u$, which for massive particles takes the form
\begin{equation}
    u= x^++\frac{1}{x^+}-\frac{i}{h}=x^-+\frac{1}{x^-}+\frac{i}{h}\,.
\end{equation}
We can therefore parametrise $\gamma^\pm(u)$ on the $u$-plane with cuts, for particles in the string or mirror theory. We get for the string $u$-plane variable
\begin{equation}
\label{eq:gammas}
\gamma^+(u)=\frac{1}{2} \log \left(\frac{u+\frac{i}{h}-2}{u+\frac{i}{h}+2}\right)- \frac{i \pi}{2} \,,\qquad \gamma^-(u)=\frac{1}{2} \log \left(\frac{u-\frac{i}{h}-2}{u-\frac{i}{h}+2}\right)+\frac{i \pi}{2}\,.
\end{equation}
For the mirror $u$-plane variable, we have instead
\begin{equation}
\label{eq:gammaum}
\tilde{\gamma}^+(u)=\frac{1}{2} \log \left(-\frac{u+\frac{i}{h}-2}{u+\frac{i}{h}+2}\right) -i\pi\,,\qquad \tilde{\gamma}^-(u)=\frac{1}{2} \log \left(-\frac{u-\frac{i}{h}-2}{u-\frac{i}{h}+2}\right)\,.
\end{equation}
As a consequence, the cuts in the string kinematics are \textit{short}, while in the mirror kinematics they are \textit{long} (\textit{i.e.}, they go through infinity, see figure~\ref{fig:uplane}). This is very reminiscent of what happens in $AdS_5\times S^5$. Here the branch points are logarithmic: analytic continuation through a cut of $\gamma^\pm(u)$ from below to the same point $u$ of the next plane decreases $\gamma^\pm(u)$ by $i\pi$.
We see that in the string region physical particles have real rapidity $u$. The same is true for mirror particles. In terms of $u$ the complex-conjugation rule reads
\begin{equation}
    (\gamma^+(u))^*= \gamma^-(u^*)\,,\qquad
    (\tilde{\gamma}^+(u))^*= \tilde{\gamma}^-(u^*)+i\pi\,.
\end{equation}

In a similar way, we may define the $u$-rapidity for massless particles,
\begin{equation}
    u = x+\frac{1}{x}\,.
\end{equation}
In analogy with the massive case, we set
\begin{equation}
    \gamma(u)=\frac{1}{2} \log \left(\frac{u-2}{u+2}\right)- \frac{i \pi}{2} \,,\qquad 
    \tilde{\gamma}(u)=\frac{1}{2} \log \left(-\frac{u-2}{u+2}\right)+ \frac{i \pi}{2}\,,
\end{equation}
so that in the string region $\gamma(u)$ has a short cut, and $\tilde{\gamma}(u)$ has  a long cut. In the string region, the cut runs from $-2$ to $+2$, and real values of $\gamma(u)$ occur just above the cut. In the mirror region, the cut runs between minus infinity and $-2$, and between $2$ and plus infinity. Real values of $\tilde{\gamma}$ occur just above the cut.
The two functions are related as follows
\begin{equation}
    \tilde{\gamma}(u+i0) = \gamma(u+i0)+\frac{i}{2}\pi \,,
\end{equation}
and any of them can be used in both regions.

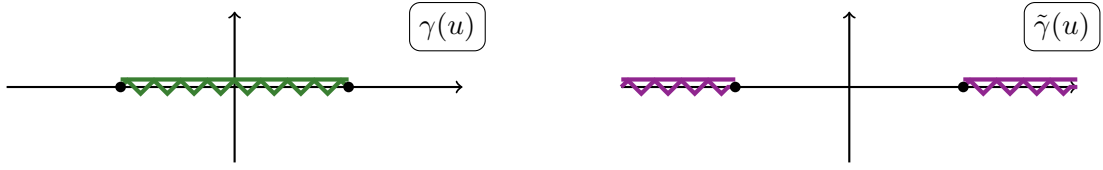
\begin{figure}
    \centering
\begin{tikzpicture}
\draw[->, thick] (-3cm,0)--(3cm,0);
\draw[->, thick] (0,-1cm)--(0,1cm);
\node[rounded corners, draw=black, fill=white, fill opacity=0.7, text opacity=1] at (2.8cm,0.8cm) {$\gamma(u)$};
\draw[OliveGreen, snake, ultra thick] (-1.5cm,0cm) -- (1.5cm,0cm);
\draw[OliveGreen, ultra thick] (-1.5cm,0.1cm) -- (1.5cm,0.1cm);
\fill[fill=black] (-1.5cm,0cm) circle (0.07cm);
\fill[fill=black] (+1.5cm,0cm) circle (0.07cm);
\end{tikzpicture} \hspace{1.5cm}
\begin{tikzpicture}
\draw[->, thick] (-3cm,0)--(3cm,0);
\draw[->, thick] (0,-1cm)--(0,1cm);
\node[rounded corners, draw=black, fill=white, fill opacity=0.7, text opacity=1] at (2.8cm,0.8cm) {$\tilde{\gamma}(u)$};
\draw[Plum, snake, ultra thick] (-3cm,0cm) -- (-1.5cm,0cm);
\draw[Plum, snake, ultra thick] (1.5cm,0cm) -- (3cm,0cm);
\draw[Plum, ultra thick] (-3cm,0.1cm) -- (-1.5cm,0.1cm);
\draw[Plum, ultra thick] (1.5cm,0.1cm) -- (3cm,0.1cm);
\fill[fill=black] (-1.5cm,0cm) circle (0.07cm);
\fill[fill=black] (+1.5cm,0cm) circle (0.07cm);
\end{tikzpicture}
    \caption{The $u$ plane for massless particles. In the left panel, the function $\gamma(u)$ has branch points at $u=\pm 2$ and the branch cut runs on the green wavy line; real momentum and positive energy corresponds to $\gamma(u)$ just above the cut. In the right panel, we make a similar drawing for $\tilde{\gamma}(u)$; this time the branch cuts are ``long'', \textit{i.e.}\ they go to infinity, and real values of the mirror momentum as well as positive values of the mirror energy correspond to $\tilde{\gamma}(u)$ just above the cut.}
    \label{fig:uplanemassless}
\end{figure}

\subsection{Crossing equations and rapidity-difference solutions}
\label{sec:rapidities:crossing}
It is convenient to rewrite the crossing equations in terms of these rapidity variables. We will see that all equations can be formally solved in terms of two functions which depend only on the \textit{difference} of the rapidities.
To write more compact formulae, let us introduce the short-hand notation for the difference of two rapidities
\begin{equation}
\label{eq:gammashorthand}
    \gamma^{ab}_{12}=\gamma_1^a-\gamma_2^b\,,\qquad a=\pm,\quad b=\pm\,,
\end{equation}
as well as
\begin{equation}
    \gamma^{\pm\circ}_{12}=\gamma_1^\pm-\gamma_2\,, \qquad
    \gamma^{\circ\pm}_{12}=\gamma_1-\gamma_2^\pm\,,\qquad
    \gamma^{\circ\circ}_{12}=\gamma_1-\gamma_2\,.
\end{equation}
 
\subsubsection{Massive sector}
To begin with, we have
\begin{equation}
    \Big(\sigmap(\bar{\gamma}_1^\pm,\gamma_2^\pm)\Big)^{-2}\Big(\sigmap(\gamma_1^\pm,\gamma_2^\pm)\Big)^{-2}
    =
    \coth\frac{\gamma_{12}^{++}}{2}
    \coth\frac{\gamma_{12}^{+-}}{2}
    \coth\frac{\gamma_{12}^{-+}}{2}
    \coth\frac{\gamma_{12}^{--}}{2}\,.
\end{equation}
Similarly, we have for the monodromy equation
\begin{equation}
    \frac{\Big(\sigmam(\bar{\gamma}_1^\pm,\gamma_2^\pm)\Big)^{-2}}{\Big(\sigmam(\gamma_1^\pm,\gamma_2^\pm)\Big)^{-2}}=
    \frac{\sinh\gamma_{12}^{+-}\ \sinh\gamma_{12}^{-+}}{\sinh\gamma_{12}^{++}\ \sinh\gamma_{12}^{--}}\,.
\end{equation}
We can solve these equations \textit{in terms of an expression of difference form} in $\gamma$ by requiring that
\begin{equation}
    \Big(\sigmap(\gamma_1^\pm,\gamma_2^\pm)\Big)^{-2}=F^+_{\text{CDD}}(\gamma_1^\pm,\gamma_2^\pm)\,\PhiSG(\gamma^{++}_{12})\,\PhiSG(\gamma^{+-}_{12})\,\PhiSG(\gamma^{-+}_{12})\,\PhiSG(\gamma^{--}_{12})\,,
\end{equation}
with
\begin{equation}
\label{eq:varPhicrossing}
    \PhiSG(\gamma-i \pi) \PhiSG(\gamma)=i \coth\frac{\gamma}{2}\,,\qquad
    \PhiSG(\gamma) \PhiSG(\gamma+i\pi)=i \tanh\frac{\gamma}{2}\,,
\end{equation}
and where $F^+_{\text{CDD}}(\gamma_1^\pm,\gamma_2^\pm)$ is an undetermined CDD factor~\cite{Castillejo:1955ed}, \textit{i.e.}\ a solution of the homogeneous (trivial) crossing equation and of the unitarity conditions (which could also be of difference form). 
Furthermore, to satisfy the unitarity conditions it must be
\begin{equation}
    \PhiSG(\gamma)\,\PhiSG(-\gamma)=1\,,\qquad
    \PhiSG(\gamma)^*=\frac{1}{\PhiSG(\gamma^*)}\,,
\end{equation}
with similar unitarity conditions holding separately for $F^+_{\text{CDD}}(\gamma_1^\pm,\gamma_2^\pm)$ too.
Similarly, the monodromy equation can be solved in terms of
\begin{equation}
    \Big(\sigmam(\gamma_1^\pm,\gamma_2^\pm)\Big)^{-2}=
    F^-_{\text{CDD}}(\gamma_1^\pm,\gamma_2^\pm)\,
    \frac{\PhiMON(\gamma_{12}^{++})\,\PhiMON(\gamma_{12}^{--})}{\PhiMON(\gamma_{12}^{+-})\,\PhiMON(\gamma_{12}^{-+})}\,,
\end{equation}
provided that it satisfies
\begin{equation}
\label{eq:varPsimonodromy}
    \frac{\PhiMON(\gamma+i \pi)}{\PhiMON(\gamma)}=2i\,\sinh\gamma\,,\qquad
    \frac{\PhiMON(\gamma-i \pi)}{\PhiMON(\gamma)}=\frac{i}{2\sinh\gamma}\,,
\end{equation}
as well as
\begin{equation}
    \PhiMON(\gamma)\,\PhiMON(-\gamma)=1\,,\qquad
    \PhiMON(\gamma)^*=\frac{1}{\PhiMON(\gamma^*)}\,,
\end{equation}
and $F^-_{\text{CDD}}(\gamma_1^\pm,\gamma_2^\pm)$ is another yet-to-be-determined CDD factor.

\subsubsection{Mixed-mass sector}
The mixed-mass crossing equations take the form
\begin{equation}
\label{eq:crossingmixedgamma}
\begin{aligned}
    \Big(\sigmabc(\bar{\gamma}_1^\pm,\gamma_2)\Big)^{-2} \Big(\sigmabc(\gamma_1^\pm,\gamma_2)\Big)^{-2}&=
    \coth\frac{\gamma_{12}^{+\circ}}{2}\coth\frac{\gamma_{12}^{-\circ}}{2}\,,
    \\
    \Big(\sigmacb(\bar{\gamma}_1,\gamma_2^\pm)\Big)^{-2} \Big(\sigmacb(\gamma_1,\gamma_2^\pm)\Big)^{-2}&=
    \tanh\frac{\gamma_{12}^{\circ+}}{2}\tanh\frac{\gamma_{12}^{\circ-}}{2}\,,\\
    \Big(\sigmacb(\gamma_1,\bar{\gamma}_2^\pm)\Big)^{-2} \Big(\sigmacb(\gamma_1,\gamma_2^\pm)\Big)^{-2}&=
    \tanh\frac{\gamma_{12}^{\circ+}}{2}\tanh\frac{\gamma_{12}^{\circ-}}{2},\\
    \Big(\sigmabc(\gamma_1^\pm,\bar{\gamma}_2)\Big)^{-2} \Big(\sigmabc(\gamma_1^\pm,\gamma_2)\Big)^{-2}&=
    \coth\frac{\gamma_{12}^{+\circ}}{2}\coth\frac{\gamma_{12}^{-\circ}}{2}\,.
\end{aligned}
\end{equation}
We start by assuming that we can solve the first equation by setting
\begin{equation}
    \Big(\sigmabc(\gamma_1^\pm,\gamma_2)\Big)^{-2}=F_{\text{CDD}}(\gamma_{12}^{\pm\circ})\,\PhiSG(\gamma_{12}^{+\circ})\,\PhiSG(\gamma_{12}^{-\circ})\,,
\end{equation}
where like before $\varPhi(\gamma)$ satisfies eq.~\eqref{eq:varPhicrossing}. By braiding unitarity we necessarily have that
\begin{equation}
    \Big(\sigmacb(\gamma_1,\gamma_2^\pm)\Big)^{-2}=\frac{1}{F_{\text{CDD}}(\gamma_{21}^{\pm\circ})}\,\PhiSG(\gamma_{12}^{\circ+})\,\PhiSG(\gamma_{12}^{\circ-})\,,
\end{equation}
which solves the relevant crossing equation consistently with our prescription
\begin{equation}
    \bar{\gamma}=\gamma+i\pi\,,
\end{equation}
in contrast with what we did for the massive case. The last two equations in~\eqref{eq:crossingmixedgamma} are then automatically satisfied.

\subsubsection{Massless sector}
Finally, for the massless-massless kinematics we have
\begin{equation}
\label{eq:masslesscrossingrapid}
\begin{aligned}
    &\Big(\sigmacc(\bar{\gamma}_1,\gamma_2)\Big)^{-2} \Big(\sigmacc(\gamma_1,\gamma_2)\Big)^{-2}
    =
    \tanh^2\frac{\gamma_{12}^{\circ\circ}}{2}\,\\
    &\Big(\sigmacct(\bar{\gamma}_1,\gamma_2)\Big)^{-2} \Big(\sigmacct(\gamma_1,\gamma_2)\Big)^{-2}
    =
    \tanh^2\frac{\gamma_{12}^{\circ\circ}}{2}\,.
\end{aligned}
\end{equation}
These equations look identical but there is an important difference in how we may solve them. In the case of opposite-chirality scattering we can simply set
\begin{equation}
    \Big(\sigmacct(\gamma_1,\gamma_2)\Big)^{-2} = i\,\Big(\PhiSG(\gamma_{12}^{\circ\circ})\Big)^2\,,
\end{equation}
as usual up to a CDD factor.
However, this is not possible for same-chirality scattering. Indeed in that case braiding unitarity forbids such a solution. Instead we need to set
\begin{equation}
\label{eq:masslesssamechirsol}
    \Big(\sigmacc(\gamma_1,\gamma_2)\Big)^{-2} = 
    a(\gamma_{12}^{\circ\circ})\,\Big(\PhiSG(\gamma_{12}^{\circ\circ})\Big)^2\,,
\end{equation}
where we have
\begin{equation}
    a(\gamma)\,a(\gamma+i \pi) = -1\,,\qquad a(\gamma)\,a(-\gamma)=1\,,\qquad
    a(\gamma)^*=\frac{1}{a(\gamma^*)}\,.
\end{equation}
It is also possible and perhaps advisable to use a solution of the form~\eqref{eq:masslesssamechirsol} also for the opposite-chirality scattering. 

In~\cite{Fontanella:2019ury}, the massless-massless crossing equation was already rewritten and solved in the difference-form parametrisation of~\cite{Fontanella:2019baq}. However their result  did not apparently feature any nontrivial function~$a(\gamma)$ (nor a factor of $\pm i$).
Yet, such a factor seems unavoidable for same-chirality scattering. Besides, the S~matrix should take specific values when one of the momenta is zero, which also constrains the normalisation~\cite{upcoming:massless}. 
We shall return to this point in section~\ref{sec:proposal:massless} when discussing the analytic properties of our proposed solutions.

\section{Building blocks of the dressing factors}
\label{sec:buildingblocks}
We have assumed that the crossing equations may be solved in terms of the BES phase (appropriately generalised to the massless kinematics where needed) together with two functions $\varPhi(\gamma)$ and $\PhiMON(\gamma)$ whose argument will be the difference of the rapidities introduced in section~\ref{sec:rapidities:massive}. Here we present the main features of these functions, relegating some details of their derivation to the appendices.

\subsection{The Beisert-Eden-Staudacher factor}
\label{sec:buildingblocks:BES}
The BES dressing factor for massive particles can be represented as
\begin{equation}
    \BES(x^\pm_1,x^\pm_2) = e^{i\theta(x^\pm_1,x^\pm_2)}\,,
\end{equation}
where
\begin{equation}
\label{eq:thetaBES}
\theta(x_1^+,x_1^-,x_2^+,x_2^-) =\chi(x_1^+,x_2^+)
-\chi(x_1^+,x_2^-)-\chi(x_1^-,x_2^+)+\chi(x_1^-,x_2^-)\,.
\end{equation}
For $|x_1|>1$ and $|x_2|>1$ the function $\chi(x_1,x_2)$ is given by the integral representation~\cite{Dorey:2007xn}
\begin{equation}
\label{eq:chiBES}
\begin{aligned}
&\chi(x_1,x_2)=\Phi(x_1,x_2)\,,
\qquad|x_1|>1\,,\quad|x_2|>1\,,\\
&\Phi(x_1,x_2)=i\oint\frac{{\rm d}w_1}{2\pi i}\oint \frac{{\rm
d}w_2}{2\pi i}\frac{1}{(w_1-x_1)(w_2-x_2)} \log\frac{\Gamma\big[1+\tfrac{ih}{
2}\big(w_1+\tfrac{1}{w_1}-w_2-\tfrac{1}{w_2}\big)\big]}{
\Gamma\big[1-\tfrac{ih}{
2}\big(w_1+\tfrac{1}{w_1}-w_2-\tfrac{1}{w_2}\big)\big]}\, ,
\end{aligned}
\end{equation}
which is skew-symmetric. As we mentioned, the BES factor satisfies the crossing equation
\begin{equation}
\label{eq:crossingBES}
\BES (x^\pm_1,x^\pm_2)\BES (\bar{x}^\pm_1,x^\pm_2)=\frac{x_2^-}{x_2^+}g(x_1^\pm,x_2^\pm)\,,
\end{equation}
see eq.~\eqref{eq:gfunction}. It is worth emphasising that the BES factor is regular as long as $|x_i|>1$. Moreover, the function~\eqref{eq:chiBES} can be continued inside the unit circle, even though in that region there are singularities whose position depends on the coupling~$h$. Still, no singularities occur outside  a disk of radius $(\sqrt{1+(2h)^{-2}}-(2h)^{-1})$, so that the continuation is straightforward in an annulus around the unit circle. See~\cite{Arutyunov:2009kf} for a discussion of the analytic properties of $\chi(x_1,x_2)$.

It is also worth recalling that the BES factor admits a well-known large-$h$ asymptotic expansion.%
\footnote{The BES factor also admits a rather well-behaved small-$h$ expansion, which is very important in AdS5/CFT4 but which will be less useful for us at least presently, as the CFT2 dual of our model is unknown.}
Specifically, the function $\Phi(x_1,x_2)$ can be expanded in terms of the AFS phase~\cite{Arutyunov:2004vx} and of the Hernandez-Lopez (HL) one~\cite{Hernandez:2006tk},
\begin{equation}
    \Phi(x_1,x_2) = \Phi_{\text{AFS}}(x_1,x_2) + \Phi_{\text{HL}}(x_1,x_2) +\cdots\,.
\end{equation}
The leading term scales with~$h$ and, when $|x_1|>1$, $|x_2|>1$, it takes the form
\begin{equation}
\label{eq:AFS}
    \Phi_{\text{AFS}}(x_1,x_2) 
    =\frac{h}{2} \left[\frac{1}{x_1}-\frac{1}{x_2}+\big(-x_1+x_2+\frac{1}{x_2}-\frac{1}{x_1}\big) \log \big(1-\frac{1}{x_1 x_2}\big)\right]\,,
\end{equation}
while the sub-leading one takes the form
\begin{equation}
\label{eq:HL}
    \Phi_{\text{HL}}(x_1,x_2) =- \frac{\pi}{2}\oint\frac{{\rm d}w_1}{2\pi i}\oint \frac{{\rm
d}w_2}{2\pi i}\frac{\text{sgn}\big(w_1+\frac{1}{w_1}-w_2-\frac{1}{w_2}\big)}{(w_1-x_1)(w_2-x_2)}\,,
\end{equation}
which can be computed explicitly in terms of dilogarithms~\cite{Arutyunov:2006iu,Beisert:2006ib}.

Finally, it is worth recalling that the BES factor does not satisfy phsyical unitarity in the mirror region. Rather it obeys~\cite{Arutyunov:2007tc, Arutyunov:2009kf}
\begin{equation}
    \big(\BES(x_{1,m}^\pm,x_{2,m}^\pm)\big)^{-2}
    \Big(\big(\BES(x_{1,m}^\pm,x_{2,m}^\pm)\big)^{-2}\Big)^*
    =
    e^{-2ip_1}e^{+2ip_2}\,.
\end{equation}
This factor precisely compensates the one arising from the normalisation~\eqref{eq:massivenorm}. In fact, this can be extended to the case of the dressing factor describing the scattering of two mirror bound states with bound state numbers $Q_1\geq1$ and $Q_2\geq1$, in which case~\cite{Arutyunov:2009kf}
\begin{equation}
\label{eq:boundstatemirrorBES}
    \big(\BES(\tilde{x}_{1}^\pm,\tilde{x}_{2}^\pm)\big)^{-2}
    \Big(\big(\BES(\tilde{x}_{1}^\pm,\tilde{x}_{2}^\pm)\big)^{-2}\Big)^*
    =
    e^{-2iQ_2p_1}e^{+2iQ_1p_2}\,,
\end{equation}
where $\tilde{x}_i^\pm$ are real mirror particles.
More generally, when considering the mirror region, it is often useful to introduce an ``improved'' BES phase~\cite{Arutyunov:2009kf}, which is unitary in the mirror region and has simple properties when considering bound states of the mirror theory (it behaves nicely under the fusion procedure for such bound states). While this improved phase has good properties in the mirror kinematics, it is not convenient to study the original (string-theory) kinematics. In this work we are mainly interested in the string-theory kinematics, hence we postpone most of the discussion of the mirror dressing factors to future work~\cite{upcoming:mirror}.

\subsubsection{Mixed-mass kinematics}
We want to define the BES factor for the case where one particle, e.g. the first one, is massless, that is when
\begin{equation}
    |x_1^+|=1\,,\qquad \mathfrak{I}[x^+_1]\geq0\,,\qquad
    x_1^-\,x_1^+=1\,.
\end{equation}
For finite $h$ we can just use the fact that if one of the circles in~\eqref{eq:chiBES} is of unit radius then the radius of the second
circle can be reduced up to $(\sqrt{1+(2h)^{-2}}-(2h)^{-1})$. Then we can place $x_1^\pm$ in  \eqref{eq:chiBES} on the unit circle, and get a representation for the mixed-mass BES phase for real momentum.
Consider now the crossing equation \textit{when we cross the massive particle}. In this case, the massless rapidity is a spectator, and we can continue it to take values on the unit circle without encountering any obstruction. This may be done in each step of the derivation of the crossing equation of ref.~\cite{Arutyunov:2009kf}, finding 
\begin{equation}
    \sigma(\bar{x}^\pm_1,x_2)\,\sigma(x^\pm_1,x_2)=\frac{1}{x_2^2}\frac{f(x_1^+,x_2)}{f(x_1^-,x_2)}\,,
\end{equation}
where the right-hand side is the limit of the right-hand-side of eq.~\eqref{eq:crossingBES} when $x_2^\pm\to (x_2)^{\pm1}$.
Things are not so straightforward when considering the crossing transformation for the \textit{massless} variable, because it is a priori not obvious which path we should follow. Let us recall that, for massive particles, the analytic continuation that yields crossing takes a path inside the unit circle of the $x$ plane. In that region, the BES phase has cuts~\cite{Arutyunov:2009kf}. Depending on the mass $M$ of the particle which we are considering, we need to follow different paths to perform the crossing transformation. More precisely, for $M$-particle bound states, the crossing equation is reproduced when crossing \textit{exactly $M$} branch cuts~\cite{Arutyunov:2009kf} --- and for a fundamental particle with $M=1$, it is reproduced when crossing exactly one cut. Postponing a more detailed discussion of the analytic properties of the mixed-mass and massless-massless BES phase to appendix~\ref{app:BES}, in this case we find that, as long as $h$ is finite, it is possible to perform the crossing transformation \textit{without crossing any of the cuts inside the unit disk}. This is consistent with the mass-$M$ particle crossing for the case $M=0$. In this way, we find that the crossing equation becomes trivial,
\begin{equation}
\label{eq:mixedmasstrivialcrossing}
    \BES(\bar{x}_1,x_2^\pm)\,\BES(x_1,x_2^\pm)=1\,.
\end{equation}

While the preceding discussion is perfectly fine for finite $h$, it is not well-suited for the asymptotic large-$h$ expansion. To have a representation well-defined for any $h$ we can analytically continue the BES phase to complex momentum keeping the relation $x_1^-=1/x_1^+$. An advantage of this way is that for complex $p$ both circles can be unit, and taking the large-$h$ limit is the same as for massive particles. The only difference is that if, for example, $|x^+_1|<1$ then $|x^-_1|>1$, which means that the integral representation of $\theta(x_1^\pm,x_2^\pm)$ in terms of $\Phi(x_1,x_2)$ --- \textit{cf.}\ eqs.~\eqref{eq:thetaBES} and \eqref{eq:chiBES} --- needs to be amended. By analytic continuation we find that in this region
\begin{equation}
\label{eq:massless-massive-BES}
    \begin{aligned}
    \theta(x_1,x_2^\pm) &=\Phi(x_1,x_2^+)
-\Phi(x_1,x_2^-)-\Phi\big(\frac{1}{x_1},x_2^+\big)+\Phi\big(\frac{1}{x_1},x_2^-\big)
\\
&+\Psi\big(\frac{1}{x_1},x_2^+\big)-\Psi\big(\frac{1}{x_1},x_2^-\big)\,,
\\&=2\Phi(x_1,x_2^+)
-2\Phi(x_1,x_2^-)-\Phi(0,x_2^+)+\Phi(0,x_2^-)
\\
&+\Psi({ x_1},x_2^+)-\Psi({ x_1},x_2^-)\,,
    \end{aligned}
\end{equation}
where
\begin{equation}
    \Psi(x_1,x_2)=i\oint\frac{{\rm d
}w}{2\pi i} \frac{1}{w-x_2}\log\frac{\Gamma\big[1+\frac{ih}{2}\big(x_1+\frac{1}{x_1}-w-\frac{1}{ w}\big)\big]}
{\Gamma\big[1-\frac{ih}{2}\big(x_1+\frac{1}{x_1}-w-\frac{1}{ w}\big)\big]}\,,
\end{equation}
and we used the identity  $\Phi(x_1,x_2)+\Phi(\tfrac{1}{x_1},x_2)=\Phi(0,x_2)$ and the properties of the $\Psi$ function to write the result in a slightly simpler way in the last equality of~\eqref{eq:massless-massive-BES}.
This formula will allow us to expand the phase at strong coupling, $h\gg1$. 
In appendix~\ref{app:BESexpansion} we discuss how to derive the asymptotic expansion  for~\eqref{eq:massless-massive-BES} when $|x_1|=1$.

We also will need to consider the dressing factor in the mirror kinematics. Following the logic of~\cite{Arutyunov:2009kf}, this can be obtained by analytic continuation from the string region to the mirror region (where for massless particles these ``regions'' are lines). To perform this continuation in the massless variable we need to give a prescription because we may potentially encounter branch points. In analogy with what was done for the crossing transformation, we perform the continuation by avoiding all cuts. The resulting expression is not much different from the one for the BES (massive) factor in the mirror-mirror region, and we present it in appendix~\ref{app:BES}. In the same appendix, using that computation we establish the behaviour of the mirror-mirror mixed-mass phase under complex conjugation, which turns out to be
\begin{equation}
\label{eq:mixedmirrorBES}
    \big(\BES(\tilde{x}_{1},\tilde{x}_{2}^\pm)\big)^{-2}
    \Big(\big(\BES(\tilde{x}_{1},\tilde{x}_{2}^\pm)\big)^{-2}\Big)^*
    =
    e^{-2ip_1}\,,
\end{equation}
where $\tilde{x}_1$ and $\tilde{x}_{2}^\pm$ are in the mirror kinematics (with real mirror momentum; we may also generalise this to complex momenta).
Interestingly, this is equivalent to the ``mass-to-zero'' limit of the mirror bound state relation~\eqref{eq:boundstatemirrorBES} obtained by formally taking $Q_1\to0$ , while $Q_2=1$.

\subsubsection{Massless kinematics}
In a similar way, we want to define the BES factor when \textit{both} particles are massless. By the same logic as above, for finite $h$ we can deform both integration contours in~\eqref{eq:chiBES}, as long as they lie outside a disk of radius $(\sqrt{1+(2h)^{-2}}-(2h)^{-1})$. It is therefore straightforward to derive the crossing equation in this kinematics. We start from the crossing equation for the mixed-mass kinematics~\eqref{eq:mixedmasstrivialcrossing}, and continue analytically $x_2^\pm\to (x_2)^{\pm1}$ where $|x_2|=1$. Like we argued before, since $x_2$ is a spectator in the equation, we can straightforwardly take the limit in the right-hand side, which in this case happens to be trivial. Hence
\begin{equation}
    \BES(\bar{x}_1,x_2)\,\BES(x_1,x_2)=1\,.
\end{equation}

Once again, for the purpose of a large-$h$ expansion this representation is not convenient. Instead, we repeat what we did for the mixed-mass sector. Starting now from eq.~\eqref{eq:massless-massive-BES} we take $x_2^\pm$ to satisfy $x_2^+=1/x_2^-$, approaching the unit circle with $|x_2^+|>1$. Therefore, we need  to analytically continue~\eqref{eq:massless-massive-BES} because $x_2^-$ is inside the circle. We have then, for $|x_1|=|x_2|=1$
\begin{equation}
\label{eq:masslessBES}
    \theta(x_1,x_2)
    =2\Phi(x_1,x_2) -2\Phi\big(x_1,\frac{1}{x_2}\big)-\Phi(0,x_2)+\Phi\big(0,\frac{1}{x_2}\big)
    +\Psi(x_1,x_2)-\Psi\big(x_1,\frac{1}{x_2}\big)\,,
\end{equation}
where we used the identity $\Phi(x_1,x_2)+\Phi(\tfrac{1}{x_1},x_2)=\Phi(0,x_2)$ to somewhat simplify the expression for $\theta(x_1,x_2)$.
We can then proceed to expand the various terms asymptotically at large~$h$, which again yields an AFS-like leading order, and an HL-like sub-leading order. We refer the reader to appendix~\ref{app:BESexpansion} for the details.

As for the behaviour in the mirror region, similarly to what we did above we can obtain it by analytic continuation from the real string line to the real mirror line (avoiding the cuts). Once again, we report the expression in appendix~\ref{app:BES:massless}, where we also check the behaviour under complex conjugations of the massless-massless BES in the mirror region. We find that
\begin{equation}
\label{eq:masslessmirrorBES}
    \big(\BES(\tilde{x}_{1},\tilde{x}_{2})\big)^{-2}
    \Big(\big(\BES(\tilde{x}_{1},\tilde{x}_{2})\big)^{-2}\Big)^*
    =
    1\,.
\end{equation}
which is compatible with formally taking $Q_1\to0$ and $Q_2\to0$ in~\eqref{eq:boundstatemirrorBES}.
 
\subsection{The Sine-Gordon factor}
\label{sec:buildingblocks:SG}
A natural candidate for $\PhiSG(\gamma)$ is the Sine-Gordon dressing factor,
\begin{equation}
\label{eq:SineGordonfactor}
\PhiSG(\gamma) = \prod_{\ell=1}^\infty \RSG\big(\ell,\gamma\big)\,,\qquad  \RSG(\ell,\gamma) =\frac{\Gamma^2(\ell -\tfrac{\gamma}{2\pi i}) \Gamma(\tfrac{1}{2}+\ell +\tfrac{\gamma}{2\pi i})\Gamma(-\tfrac{1}{2}+\ell +\tfrac{\gamma}{2\pi i})}{\Gamma^2(\ell +\tfrac{\gamma}{2\pi i})\Gamma(\tfrac{1}{2}+\ell -\tfrac{\gamma}{2\pi i})\Gamma(-\tfrac{1}{2}+\ell -\tfrac{\gamma}{2\pi i})}\,.
\end{equation}
The function so defined satisfies
\begin{equation}
    \PhiSG(\gamma)\PhiSG(-\gamma)=1\,,\qquad
    \PhiSG(\gamma)^*=\frac{1}{\PhiSG(\gamma^*)}\,,\qquad
    \PhiSG(\gamma)\PhiSG(\gamma+i\pi)=i\tanh\frac{\gamma}{2}\,,
\end{equation}
as it is possible to verify from~\eqref{eq:SineGordonfactor}.
It should be stressed that this is by no means \emph{the only} solution of the above equations. Indeed, we can multiply this solution by any solution~$F(\gamma)$ of the homogenous equation
\begin{equation}
    F(\gamma)F(-\gamma)=1\,,\qquad
    F(\gamma)^*=\frac{1}{F(\gamma^*)}\,,\qquad
    F(\gamma)F(\gamma+i\pi)=1\,,
\end{equation}
and thus obtain a new solution $F(\gamma)\PhiSG(\gamma)$. We will discuss later the choice of this CDD factor.

It is convenient to consider the logarithmic derivative $\varphi{}'(\gamma)$ with
\begin{equation}
    \phiSG(\gamma) = \log\PhiSG(\gamma)\,,
\end{equation}
which can be expressed as an integral by means of the integral representation of the Digamma function,
\begin{equation}
    \psi(z)\equiv\frac{d}{dz}\log\Gamma(z) = \int\limits_{0}^{+\infty}\left(\frac{e^{-t}}{t}-\frac{e^{-z\,t}}{1-e^{-t}}\right)dt\,,\qquad\mathfrak{R}[\gamma]>0\,.
\end{equation}
Then we have that~$\phiSG{}'(\gamma)$ is simply
\begin{equation}
    \phiSG{}'(\gamma)=\frac{i}{4\pi}\int\limits_0^{+\infty}\frac{\cos({\gamma t}/{2\pi})}{\cosh^2(t/{4})}dt = \frac{i\gamma}{\pi\sinh\gamma}\,,\qquad -\pi<\mathfrak{I}[\gamma]<\pi\,.
\end{equation}
From this we can derive the integral representation for $\phiSG(\gamma)$
\begin{equation}
    \phiSG(\gamma)=\frac{i}{2}\int\limits_0^{+\infty}\frac{\sin({\gamma t}/{2\pi})}{t\cosh^2(t/4)}dt\,,\qquad-\pi<\mathfrak{I}[\gamma]<\pi\,,
\end{equation}
as well as the explicit expression
\begin{equation}
\label{eq:varphiexplicit}
\begin{aligned}
\phiSG(\gamma)&=
\frac{i}{\pi}\text{Li}_2(-e^{-\gamma})-\frac{i}{\pi}\text{Li}_2(e^{-\gamma})+\frac{i\gamma}{\pi}\log(1-e^{-\gamma})-\frac{i\gamma}{\pi}\log(1+e^{-\gamma})+\frac{i\pi}{4}\,,
\end{aligned}
\end{equation}
again valid in the region $-\pi<\mathfrak{I}[\gamma]<\pi$. For real $\gamma$, $\varphi(\gamma)$ is purely imaginary. Three notable values of $\varphi(\gamma)$ are
\begin{equation}
    \varphi(\pm\infty) = \pm \frac{i\pi}{4}\,,\qquad
    \varphi(0) = 0\,.
\end{equation}

\subsection{The monodromy factor}
\label{sec:buildingblocks:monodromy}
We may define a function
\begin{equation}
\label{eq:phihatproduct}
    \PhiMON(\gamma) = e^{\frac{\gamma}{i\pi}} \prod_{\ell=1}^\infty \frac{\Gamma(\ell +\tfrac{\gamma}{i\pi})}{\Gamma(\ell -\tfrac{\gamma}{i\pi})} \,e^{\frac{2i}{\pi}\,\psi(\ell)\,\gamma } \,,
\end{equation}
where $\psi(z)$ is the Digamma function.%
\footnote{The factor involving the Digamma function is needed to make the product convergent, while the exponential term in front of the product is a convenient normalisation.}
By its definition this satisfies what we want
\begin{equation}
    \frac{\PhiMON(\gamma\pm i\pi)}{\PhiMON(\gamma)}=i(2\sinh\gamma)^{\pm1}\,,\qquad
        \PhiMON(z)\,\PhiMON(-z)=1\,,\qquad
    \PhiMON(z)^*=\frac{1}{\PhiMON(z^*)}\,.
\end{equation}
Once again, this is not \textit{the only} solution to the above equations.

Defining $\phiMON(\gamma)=\log\PhiMON(z)$ we have that
\begin{equation}
    \phiMON{}'(\gamma)=\frac{\gamma\coth\gamma}{i\pi}\,,
\end{equation}
and
\begin{equation}
\label{eq:varphihatexplicit}
    \phiMON(\gamma) =
    \frac{i}{2 \pi }\text{Li}_2\left(e^{-2\gamma }\right)-\frac{i}{2 \pi}\gamma^2 -\frac{i}{\pi}  \log
   \left(1-e^{-2\gamma }\right)-\frac{i\pi }{12}\,.
\end{equation}
This solution is ``minimal'' in the sense that we may derive it under the assumption that it does not have singularities in the strip between zero and $i\pi$ using the Fourier transform, see appendix~\ref{app:Fourier}. Let us also introduce
\begin{equation}
    \RMON(\ell,\gamma) =\frac{\Gamma({\ell} +\tfrac{\gamma}{2\pi i})^2\, \Gamma({\ell} +\tfrac{1}{2}+\tfrac{\gamma}{2\pi i})\, \Gamma({\ell} -\tfrac{1}{2}+\tfrac{\gamma}{2\pi i})}
    {\Gamma({\ell} -\tfrac{\gamma}{2\pi i})^2\, \Gamma({\ell} +\tfrac{1}{2}-\tfrac{\gamma}{2\pi i}) \,
\Gamma({\ell} -\tfrac{1}{2}-\tfrac{\gamma}{2\pi i})  } \,,
\end{equation}
and observe that we may write
\begin{equation}
\label{eq:Rhatrepresentation}
    \frac{\PhiMON(\gamma^{++}_{12})\,\PhiMON(\gamma^{--}_{12})}{\PhiMON(\gamma^{+-}_{12})\,\PhiMON(\gamma^{-+}_{12})}
    =
    \prod_{\ell=1}^\infty\frac{ \RMON(\ell,\gamma^{++}_{12})\, \RMON(\ell,\gamma^{--}_{12})}{ \RMON(\ell,\gamma^{+-}_{12})\, \RMON(\ell,\gamma^{-+}_{12})}\,.
\end{equation}
Note that the infinite product over a single factor, $\prod_\ell^\infty \RMON(\ell,\gamma)$, does not converge, while the expression~\eqref{eq:Rhatrepresentation} does.

\subsection{The auxiliary function \texorpdfstring{$a(\gamma)$}{a(gamma)}}
\label{sec:buildingblocks:a}
To solve the massless scattering we introduced an auxiliary function $a(\gamma)$ that must satisfy
\begin{equation}
    a(\gamma)\,a(\gamma+i \pi) = -1\,,\qquad a(\gamma)\,a(-\gamma)=1\,,\qquad
    a(\gamma)^*=\frac{1}{a(\gamma^*)}\,.
\end{equation}
One such function is
\begin{equation}
    a(\gamma) = -i\,\tanh\left(\frac{\gamma}{2}-\frac{i\pi}{4}\right)\,,
\end{equation}
and we note for later convenience that 
\begin{equation}
    a(\mp\infty)=\pm i\,,\qquad
    a(0)=-1\,.
\end{equation}
It is worth noting that this function has appeared before in the context of massless integrable quantum field theories, see for instance~\cite{Zamolodchikov:1991vx}.

\section{Proposal for the dressing factors}
\label{sec:proposal}
Using the functions introduced above, we can write down solutions of the crossing equations for the various dressing factors.

\subsection{Massive sector}
\label{sec:proposal:massive}
Here the building blocks are the BES phase~$\BES(x_1^\pm,x_2^\pm)$ and the functions $\varPhi(\gamma)$ and $\PhiMON(\gamma)$, which appear in the equations for the product and ratio of the dressing factors, respectively.
Moreover, it turns out that we need to include a particular solution of the homogeneous equations, which will provide us with the correct pole structure and match the expected perturbative result.
\begin{equation}
\begin{aligned}
&\left(\sigmap (x_1^\pm,x_2^\pm)\right)^{-2} =-\frac{\tanh\frac{\gamma^{-+}_{12}}{2}}{\tanh\frac{\gamma^{+-}_{12}}{2}}\, \PhiSG(\gamma_{12}^{--})\PhiSG(\g_{12}^{++}) \PhiSG(\gamma_{12}^{-+})\PhiSG(\g_{12}^{+-})\,,\\
&\left(\sigmam (x_1^\pm,x_2^\pm)\right)^{-2} =-\frac{\sinh \gamma^{-+}_{12}}{\sinh\gamma^{+-}_{12}}\, \frac{\PhiMON(\gamma^{++}_{12})\,\PhiMON(\gamma^{--}_{12})}{\PhiMON(\gamma^{+-}_{12})\,\PhiMON(\gamma^{-+}_{12})}\,.
\end{aligned}
\end{equation}
By manipulating the product representation for these functions we may then introduce
\begin{equation}
\label{eq:Rplusminus}
    R_+(\ell,\gamma)=  \frac{ \Gamma(\ell +\tfrac{1}{2}+\tfrac{\gamma}{2\pi i})\, \Gamma(\ell -\tfrac{1}{2}+\tfrac{\gamma}{2\pi i})}
    { \Gamma(\ell +\tfrac{1}{2}-\tfrac{\gamma}{2\pi i})\, \Gamma(\ell -\tfrac{1}{2}-\tfrac{\gamma}{2\pi i})}\,,\qquad
    R_-(\ell,\gamma)=\frac{\Gamma^2(\ell-\frac{\gamma}{2\pi i})}{\Gamma^2(\ell+\frac{\gamma}{2\pi i})}\,,
\end{equation}
which satisfy
\begin{equation}
    R_+(\ell,\gamma)\,R_-(\ell,\gamma)=
    \RSG(\ell,\gamma)\,,\qquad
    \frac{R_+(\ell,\gamma)}{R_-(\ell,\gamma)}=
    \RMON(\ell,\gamma)\,.
\end{equation}
Additionally, these factors satisfy a crossing equation of sorts,
\begin{equation}
\label{eq:crossingregularisation}
\begin{aligned}
    \prod_{\ell=1}^\infty R_{+}(\ell,\gamma)R_{-}(\ell,\gamma+i\pi)=\frac{1}{\cosh\tfrac{\gamma}{2}}\,,\qquad
    \prod_{\ell=1}^\infty R_{+}(\ell,\gamma)R_{-}(\ell,\gamma-i\pi)= \cosh\tfrac{\gamma}{2}\,,\\
    \prod_{\ell=1}^\infty R_{-}(\ell,\gamma)R_{+}(\ell,\gamma+i\pi)=i\sinh\tfrac{\gamma}{2}\,,\qquad
    \prod_{\ell=1}^\infty R_{-}(\ell,\gamma)R_{+}(\ell,\gamma-i\pi)= \frac{i}{\sinh\tfrac{\gamma}{2}}\,,
\end{aligned}
\end{equation}
where strictly speaking the left-hand side does not converge and the right-hand side is the result of a regularisation.%
\footnote{Namely, we consider the series for the logarithmic derivative of the left-hand-side, which is convergent. This does leave an ambiguity in fixing an overall multiplicative factor in the crossing equations.
We could also amend the definition of $R_\pm(\ell,\gamma)$  by a Digamma term similar to~\eqref{eq:phihatproduct} to make the products over~$\ell$ convergent. This would modify the crossing equations~\eqref{eq:crossingregularisation} by constant factors which drop out when considering the full dressing factors.
}
We then write
\begin{equation}
\label{eq:massiveproductrepr}
\begin{aligned}
&\left(\sigmabb(x^\pm_1,x^\pm_2)\right)^{-2} =
-\frac{\sinh \frac{\gamma_{12}^{-+}}{2}}{\sinh \frac{\gamma_{12}^{+-}}{2}}\, 
\prod_{\ell=1}^\infty { R_+(\ell,\gamma_{12}^{--})R_+(\ell,\gamma_{12}^{++}) R_-(\ell,\gamma_{12}^{-+})R_-(\ell,\gamma_{12}^{+-})}\,,\\
&\left(\sigmabbt(x^\pm_1,x^\pm_2)\right)^{-2} =
+\frac{\cosh \frac{\gamma_{12}^{+-}}{2}}{ \cosh \frac{\gamma_{12}^{-+}}{2}}\,
\prod_{\ell=1}^\infty { R_-(\ell,\gamma_{12}^{--})R_-(\ell,\gamma_{12}^{++}) R_+(\ell,\gamma_{12}^{-+})R_+(\ell,\g_{12}^{+-})}\,.
\end{aligned}
\end{equation}
The products have the following representations
\begin{equation}
    \begin{aligned}
    &\prod_{\ell=1}^\infty { R_+(\ell,\g_{12}^{--})R_+(\ell,\g_{12}^{++}) R_-(\ell,\g_{12}^{-+})R_-(\ell,\g_{12}^{+-})} =
e^{\varphi^{\bullet\bullet}(\gamma_1^\pm,\gamma_2^\pm)}\,,\\
    &\prod_{\ell=1}^\infty { R_-(\ell,\g_{12}^{--})R_-(\ell,\g_{12}^{++}) R_+(\ell,\g_{12}^{-+})R_+(\ell,\g_{12}^{+-})} =
e^{\tilde\varphi^{\bullet\bullet}(\gamma_1^\pm,\gamma_2^\pm)}\,,
    \end{aligned}
\end{equation}
with
\begin{equation}
\begin{aligned}
   &\varphi^{\bullet\bullet}(\gamma_1^\pm,\gamma_2^\pm)
   =\varphi_+^{\bullet\bullet}(\g_{12}^{--}) +\varphi_+^{\bullet\bullet}(\g_{12}^{++})+\varphi_-^{\bullet\bullet}(\g_{12}^{-+})+\varphi_-^{\bullet\bullet}(\g_{12}^{+-})\,,
   \\
  &\tilde\varphi^{\bullet\bullet}(\gamma_1^\pm,\gamma_2^\pm)
  =\varphi_-^{\bullet\bullet}(\g_{12}^{--}) +\varphi_-^{\bullet\bullet}(\g_{12}^{++})+\varphi_+^{\bullet\bullet}(\g_{12}^{-+})+\varphi_+^{\bullet\bullet}(\g_{12}^{+-})\,,
\end{aligned}
\end{equation}
The right-hand side can be evaluated explicitly in terms of
\begin{equation}
\label{eq:massivesoldilog}
\begin{aligned}
    &\varphi_-^{\bullet\bullet}(\g) =+ \frac{ i}{\pi} \text{Li}_2\left(+e^{\gamma}\right)- \frac{i}{4\pi} \gamma^2+\frac{i}{\pi} \gamma\,  \log
   \left(1-e^{\gamma }\right)-\frac{i \pi }{6}\,,
   \\
   &\varphi_+^{\bullet\bullet}(\g) =-\frac{ i}{\pi} \text{Li}_2\left(-e^{\gamma}\right)+ \frac{i}{4\pi}  \gamma^2-\frac{i}{\pi}\gamma \,  \log
   \left(1+e^{\gamma}\right)-\frac{i \pi }{12}\,.
\end{aligned}
\end{equation}

In conclusion, we can write the full dressing factors $\sigma^{\bullet\bullet}(x_1^\pm,x_2^\pm)$ and $\widetilde{\sigma}^{\bullet\bullet}(x_1^\pm,x_2^\pm)$ in terms of the BES phase $\BES(x_1^\pm,x_2^\pm)$ as
\begin{equation}
\begin{aligned}&\left(\sigma^{\bullet\bullet}(x^\pm_1,x^\pm_2)\right)^{-2} =
-\frac{\sinh \frac{\gamma_{12}^{-+}}{2}}{ \sinh \frac{\gamma_{12}^{+-}}{2}}\, 
e^{\tilde\varphi^{\bullet\bullet}(\gamma_1^\pm,\gamma_2^\pm)}\,\left(\BES(x_1^\pm,x_2^\pm)\right)^{-2},\\
&\left(\widetilde{\sigma}^{\bullet\bullet}(x^\pm_1,x^\pm_2)\right)^{-2} =
+\frac{\cosh \frac{\gamma_{12}^{+-}}{2}}{\cosh \frac{\gamma_{12}^{-+}}{2}}\,
e^{\varphi^{\bullet\bullet}(\gamma_1^\pm,\gamma_2^\pm)}\,\left(\BES(x_1^\pm,x_2^\pm)\right)^{-2}.
\end{aligned}
\end{equation}

\subsubsection{Properties}
We start by discussing the singularity structure of these dressing factors in the physical region for string theory and for the mirror theory. The location of these regions for  the $\gamma^\pm$ parametrisation was described in section~\ref{sec:rapidities:massive}. Since we have stripped of the poles expected for the bound states of the theory in the normalisation~\eqref{eq:massivenorm}, we do not expect any additional poles to appear.

\paragraph{Singularities in the string region.}
In principle, we may expect singularities at any of the points  $\gamma_{1}^\pm=\gamma^\pm_2$ mod$\,\pi$ and $\gamma_{1}^\pm=\gamma^\mp_2$ mod$\,\pi$, \textit{cf.}\ table~\ref{table:zeros}. Most of these configurations are not in the physical region. Indeed, the only physical configurations are $\gamma_1^+=\gamma_2^- -i\pi$, $\gamma_1^-=\gamma_2^+ +i\pi$, and $\gamma_1^\pm=\gamma_2^\pm$. Let us start from $\gamma_1^+=\gamma_2^- -i\pi$, and look at the product representation~\eqref{eq:massiveproductrepr} (we do not need to worry about the BES factor, which is regular in this region). For $\sigmabb(x_1^\pm,x_2^\pm)^{-2}$, the factor $\sinh\tfrac{\gamma^{+-}_{12}}{2}=-i$ is regular. Possible singularities might in principle come from the terms $R_-(\ell,\gamma^{-+}_{12})$, but recalling eq.~\eqref{eq:Rplusminus} we see that never happens. For $\sigmabbt(x_1^\pm,x_2^\pm)^{-2}$ we get a simple zero from $\cosh\tfrac{\gamma^{+-}_{12}}{2}=0$. However, we also have
\begin{equation}
    R_{+}(\ell,\gamma^{+-}_{12})\Big|_{\gamma^{+-}_{12}=-i\pi}=\frac{1}{(\ell-1)\ell}\,,\qquad \ell=1,2,\dots\,,
\end{equation}
so that the pole for $\ell=1$ precisely compensates the zero.
The discussion for $\gamma_1^-=\gamma_2^+ +i\pi$ is similar, with the only caveat that for $\sigmabbt(x_1^\pm,x_2^\pm)^{-2}$ now $\cosh\tfrac{\gamma^{-+}_{12}}{2}=0$ appears in the denominator, yielding a pole, which is compensated by a zero of  $R_{+}(\ell,\gamma^{-+}_{12})=(\ell-1)\ell$.
Finally, for $\gamma_1^+=\gamma_2^+$ or $\gamma_1^-=\gamma_2^-$ no singularity occurs because $R_\pm(\ell,0)=1$, \textit{cf.}~\eqref{eq:Rplusminus}.

\paragraph{Singularities in the mirror region.}
Even though the string and mirror region lie in different strips of $\gamma^\pm$ plane, the differences $\gamma_1^\pm-\gamma_2^\mp$ are still physical only if $\gamma_1^+=\gamma_2^- -i\pi$ or $\gamma_1^-=\gamma_2^+ +i\pi$. Additionally, we always have the possibility $\gamma_1^\pm=\gamma_2^\pm$. These are precisely the same configurations discussed for the string region. It follows from the above discussion that there are no poles in the physical mirror region.

\paragraph{String bound states and fusion.}
Two massive ``left'' particles, or two massive ``right'' particles, may form a supersymmetric bound state~\cite{Borsato:2013hoa}. Compatibly with the pole in~\eqref{eq:massivenorm}, this happens when
\begin{equation}
\label{eq:stringboundstate}
    x_1^+ =x_2^- \qquad \Leftrightarrow\qquad  \gamma_1^+=\gamma_2^- - i\pi\,.
\end{equation}
Given such a bound state of particles with rapidities $\gamma_1^\pm$ and $\gamma_2^\pm$, we may ask what are the S-matrix elements when scattering a particle of rapidity $\gamma_3^\pm$. For the matrix part of the S~matrix, this is fixed by representation theory. For the dressing factor, we expect the result to depend on $x_1^-$ and $x_2^+$ only---we say in this case that the dressing factor is \textit{fused}, in which case the argument can be iterated to study the scattering of $M$-particle bound states. This is what happens for the BES phase~\cite{Chen:2006gq}.
We now want to check that this is the case for $\sigmabb(\gamma_1^\pm,\gamma_2^\pm)^{-2}$. We have, schematically,%
\footnote{To prove this identity we impose the condition~\eqref{eq:stringboundstate} and may use the regularised expressions~\eqref{eq:crossingregularisation}.
Alternatively, we may work with the square of the equation and use the crossing equations, which gives the same result up to a possible sign ambiguity.} 
\begin{equation}
\begin{aligned}
    \left(\sigmabb_{13}\right)^{-2}\left(\sigmabb_{23}\right)^{-2}&=-\frac{\sinh\tfrac{\gamma^{-+}_{13}}{2}}{\sinh\tfrac{\gamma^{+-}_{13}}{2}}
    \prod_{\ell=1}^\infty R_+(\ell,\gamma^{++}_{13})R_+(\ell,\gamma^{--}_{13})
    R_-(\ell,\gamma^{+-}_{13})R_+(\ell,\gamma^{-+}_{13})\\
    &\times\left(-\frac{\sinh\tfrac{\gamma^{-+}_{23}}{2}}{\sinh\tfrac{\gamma^{+-}_{23}}{2}}
    \prod_{\ell=1}^\infty R_+(\ell,\gamma^{++}_{23})R_+(\ell,\gamma^{--}_{23})
    R_-(\ell,\gamma^{+-}_{23})R_+(\ell,\gamma^{-+}_{23})\right)\\
    &=-\frac{\sinh\tfrac{\gamma^{-+}_{13}}{2}}{\sinh\tfrac{\gamma^{+-}_{23}}{2}}
    \prod_{\ell=1}^\infty R_+(\ell,\gamma^{++}_{23})R_+(\ell,\gamma^{--}_{13})
    R_-(\ell,\gamma^{+-}_{23})R_+(\ell,\gamma^{-+}_{13})\,,
\end{aligned}
\end{equation}
In a similar way we find
\begin{equation}
    \left(\sigmabbt_{13}\right)^{-2}\left(\sigmabbt_{23}\right)^{-2}=
    \frac{\cosh \frac{\gamma_{23}^{+-}}{2}}{\cosh \frac{\gamma_{13}^{-+}}{2}}\, 
\prod_{\ell=1}^\infty { R_-(\ell,\gamma_{13}^{--})R_-(\ell,\gamma_{23}^{++}) R_+(\ell,\gamma_{13}^{-+})R_+(\ell,\gamma_{23}^{+-})}\,.
\end{equation}

\paragraph{Mirror bound states and fusion.}
Next, we look at fusion for mirror bound states, and consider
\begin{equation}
    x_1^- = x_2^+
    \qquad \Leftrightarrow\qquad
    \gamma_1^- =\gamma_2^+ +i\pi\,.
\end{equation}
Now we expect the fused result to depend only on $\gamma_1^+$ and $\gamma_2^-$. Indeed we find
\begin{equation}
    \left(\sigmabb_{13}\right)^{-2}\left(\sigmabb_{23}\right)^{-2}=
    -\frac{\sinh \frac{\gamma_{23}^{-+}}{2}}{ \sinh \frac{\gamma_{13}^{+-}}{2}}\,
    \prod_{\ell=1}^\infty { R_+(\ell,\gamma_{23}^{--})R_+(\ell,\gamma_{13}^{++}) R_-(\ell,\gamma_{23}^{-+})R_-(\ell,\gamma_{13}^{+-})}\,,
\end{equation}
and
\begin{equation}
    \left(\sigmabbt_{13}\right)^{-2}\left(\sigmabbt_{23}\right)^{-2}=
    \frac{\cosh \frac{\gamma_{13}^{+-}}{2}}{\cosh \frac{\gamma_{23}^{-+}}{2}}\, 
\prod_{\ell=1}^\infty { R_-(\ell,\gamma_{23}^{--})R_-(\ell,\gamma_{13}^{++}) R_+(\ell,\gamma_{23}^{-+})R_+(\ell,\gamma_{13}^{+-})}\,.
\end{equation}

\paragraph{Physical unitarity in the string and mirror regions.}
We need to check that the dressing factors have modulus one for real values of the momenta. For string particles, recall that we have
\begin{equation}
\label{eq:conjstring}
    (x^\pm)^* = x^\mp\,,\qquad
    (\gamma^\pm)^* = \gamma^\mp\,,
\end{equation}
for real momenta, while for mirror particles recalling~\eqref{eq:massivemirrorconj} we have
\begin{equation}
\label{eq:conjmirror}
    (\tilde{x}^\pm)^*=\frac{1}{\tilde{x}^\mp}\,,\qquad
    (\tilde{\gamma}^-)^*=\tilde{\gamma}^++i\pi\,,\quad (\tilde{\gamma}^+)^*=\tilde{\gamma}^-+i\pi\,.
\end{equation}
Let us start from the rational prefactors and the BES term. In the string region, they are both unitary. In the mirror region, instead, the rational prefactor in~\eqref{eq:massivenorm} is not unitary by itself; the offending term is the  $e^{i (p_1-p_2)}$ factor in front of it. That term precisely compensates the non-unitarity of the BES factor in the mirror region, see eq.~\eqref{eq:boundstatemirrorBES}. As for the rest of the dressing factors, physical unitarity in the string region follows from the behaviour of $\PhiSG(\gamma)$ and $\PhiMON(\gamma)$ under complex conjugation. In the mirror region the conclusion is the same, given that the mirror rapidities undergo the same constant shift in~\eqref{eq:conjmirror} and  $\PhiSG(\gamma),\PhiMON(\gamma)$ depend only on the difference of rapidities.

\paragraph{Zero-momentum limit.}
We expect the S~matrix to simplify when the momentum of either particle is zero. Indeed, given a state represented in the Bethe-Yang equations by a set of $N$ momenta we generally should be able to construct \textit{a new state}, with one more momentum $p_{N+1}=0$, which still solves the Bethe-Yang equations. Such a state is then a symmetry descendant of the original one. We therefore expect that for any dressing factor $\sigma(p_1,p_2)$, $\sigma(p_1,0)=1$, possibly up to an integer power of $\exp(\tfrac{i}{2}p_1)$.
This is the case for the rational prefactor in~\eqref{eq:massivenorm} and for the BES factor. Let us now inspect the new ingredients in our construction. Recalling eq.~\eqref{eq:zeromomentummassive}, we set
\begin{equation}
    \gamma_2^\pm=\mp\frac{i\pi}{2}\,.
\end{equation}
Using this we compute
\begin{equation}
\begin{aligned}
    (\widetilde{\sigma}^{\bullet\bullet}_{12})^{-2}&=\frac{\cosh\tfrac{1}{2}(\gamma_1^+-\tfrac{i\pi}{2})}{\cosh\tfrac{1}{2}(\gamma_1^-+\tfrac{i\pi}{2})}
    \prod_{\ell=1}^\infty [R_-(\ell,\gamma^{--}_{12})R_+(\ell,\gamma^{-+}_{12})][R_-(\ell,\gamma^{++}_{12})R_+(\ell,\gamma^{+-}_{12})]\\
    &=\frac{\cosh\tfrac{1}{2}(\gamma_1^+-\tfrac{i\pi}{2})}{\cosh\tfrac{1}{2}(\gamma_1^-+\tfrac{i\pi}{2})} \frac{\cosh\tfrac{1}{2}\gamma_{12}^{-+}}{\cosh\tfrac{1}{2}\gamma_{12}^{+-}}=1\,,
\end{aligned}
\end{equation}
where in the second equality we used~\eqref{eq:crossingregularisation} for the terms in the square brackets.
In a  similar way we find
\begin{equation}
    (\sigma^{\bullet\bullet}_{12})^{-2}
    =
    -\frac{\sinh\tfrac{1}{2}(\gamma_1^- +\tfrac{i\pi}{2})}{\sinh\tfrac{1}{2}(\gamma_1^+ -\tfrac{i\pi}{2})}
    \frac{\cosh\tfrac{1}{2}(\gamma_1^+ +\tfrac{i\pi}{2})}{\cosh\tfrac{1}{2}(\gamma_1^- -\tfrac{i\pi}{2})} =1\,,
\end{equation}
as it should be.

\subsubsection{Perturbative expansion}
It is useful to work out the large-$h$ expansion of the dressing factors, which can  in principle be used to compare with string-NLSM computations. For the BES phase~$\sigma(x_1^\pm,x_2^\pm)$, this expansion is well-known, being given at leading order by the AFS phase~\cite{Arutyunov:2004vx} and at next-to-leading-order by the HL phase~\cite{Hernandez:2006tk} of eqs.~\eqref{eq:AFS} and~\eqref{eq:HL}, respectively.
The other pieces of our proposal can be expanded quite straightforwardly by starting from their representation in terms of polylogarithms, see eqs.~\eqref{eq:varphiexplicit} and~\eqref{eq:varphihatexplicit}. In this way we can find  the near-BMN~\cite{Berenstein:2002jq} expansion of the massive dressing factors, see  appendix~\ref{app:BESexpansion}. This should be compared with the perturbative expansion of the existing proposal~\cite{Borsato:2013hoa}, as well as with the perturbative computations of ref.~\cite{Rughoonauth:2012qd,Sundin:2013ypa,Engelund:2013fja,Roiban:2014cia,Bianchi:2014rfa,Sundin:2016gqe}.

The perturbative computations for the $AdS_3\times S^3\times T^4$ background have been performed by different techniques and in different kinematics regimes. The tree-level S~matrix was first worked out in~\cite{Rughoonauth:2012qd}. In refs.~\cite{Sundin:2013ypa} the dressing factors were computed at one-loop in the near-flat-space limit~\cite{Maldacena:2006rv}. In~\cite{Engelund:2013fja}, the cut-constructible part of the one- and two-loop dressing factor was worked out by unitarity. Later, the complete one-loop dressing factors were considered in~\cite{Roiban:2014cia} (see also~\cite{Sundin:2016gqe}) by evaluating the Feynman graphs by means of integral identities and dimensional regularisation. The same calculation was also done in~\cite{Bianchi:2014rfa} by exploiting both unitarity as well as symmetry and regularisation of certain divergent integrals. All these results are compatible with the proposal of~\cite{Borsato:2013hoa}.

Comparing our result with the perturbative results we find what follows:
\begin{enumerate}
    \item A tree level, our proposal agrees with~\cite{Rughoonauth:2012qd} and subsequent computations.
    \item At one-loop in the near-flat-space kinematics, our proposal agrees with the perturbative results~\cite{Sundin:2013ypa}. Note that the near-flat-space kinematics is more restricted than the near-BMN, so that the near-flat-space expansion contains less information than the near-BMN one.
    \item At one- and two-loops, our proposal agrees with the cut-constructible part of the dressing factors~\cite{Engelund:2013fja}.
    \item At one-loop in the near-BMN limit, our proposal \textit{does not agree} with~\cite{Roiban:2014cia,Bianchi:2014rfa}. In particular, while the sum of the phases $\sigmap_{12}$ agrees, we find a mismatch for the difference~$\sigmam_{12}$, namely
\begin{equation}
\label{eq:massivediscrepacy}
    \log (\sigmam_{12})^{-2}\Big|_{\text{ours}}-
    \log (\sigmam_{12})^{-2}\Big|_{\text{theirs}}=
    \frac{i}{2\pi h^2}\left(\omega_1p_2-\omega_2p_1\right)p_1p_2+O(h^{-3})\,.
\end{equation}
Compatibly with the above observation, this term is not cut-constructible and it is zero in the near-flat-space limit. It is also interesting to note that this term could be interpreted as arising from a local counterterm.
\end{enumerate}
This could look troublesome, especially in view of the fact that the original proposal of~\cite{Borsato:2013hoa} matches all known perturbative computations. However, a closer inspection of the monodromy factor of~\cite{Borsato:2013hoa} shows that it violates parity invariance of the model, see appendix~\ref{app:monodromy}. Hence, it cannot be the correct solution.
Having made these observations, we postpone their discussion to the section~\ref{sec:proposal:relations} and to the conclusions. 

\subsection{Mixed-mass sector}
\label{sec:proposal:mixed}
In the mixed-mass sector we define
\begin{equation}
\begin{aligned}
    &\big(\sigmabc(\gamma_1^\pm,\gamma_2)\big)^{-2} =
    +i\, \frac{\tanh\tfrac{\gamma_{12}^{-\circ}}{2}}{\tanh\tfrac{\gamma_{12}^{+\circ}}{2}}\,
    \PhiSG(\gamma_{12}^{+\circ})\,\PhiSG(\gamma_{12}^{-\circ})\,,\\    &\big(\sigmacb(\gamma_1,\gamma_2^\pm)\big)^{-2} =
    -i\, \frac{\tanh\tfrac{\gamma_{12}^{\circ+}}{2}}{\tanh\tfrac{\gamma_{12}^{\circ-}}{2}}\,
    \PhiSG(\gamma_{12}^{\circ+})\,\PhiSG(\gamma_{12}^{\circ-})\,,\\
\end{aligned}
\end{equation}
In most of what follows, it will be sufficient to focus our attention on one of the two phases (say, $\sigmabc_{12}$) since they are related by braiding unitarity~\eqref{eq:braidunit}.
To complete our definition we now need the BES dressing factor in the mixed-mass sector $\sigma(x_1^\pm,x_2)$, which we discussed above, see eq.~\eqref{eq:massless-massive-BES}. Using this we have finally
\begin{equation}
    \big(\sigma^{\bullet\circ}(x^\pm,x_2)\big)^{-2} =
    +i\, \frac{\tanh\tfrac{\gamma_{12}^{-\circ}}{2}}{\tanh\tfrac{\gamma_{12}^{+\circ}}{2}}\,
    \PhiSG(\gamma_{12}^{+\circ})\,\PhiSG(\gamma_{12}^{-\circ})\big(\sigma(x_1^\pm,x_2)\big)^{-2}\,,
\end{equation}
and we recall that the whole S-matrix elements are given in eq.~\eqref{eq:mixednormalisation}.

\subsubsection{Properties}
Let us analyze some properties of the dressing factor that we just introduced. Unlike what we did for the massive sector, we will not restrict our discussion to the $\sigmabc(p_1,p_2)$ piece: here the properties of the (mixed-mass) BES factor are less obvious.

\paragraph{Poles and (absence of) branch points.}
The dressing factors, taken together with their rational prefactors~\eqref{eq:mixednormalisation}, show a number of apparent poles at
\begin{equation}
\label{eq:mixedpoles}
    x_1^+=x_2\,,\quad x_1^+=\frac{1}{x_2}\,,\quad
    x_1^-=x_2\,,\quad
    x_1^+=\frac{1}{x_2}\,.
\end{equation}
Some of these poles, in particular those appearing at $x_1^\pm=(x_2)^{\mp1}$, appear also in the matrix part of the S-matrix and as such cannot be removed by CDD factors; note also that such poles were also present in the proposal of~\cite{Borsato:2016xns}. As we discussed, none of the four conditions~\eqref{eq:mixedpoles} may be used to describe a bound state with real momentum and real (and positive) energy, neither in the string region nor in the mirror one. Therefore, we shall not consider them.
Unlike the proposal of~\cite{Borsato:2016xns}, our solution is manifestly free from  square-root branch points at the positions~\eqref{eq:mixedpoles}. This simplifies considerably the discussion of the analytic properties of the dressing factor. We will see in section~\ref{sec:proposal:relations} that, as a matter of fact, the proposal of~\cite{Borsato:2016xns} was actually free from branch-points at~\eqref{eq:mixedpoles} due to a quite non-trivial cancellation between the dressing factor and its normalisation. To the best of our knowledge, this fact was not previously appreciated.%
\footnote{Some analytic properties of the proposal~\cite{Borsato:2016xns} have been discussed in ref.~\cite{OhlssonSax:2019nlj}, but unfortunately the branch cuts of the mixed-mass factors have not been discussed there.} 
Nonetheless, the proposal of~\cite{Borsato:2016xns} manifestly has the AFS singularities, see~\eqref{eq:AFS}, which make its analytic continuation rather subtle if not ill defined.

\paragraph{String bound states and fusion.}
Let us stress that, even if we are considering the mixed-mass scattering, when discussing fusion we refer of \textit{the fusion of two massive particles} parametrised by $x_1^\pm$ and $x_2^\pm$ (say, both of charge $M=1$) to create a bound state (say, of mass $M=2$); this bound state may then scatter with a massless particle parametrised by~$x_3$. This setup should not be confused with an attempt to construct bound states of one massive and one massless particles, which as recalled above does not seem to yield a physical (massless) excitation.
Like before, the condition for bound states in the string region is
\begin{equation}
\label{eq:fusionstring}
    x_1^+ ={ x_2^-} \qquad \Leftrightarrow\qquad  \g_1^+=\g_2^- -i\pi\,.
\end{equation}
We now need to analyze various contributions to fusion. We start by considering $\sigmabc(x_1^\pm,x_3)$ and $\sigmabc(x_2^\pm,x_3)$ and using~\eqref{eq:varPhicrossing},
\begin{equation}
\begin{aligned}
(\sigmabc_{13})^{-2}(\sigmabc_{23})^{-2} &=-
\frac{\tanh \frac{\g_{13}^{-\circ}}{2}}{\tanh \frac{\g_{13}^{+\circ}}{2}}
\frac{\tanh \frac{\g_{23}^{-\circ}}{2}}{\tanh \frac{\g_{23}^{+\circ}}{2}}\,
\PhiSG(\g_{13}^{-\circ})\,\PhiSG(\g_{13}^{+\circ})\,\PhiSG(\g_{23}^{-\circ})\,\PhiSG(\g_{23}^{+\circ})
\\
&=-
\frac{\tanh \frac{\g_{13}^{-\circ}}{2}}{\tanh \frac{\g_{13}^{+\circ}}{2}}
\frac{\coth \frac{\g_{13}^{+\circ}}{2}}{\tanh \frac{\g_{23}^{+\circ}}{2}}\,
\,i\tanh \frac{\g_{13}^{+\circ}}{2}\,\PhiSG(\g_{13}^{-\circ})\, \PhiSG(\g_{23}^{+\circ})
\\
&=-i\coth\frac{\g_{13}^{+\circ}}{2}\,\frac{\tanh \frac{\g_{13}^{-\circ}}{2}}{\tanh \frac{\g_{23}^{+\circ}}{2}}
\,\PhiSG(\g_{13}^{-\circ})\, \PhiSG(\g_{23}^{+\circ})\,.
\end{aligned}
\end{equation}
We see that we are left with a rather unpleasant dependence on $\gamma_1^+$ through $\coth \frac{\g_{13}^{+\circ}}{2}$.
We have also to consider the BES factor and any rational prefactor. The BES factor $\sigma(x_1^\pm, x_2)$ has the same fusion properties as its massive progenitor $\sigma(x_1^\pm, x_2^\pm)$, \textit{i.e.}\ it can be fused over the massive particles without problems following ref.~\cite{Arutyunov:2009kf}. For the prefactor, we should now consider carefully which process to pick for the normalisation. The simplest picture, since the condition~\eqref{eq:fusionstring} identifies bound states in the symmetric representation~\cite{Borsato:2013hoa} (whose highest-weight state is $Y(p_1)Y(p_2)$), should arise when we are dealing with left-particles ($M_1=M_2=+1$). As a result, we should consider the normalisation appearing in the first line of~\eqref{eq:mixednormalisation}, \textit{i.e.}\ the prefactor
\begin{equation}
    \sqrt{\frac{x_1^+}{x_1^-}}e^{-i p_3}    \frac{x^-_1-x_3}{1-x^+_1x_3}
    \sqrt{\frac{x_2^+}{x_2^-}}e^{-i p_3}   \frac{x^-_1-x_3}{1-x^+_1x_3}
    =
    -\sqrt{\frac{x_2^+}{x_1^-}}e^{-2i p_3}\frac{x^-_1-x_3}{1-x^+_2x_3}\,\tanh\frac{\gamma^{+\circ}_{13}}{2}\,.
\end{equation}
We see that the last term cancels the $\coth \frac{\g_{13}^{+\circ}}{2}$ contribution as we wanted, leaving 
\begin{equation}
\label{eq:leftmassivefusion}
    +i\,
    \sqrt{\frac{x_2^+}{x_1^-}}e^{-2i p_3}\frac{x^-_1-x_3}{1-x^+_2x_3}\,
    \frac{\tanh \frac{\g_{13}^{-\circ}}{2}}{\tanh \frac{\g_{23}^{+\circ}}{2}}
\,\PhiSG(\g_{13}^{-\circ})\, \PhiSG(\g_{23}^{+\circ})\,
(\sigma_{13})^{-2}(\sigma_{23})^{-2}.
\end{equation}
 We could have fused two \textit{right-massive particles} ($M_1=M_2=-1$). Here the highest-weight state is $\bar{Z}(p_{1,2})$. Repeating the same treatment much of the discussion would go through in the same way, except that at the very end we would encounter a different rational pre-factor. If we look at $\bar{Z}(p_{1,2})$, the normalisation is fixed by the second line of~\eqref{eq:mixednormalisation}. The overall difference with the case we just considered is a rational term (whose form follows from representation theory and left-right symmetry~\cite{Borsato:2014hja}). The new term, which multiplies the fused S~matrix~\eqref{eq:leftmassivefusion}, is
\begin{equation}
    \frac{x_1^-}{x_1^+}\frac{(1-x_1^+x_3)^2}{(x_1^- - x_3)^2}
    \frac{x_2^-}{x_2^+}\frac{(1-x_2^+x_3)^2}{(x_2^- - x_3)^2}
    =\frac{x_1^-}{x_2^+}\frac{(1-x_2^+x_3)^2}{(x_1^- - x_3)^2}
    \frac{(1-x_1^+x_3)^2}{(x_1^+ - x_3)^2}\,.
\end{equation}
We see that the new prefactor does not fuse nicely (due to the last term in the last equation). This is not surprising because we are looking at a symmetric-representation bound state in the antisymmetric sector. To fuse this properly we would need (in the language of the Bethe-Yang equations) to also consider auxiliary Bethe roots which would remove this additional contribution.

\paragraph{Mirror bound states and fusion.}
Let us now consider massive bound states of the mirror theory, which will satisfy
\begin{equation}
    x_1^-=x_2^+\,,\qquad \gamma_1^- = \gamma_2^+ +i\pi\,,
\end{equation}
and transform in the antisymmetric representation. We start by looking at the fusion properties of $\sigmabc(\gamma_1^\pm,\gamma_2)^{-2}$. We find
\begin{equation}
\begin{aligned}
(\sigmabc_{13})^{-2}(\sigmabc_{23})^{-2} &=-
\frac{\tanh \frac{\g_{13}^{-\circ}}{2}}{\tanh \frac{\g_{13}^{+\circ}}{2}}
\frac{\tanh \frac{\g_{23}^{-\circ}}{2}}{\tanh \frac{\g_{23}^{+\circ}}{2}}\,
\PhiSG(\g_{13}^{-\circ})\,\PhiSG(\g_{13}^{+\circ})\,\PhiSG(\g_{23}^{-\circ})\,\PhiSG(\g_{23}^{+\circ})
\\
&=-
\frac{\coth \frac{\g_{23}^{+\circ}}{2}}{\tanh \frac{\g_{13}^{+\circ}}{2}}
\frac{\tanh \frac{\g_{23}^{-\circ}}{2}}{\tanh \frac{\g_{23}^{+\circ}}{2}}\,
\,i\tanh \frac{\g_{23}^{+\circ}}{2}\,\PhiSG(\g_{13}^{+\circ})\, \PhiSG(\g_{23}^{-\circ})
\\
&=-i\coth\frac{\g_{23}^{+\circ}}{2}\,\frac{\tanh \frac{\g_{13}^{+\circ}}{2}}{\tanh \frac{\g_{23}^{-\circ}}{2}}
\,\PhiSG(\g_{13}^{+\circ})\, \PhiSG(\g_{23}^{-\circ})\,.
\end{aligned}
\end{equation}
Let us now look at the rational pre-factor for the scattering of $\bar{Z}$ particles, which is what should give a simple result for the anti-symmetric bound state. We get an additional factor of
\begin{equation}
    \sqrt{\frac{x_1^-}{x_1^+}\frac{x_2^-}{x_2^+}}e^{-2i p_3}\frac{1- x_1^+ x_3}{x_1^--x_3}\frac{1- x_2^+ x_3}{x_2^--x_3}=\sqrt{\frac{x_1^-}{x_1^+}\frac{x_2^-}{x_2^+}}\frac{1- x_1^+ x_3}{x_2^--x_3}e^{-2i p_3} \coth\frac{\gamma_{23}^{+\circ}}{2}\,,
\end{equation}
which does not cancel the unwanted term --- rather, it produces its square. In fact, the term $\coth^2\tfrac{\gamma^{+\circ}_{23}}{2}$ is cancelled by terms coming from the fusion of the BES factor. In fact, much like in the ordinary (massive) case, the BES factor does not fuse well. It can actually be convenient to defined an ``improved'' dressing factor for the mirror region (which fuses well there, but not in the string region). In any case, considering all this, we find that the $\bar{Z}\bar{Z}$-scattering fuses well in the mirror region. The same will not be true for the scattering of $YY$, which will fuse up to terms involving the auxiliary Bethe-Yang roots.

\paragraph{Physical unitarity in the string and mirror regions.}
Let us consider the dressing factor~$\sigmabc (\g_1^\pm,\g_2)$ for particles with real momenta in the string region, where we have $(x^\pm)^*=x^\mp$ and $x^*=1/x$, as well as $(\gamma^\pm)^*=\gamma^\mp$ and $\gamma^*=\gamma$. We find
\begin{equation}
(\sigmabc_{12})^{-2} \left((\sigmabc_{12})^{-2}\right)^{*}= {\tanh {\g_{12}^{-\circ}\ov2}\ov \tanh {\g_{12}^{+\circ}\ov2}}
\varPhi(\g_{12}^{-\circ}) \varPhi(\g_{12}^{+\circ})
{\tanh {\g_{12}^{+\circ}\ov2}\ov \tanh{\g_{12}^{-\circ}\ov2}}
{1\ov\varPhi(\g_{12}^{+\circ}) \varPhi(\g_{12}^{-\circ})} 
=1\,.
\end{equation}
It can be checked that the prefactors in~\eqref{eq:mixednormalisation}, as well as the BES factor, are also unitary in the string region.
Coming now to the mirror region, we have that the dressing factor now takes the form
\begin{equation}
    \widetilde{(\sigmabc_{12})}^{-2}=i
    \frac{\tanh\frac{\tilde{\gamma}_{12}^{-\circ}-i\pi/2}{2}}{\tanh\frac{\tilde{\gamma}_{12}^{+\circ}-i\pi/2}{2}}\,\varPhi(\tilde{\gamma}^{+\circ}_{12}-\tfrac{i}{2}\pi)\,\varPhi(\tilde{\gamma}^{-\circ}_{12}-\tfrac{i}{2}\pi)\,,
\end{equation}
where the wide tilde denotes that the whole expression has been analytically continued to the mirror-mirror region.
Recall that  for real mirror momenta $(\tilde{x}^\pm)^*=1/\tilde{x}^\mp$ and $\tilde{x}^*=\tilde{x}$. In terms of the rapidities, this gives $(\tilde{\gamma}^\pm)^*=\gamma^\mp+i\pi$ and $\tilde{\gamma}^*=\tilde{\gamma}$. Thus,
\begin{equation}
\begin{aligned}
    \widetilde{(\sigmabc_{12})}^{-2}\Big(\widetilde{(\sigmabc_{12})}^{-2}\Big)^*&=
    \frac{\tanh\frac{\tilde{\gamma}_{12}^{-\circ}-i\pi/2}{2}}{\tanh\frac{\tilde{\gamma}_{12}^{+\circ}-i\pi/2}{2}}\frac{\tanh\frac{\tilde{\gamma}_{12}^{+\circ}+3i\pi/2}{2}}{\tanh\frac{\tilde{\gamma}_{12}^{-\circ}+3i\pi/2}{2}}
    \,\frac{\varPhi(\tilde{\gamma}^{+\circ}_{12}-\frac{i}{2}\pi)\,\varPhi(\tilde{\gamma}^{-\circ}_{12}-\frac{i}{2}\pi)}{\varPhi(\tilde{\gamma}^{-\circ}_{12}+\frac{3i}{2}\pi)\,\varPhi(\tilde{\gamma}^{+\circ}_{12}+\frac{3i}{2}\pi)}\,,\\
    &=\tanh^2\left(\tfrac{\tilde{\gamma}^{+\circ}_{12}-\frac{i}{2}\pi}{2}\right)\,\tanh^2\left(\tfrac{\tilde{\gamma}^{-\circ}_{12}-\frac{i}{2}\pi}{2}\right)\\
    &=\left(\frac{\tilde{x}^+_1 - \tilde{x}_2}{1-\tilde{x}^+_1\tilde{x}_2}\right)^2\left(\frac{1-\tilde{x}^-_1\tilde{x}_2}{\tilde{x}^-_1 - \tilde{x}_2}\right)^2.
\end{aligned}
\end{equation}
In the second equality we have used the periodicity of the hyperbolic tangent and the $2\pi i$-monodromy of the Sine-Gordon factor,
\begin{equation}
    \frac{\varPhi(z)}{\varPhi(z+2\pi i)}=
    \frac{\varPhi(z)\,\varPhi(z+i\pi)}{\varPhi(z+i\pi)\,\varPhi(z+2\pi i)}=\tanh^2\Big(\frac{z}{2}\Big)\,.
\end{equation}
The rational prefactor also is  not invariant under physical unitarity,
\begin{equation}
    \sqrt{\frac{\tilde{x}^+_1}{\tilde{x}^-_1}}\frac{1}{\tilde{x}_2}\frac{\tilde{x}^-_1-\tilde{x}_2}{1-\tilde{x}^+_1\tilde{x}_2}
    \Bigg(\sqrt{\frac{\tilde{x}^+_1}{\tilde{x}^-_1}}\frac{1}{\tilde{x}_2}\frac{\tilde{x}^-_1-\tilde{x}_2}{1-\tilde{x}^+_1\tilde{x}_2}\Bigg)^*=\frac{1}{(\tilde{x}_2)^4}\,.
\end{equation}
It remains to check the contribution of the BES dressing factor in the mirror-mirror region. As we discuss in appendix~\ref{app:BES:mixed}, we have
\begin{equation}
    \big(\BES^{\bullet\circ}(\tilde{x}_1^\pm,\tilde{x}_2)\big)^{-2}
    \Big(\big(\BES^{\bullet\circ}(\tilde{x}_1^\pm,\tilde{x}_2)\big)^{-2}\Big)^*=(\tilde{x}_2)^4
    \Big(\frac{1-\tilde{x}_1^+\tilde{x}_2}{\tilde{x}^+_1-\tilde{x}_2}\Big)^2\Big(\frac{\tilde{x}_1^--\tilde{x}_2}{1-\tilde{x}^-_1\tilde{x}_2}\Big)^2\,.
\end{equation}
We see a very non-trivial cancellation between the contributions of the Sine-Gordon factor and of the mixed-mass BES factor.

\paragraph{Zero-momentum limit.}
We may consider two distinct zero-momentum limits. One, which is conceptually simpler, is when we add a massive mode with zero momentum in the Bethe-Yang equations. This should represent a supersymmetry descendant.  In that case the Zhukovsky variables~$x^\pm(p_1)$ both go either to plus infinity or minus infinity, and we have
\begin{equation}
    \gamma^{\pm}_1=\mp \frac{i}{2}\pi\,.
\end{equation}
Hence we have
\begin{equation}
    (\sigmabc_{12})^{-2}=i\frac{\tanh\tfrac{+\frac{i}{2}\pi-\gamma_2}{2}}{\tanh\tfrac{-\frac{i}{2}\pi-\gamma_2}{2}}\,\varPhi(+\tfrac{i}{2}\pi-\gamma_2)\,\varPhi(+\tfrac{i}{2}\pi-\gamma_2)
    =
    \tanh\big(\frac{\gamma_2}{2}-\frac{i\pi}{4}\big)= - x_2\,.
\end{equation}
Observing that the BES phase does not contribute in this limit, this gives the scattering elements
\begin{equation}
    \gen{S}|Y_0\chi_{p_2}\rangle= -e^{-\tfrac{3}{2}ip_2}x_2|\chi_{p_2}Y_0\rangle\,,\qquad
    \gen{S}|\bar{Z}_0\chi_{p_2}\rangle= -e^{-\tfrac{1}{2}ip_2}x_2|\chi_{p_2}\bar{Z}_0\rangle\,,
\end{equation}
which can be further simplified depending on the chirality of $p_2$. In the physical region where $p_2$ is anti-chiral, $x_2=-e^{\frac{i}{2}p_2}$ and we get that the scattering reduces to $e^{-i J_1p_2}$, where $J_1$ is the R-charge of the first particle.
We may also consider the case where the massless particle has zero momentum, which means $\gamma=\pm\infty$. Recall that $\varphi(\pm\infty)=\pm i \pi/4$. As a consequence, when considering the dressing factor in presence of a zero-momentum massless particle in the physical region we have
\begin{equation}
    \sigmabc(-\infty) = i \varPhi(-\infty)^{2}=1\,.
\end{equation}
While the BES phase does not contribute in this limit, it is interesting to observe that the rational prefactor in the highest-weight scattering elements is not trivial; rather, it takes the form \textit{e.g.}
\begin{equation}
    e^{i p_1/2}\frac{1+x^-_1}{1+x^+_1}\,,
\end{equation}
which can be compensated by adding an auxiliary root in the Bethe-Yang equations (see section~\ref{sec:BYE}).

\subsubsection{Perturbative expansion}
Given that there is no standard definition of the normalisation of the dressing factors in the mixed-mass sector, we shall consider a whole S-matrix element in the $a=0$ uniform light-cone gauge~\cite{Arutyunov:2005hd}, see appendix~\ref{app:lcgauge}. Focusing on the scattering of $Y(p_1)$ and $\chi(p_2)$ in the kinematics region where $p_2<0$ we have, in the near-BMN expansion,
\begin{equation}
\label{eq:mixedmassexpansion}
\begin{aligned}
    \log\langle Y_1\chi^{\dot{\alpha}}_2|\mathbf{S}|Y_1\chi^{\dot{\alpha}}_2\rangle
    &=
    -\frac{i}{2h}\big(p_1(\omega_1-p_1)+2p_2\omega_1\big)\\
    &\quad-
    \frac{ip_1^2}{2\pi h^2}
    (\omega_1-p_1)p_2\,\log\Big(\tfrac{-(\omega_1-p_1)p_2}{4h}\Big)
    +O(h^{-3})
    \,,
\end{aligned}
\end{equation}
which is worked out in appendix~\ref{app:BMNexpansion}. 
A first observation is that this justifies the normalisation of  $\sigmabc(x_1^\pm,x_2)$ by an explicit factor of~$i$. In fact, this diagonal S-matrix element is correctly normalised so that $\langle Y_1\chi^{\dot{\alpha}}_2|\mathbf{S}|Y_1\chi^{\dot{\alpha}}_2\rangle=1+O(h^{-1})$, without any spurious signs.

This result should be compared with the existing perturbative computations. Unfortunately, for processes involving massive and massless external legs, not many results are known. Results have been computed up to one loop in ref.~\cite{Sundin:2016gqe}, and we report here their results, translated in the $a=0$ uniform lightcone gauge:
\begin{equation}
\begin{aligned}
    \log\langle Y_1\chi^{\dot{\alpha}}_2|\mathbf{S}_{\text{SW}}|Y_1\chi^{\dot{\alpha}}_2\rangle
    &=
    -\frac{i}{2h}\big(p_1(\omega_1-p_1)+2p_2\omega_1\big)\\
    &\quad-
    \frac{ip_1^2}{2\pi h^2}
    (\omega_1-p_1)p_2\,\left[1+\log\Big(\tfrac{\omega_1-p_1}{-2p_2}\Big)
    \right]+O(h^{-3})
    \,.
\end{aligned}
\end{equation}
We observe the following matches and mismatches:
\begin{enumerate}
    \item The tree-level result matches.
    \item The rational part of the one-loop result does not match. The difference is given by
\begin{equation}
    \frac{i}{2\pi h^2}(\omega_1-p_1)p_1^2p_2\,,
\end{equation}
which is actually quite reminiscent of~\eqref{eq:massivediscrepacy}. Like that term, this could be interpreted as arising from a local counter-term.
\item The logarithmic piece is actually quite different. In fact, the discrepancy of the logarithmic part was also noticed when comparing the proposal of~\cite{Borsato:2016xns} with ref.~\cite{Sundin:2016gqe}. The expansion of \cite{Borsato:2016xns} as well as ours naturally produces a one-loop result where the argument of the logarithm is of the form~\eqref{eq:mixedmassexpansion} and depends on~$h$. As far as we can see, the reason for the disagreement lies in the order of limits. In~\cite{Sundin:2016gqe} a UV regulator was  
removed first, and then the IR regulator, which was chosen to be a small mass of a particle, was taken to zero. The correct order is in fact opposite, and, moreover, one does not have to remove UV divergent terms because a natural UV regularisation for the model is a lattice one with 
the propagator replaced by $1/(m^2 + 4h^2\sin^2p/2h)$ so that the UV regulator, the inverse  lattice step, is identified with the coupling constant $h$. This follows from the form of the exact dispersion relation, and at one-loop order would naturally lead to the appearance of a $\log h$ term.
\end{enumerate}

\subsection{Massless sector}
\label{sec:proposal:massless}
We now come to the massless dressing factors, starting from the one scattering particles of the same chirality, see table~\ref{table:dressing}. Here we set
\begin{equation}
    \big(\sigmacc(\gamma_1,\gamma_2)\big)^{-2}
    =a(\gamma_{12})\, \big(\varPhi(\gamma_{12})\big)^2\,.
\end{equation}
As emphasised in section~\ref{sec:smatrix:symmetries} we need $a(\gamma)$, which satisfies the crossing equation $a(\gamma)a(\bar{\gamma})=-1$, in order to solve eq.~\eqref{eq:masslesscrossingrapid}. It is not unusual to encounter this function in the context of relativistic massless integrable QFTs, see \textit{e.g.}~\cite{Zamolodchikov:1991vx}. We stress once more that could have not obtained the minus sign by multiplying the square of the Sine-Gordon dressing factor $\varPhi^2(\gamma)$ by $\pm i$, as that would have spoiled braiding unitarity.%
\footnote{%
The pre-factor $a(\gamma)$ was not considered in ref.~\cite{Fontanella:2019ury} where the connection between the Sine-Gordon dressing factor and $AdS_3\times S^3\times T^4$ was first noticed.
} 
We now consider the case of opposite-chirality scattering. As we have commented in section~\ref{sec:smatrix:normalisation}, here \textit{we could} obtain the correct sign in the crossing equations~\eqref{eq:masslesscrossingrapid} by multiplying the dressing factor by $\pm i$. This is because braiding unitarity is not a constraint on a single dressing factor in the case of opposite chirality, but rather a relation between two distinct functions that have some freedom in their overall normalisation. Regardless, \textit{we shall assume} that the solution for opposite chirality scattering is the same as above, namely
\begin{equation}
\label{eq:oppositechiralitysol}
    \big(\sigmacct(\gamma_1,\gamma_2)\big)^{-2}
    =a(\gamma_{12})\, \big(\varPhi(\gamma_{12})\big)^2\,.
\end{equation}
The main reason for doing so is that it seems more natural not to have the dressing factor ``jump'' abruptly when changing the chirality of one of the two particles. We could always remove $a(\gamma)$ by multiplying eq.~\eqref{eq:oppositechiralitysol} by the CDD factor $i/a(\gamma)$. Ultimately, its perturbation theory will tell us which is the correct result (see below for the perturbative expansion of the dressing factors).
In conclusion we propose that all massless-massless dressing factors are given in terms of a single expression,
\begin{equation}
    \big(\widetilde{\phase}^{\circ\circ}(\gamma_1,\gamma_2)\big)^{-2}
    =\big(\phase^{\circ\circ}(\gamma_1,\gamma_2)\big)^{-2}= a(\gamma_{12})\, \big(\varPhi(\gamma_{12})\big)^2\,\big(\BES(x_1,x_2)\big)^{-2}\,.
\end{equation}
It is straightforward to verify that this satisfies the parity constraints of section~\ref{sec:smatrix:symmetries}.

\subsubsection{Properties}
Let us now analyze the main features of our proposal. Because the massless kinematics is quite restrictive --- the only physical values of the momenta, both in the mirror and in the string theory, are the real ones --- and because massless particles do not form bound states, the discussion will be a bit simpler than in the cases above.

\paragraph{Poles and branch points.}
The discussion of poles here is quite simple because the physical ``region'' is just given by the real~$\gamma$ line (either in the string or in the mirror theory). It is however worth remarking that the massless limit of the BES dressing factors has branch points inside and outside the unit disk, see section~\ref{sec:buildingblocks:BES}. While these points are outside of the physical region, we have to be careful when analytically continuing the dressing factor from the real string line (the upper-half circle in the $x$-plane) to the real mirror line (the segment $-1<x<1$). We will take any path that does not cross the cuts (such as a path that runs closely to the upper-half circle).

\paragraph{Physical unitarity in the string and mirror regions.}
The physical unitarity of $\sigmacc(\gamma_{12})$ in the string region follows straightforwardly from the properties of $a(\gamma)$ and $\varPhi(\gamma)$. Moreover, this also applies to the mirror-mirror region because, since $\sigmacc(\gamma_{12})$ depends on the difference of two massless rapidities, it does not change when both are taken to the mirror region. As for the massless-massless BES phase, by the results of appendix~\ref{app:BES:massless}, it is unitary by itself.

\paragraph{Zero-momentum limit.}
Let us now consider the case where either particle has zero momentum. We shall always order the particles so that if one particle has momentum equal to $0^+$, so that its velocity is positive (and maximum), it is the first particle; vice-versa, if the momentum is $0^-$, and the velocity is negative (and minimum), then this must be the second particle. This is the physical setup, and any other configuration can be related to this by using braiding unitarity. Hence, either $p_1=0^+$ and $\gamma_1=-\infty$, or $p_2=0^-$ and $\gamma_2=+\infty$. Either way, $\gamma_{12}=-\infty$. We can immediately verify that
\begin{equation}
    \big(\sigmacc(-\infty)\big)^{-2}=1\,,
\end{equation}
where the contribution of $a(-\infty)=i$ is important.
It remains to compute the contribution of the BES phase. From~\eqref{eq:masslessBES} for instance, it is manifest that the phase vanishes if $x_2=\pm1$; though that expression is not manifestly antisymmetric, the phase is antisymmetric by construction, so that it must also vanish if $x_1=\pm1$. In conclusion,
\begin{equation}
    \big(\phase^{\circ\circ}(0^+,p)\big)^{-2}=
    \big(\phase^{\circ\circ}(p,0^-)\big)^{-2}=1\,.
\end{equation}
It is also easy checking what is the form of the dressing factor for coincident momenta, as the BES and Sine-Gordon pieces are (separately) trivial. The only non-trivial contribution comes from $a(0)=-1$, so that
\begin{equation}
    \big(\phase^{\circ\circ}(p,p)\big)^{-2}=-1\,.
\end{equation}
This matched with what expected by general considerations in ref.~\cite{upcoming:massless}.

\subsubsection{Perturbative expansion}
The near-BMN expansion of the massless-massless dressing factor is reported in appendix~\ref{app:BMNexpansion}.  Working in the perturbative regime where $p_1>0$ and $p_2<0$ we find
\begin{equation}
\label{eq:masslessmasslessexp}
    \log\Big(\phase^{\circ\circ}(p_1,p_2)\Big)^{-2}
    =
    -\frac{i}{h}p_1p_2-\frac{i}{h^2}\frac{p_1p_2}{8}
    -i\frac{p_1p_2}{4\pi h^2} \left(\log\frac{-p_1p_2}{16h^2}-1\right)+O(h^{-3})\,.
\end{equation}
This result should be compared with the massless-massless perturbative computation of Sundin and Wulff~\cite{Sundin:2016gqe}, which gives
\begin{equation}
    \log\langle \chi^{\dot{\alpha}}_1\chi^{\dot{\beta}}_2|\mathbf{S}_{\text{SW}}|\chi^{\dot{\alpha}}_1\chi^{\dot{\beta}}_2\rangle
    =-\frac{i}{h}p_1p_2
    +
    i\frac{p_1p_2}{4\pi h^2}
    \Big(\log(-4p_1p_2)-1\Big)+O(h^{-3})
    \,.
\end{equation}
We observe the following matches and mismatches:
\begin{enumerate}
    \item At tree-level, our result matches with the perturbative computation of~\cite{Sundin:2016gqe}. In particular, the relevant term arises from the AFS order of the BES phase.
    \item At one-loop, the coefficient of the logarithm does not match; the sign is opposite. This is perhaps not entirely surprising in light of what we have encountered in the mixed-mass region.
    \item The argument of the logarithm is sensitive to the UV cutoff, which in our case is provided by the coupling constant~$h$ and in the case of Sundin and Wulff has been removed. As such, the finite pieces of the one-loop result can be removed by a change in the UV cutoff.
    \item It is interesting to notice that the one-loop result of Sundin and Wulff comes with a $1/\pi$ coefficient, while ours also involves a rational term. This discrepancy is reminiscent of the two-loop mismatch in the dispersion relation for massless modes, which in that case involves a relative factor of~$\pi^2$~\cite{Sundin:2015uva}. Nonetheless, it is interesting to note that the first $O(h^{-2})$ term in~\eqref{eq:masslessmasslessexp} is precisely due to $-i a(\gamma)$, which we may eliminate in the opposite-chirality scattering (at the price of having different dressing factors for same-chirality and opposite-chirality scattering).
\end{enumerate}

\subsection{Relations with earlier proposals}
\label{sec:proposal:relations}
We are now able to comment in more detail on the similarities and differences between our proposal and the solutions of the crossing equations found in~\cite{Borsato:2013hoa} for massive-massive scattering and in~\cite{Borsato:2016xns}  for mixed-mass and massless-massless scattering. We will see that such a comparison will result in a some precise relation between pieces of the two proposals.

In the massive sector, both our solution and that of ref.~\cite{Borsato:2013hoa} featured the BES phase, supplemented by an additional function at HL  order, \textit{i.e.}\ at order $O(h^0)$. Hence it is sufficient to compare those additional functions, which we shall do below. For mixed-mass and massless-massless scattering, instead, the difference between our proposal and that of~\cite{Borsato:2016xns} is much more fundamental: in the latter, the BES phase did not appear at all, but only its AFS and HL  orders did. One concern is that, at finite~$h$, the analytic properties of individual orders of the BES phase are less transparent than that of the whole function. In particular, the dressing factors of~\cite{Borsato:2016xns} feature the AFS phase, whose branch points depend on the relative rapidities of the two particles, see eq.~\eqref{eq:AFS}, like $x_1=1/x_2^\pm$, and so on. This makes it hard to perform the analytic continuation necessary, for instance, to define the dressing factors in the mirror theory. Physically, it is not hard to explain why (a limit of) the BES phase might appear in the mixed-mass and massless-massless scattering. In fact, massive particles can and do appear in the internal lines of mixed-mass and massless processes; the BES phase may be result of all such processes at all-loop order.

Let us now consider in more detail the $O(h^0)$ terms appearing in the various dressing factors.
We start by recalling that, by an asymptotic expansion of $\Phi(x,y)$ in $h\gg1$, see~\eqref{eq:chiBES}, we obtain at $O(h^0)$ order the HL integral $\Phi_{\text{HL}}$, see~\eqref{eq:HL}.
We can then define $\chi_{\text{HL}}(x_1,x_2) = \Phi_{\text{HL}}(x_1,x_2)$ in the region where $|x_1|>1$ and $|x_2|>1$. By means of this we may define the massive-massive HL phase
\begin{equation}
    \theta_{\text{HL}}(x^\pm_1,x^\pm_2)=
    \chi_{\text{HL}}(x_1^+,x_2^+)- \chi_{\text{HL}}(x_1^+,x_2^-)- \chi_{\text{HL}}(x_1^-,x_2^+)+ \chi_{\text{HL}}(x_1^-,x_2^-)\,.
\end{equation}
The mixed-mass limit of this expression, which we will denote simply by $\theta_{\text{HL}}(x^\pm_1,x_2)$, is defined by taking $x^+_2\to x_2$ on the upper-half circle and $x_2^-\to 1/x_2$~\cite{Borsato:2016xns}. The massless-massless limit is obtained by doing the same for $x_1^\pm$ and it will be denoted by $\theta_{\text{HL}}(x_1,x_2)$.

\paragraph{Sum of the massive dressing factors.}
In this case we have
\begin{equation}
\label{eq:massiveHLmatch}
    \Big(\sigmap(x_1^\pm,x_2^\pm)\Big)^{-2}=
    -{\tanh{\g_{12}^{-+}\ov2}\ov \tanh{\g_{12}^{+-}\ov2}}\,{ \varPhi(\g_{12}^{--})\varPhi(\g_{12}^{++}) \varPhi(\g_{12}^{-+})\varPhi(\g_{12}^{+-})}=e^{2i\theta_{\text{HL}}(\g_1^\pm,\g_2^\pm)}\,,
\end{equation}
where the last equation has been verified numerically. 
This means that, as far as the sum of the phases is concerned, our solution matches that of~\cite{Borsato:2013hoa}. Quite conveniently, the expression in terms of $\gamma_{\pm}$'s is of difference form which, as we have seen in the sections above, makes it much easier to understand its analytic properties and to consider bound states and fusion.

\paragraph{Difference of the massive dressing factors.}
In this case, we should not be comparing with the HL phase but rather with the difference phase of~\cite{Borsato:2013hoa}. We have  already seen in the near-BMN limit that our phase $\sigmam(x_1,x_2)$ is genuinely different from the previous proposal. While $\sigmam(x_1,x_2)$ is designed to take a simple (difference) form in terms of the $\gamma^\pm$ rapidities, the proposal of~\cite{Borsato:2013hoa} takes a simple form in the $x$-plane, or more precisely, on the $u$-plane.
However by looking more closely at the proposal of~\cite{Borsato:2013hoa}, see appendix~\ref{app:monodromy}, we find that it is not invariant under the parity transformation which indicates that, at least as defined, it cannot be correct. 

\paragraph{Mixed-mass dressing factor.}
By taking the massless limit of~\eqref{eq:massiveHLmatch} we can easily obtain an identity that involves the mixed-mass dressing factor, namely
\begin{equation}
    -{\tanh{\g_{12}^{-\circ}\ov2}\ov \tanh{\g_{12}^{+\circ}\ov2}}\, \varPhi(\g_{12}^{-\circ})^2\varPhi(\g_{12}^{+\circ})^2=e^{2i\theta_{\text{HL}}(x_1^\pm,x_2)}\,.
\end{equation}
In terms of our dressing factor~$\sigmabc(x_1^\pm,x_2)$ this means
\begin{equation}
    \frac{\tanh\frac{\gamma_{12}^{+\circ}}{2}}{\tanh\frac{\gamma_{12}^{-\circ}}{2}}\Big(\sigmabc(x_1^\pm,x_2\Big)^{-4}=
    e^{2i\theta_{\text{HL}}(x_1^\pm,x_2)}\,.
\end{equation}
If we were to consider a single power of $e^{i\theta_{\text{HL}}(x_1^\pm,x_2)}$, as it was done in~\cite{Borsato:2016xns}, this would yield a function with square-root branch points at $x_1^\pm=x_2$ and $x_1^\pm=1/x_2$, which would then recombine with the explicit square-root term in the normalisation of the S-matrix elements in~\cite{Borsato:2016xns}. It is immediate to see that this  removes all putative square-root branch points. Still, we remark that our proposal differs from theirs by the presence of the BES factor. The argument used in~\cite{Borsato:2016xns} to rule out the presence of the BES factor was the fact that, if one first performs the $h\gg1$ expansion of the massive factor and then takes the limit to the massless kinematics, only the AFS and HL orders survive. In hindsight, the problem with that argument is that taking the $h\gg1$ expansion and going to the massless kinematics are non-commuting operations. Moreover, when performing them in the order used in~\cite{Borsato:2016xns} (by expanding at $h\gg1$ under the integral sign), one finds that the resulting integrals are divergent in the massless kinematics and hence need to be regularised. We take this as a further indication that it is more appropriate to continue the whole BES factor to the massless kinematics, before taking any limit.

\paragraph{Massless dressing factor.}
It is also easy to take one further massless limit, in the above formulae, obtaining
\begin{equation}
    -\varPhi(\gamma_{12})^4=e^{2i\theta_{\text{HL}}(x_1,x_2)}\,.
\end{equation}
Taking the square root of this expression gives the HL order of the phase~\cite{Borsato:2016xns}. At this order, our solution differs by the $-a(\gamma_{12})$ factor; as we remarked, this factor is necessary if we insist that we have the same dressing factor both for same-chirality scattering and opposite-chirality scattering.
Once again, at all order our proposal involves the complete BES factor in the appropriate kinematics as opposed to the AFS~one.

\subsection{Application to mixed-flux backgrounds}
\label{sec:proposal:mixedflux}
The $AdS_3\times S^3\times T^4$ background can be supported by a mixture of Ramond-Ramond and Neveu-Schwarz-Neveu-Schwarz background fluxes, which does not spoil integrability~\cite{Cagnazzo:2012se}. In fact, the S~matrix for such backgrounds is remarkably similar to that of the pure-RR background~\cite{Hoare:2013pma}, even though the kinematics is rather different~\cite{Hoare:2013lja}. If the parameter~$h$ identifies  the strength RR background fluxes, and we introduce a new (quantised) parameter $k=1,2,\dots$ to indicate the amount of NSNS flux, the dispersion relation is~\cite{Hoare:2013lja,Lloyd:2014bsa}
\begin{equation}
    E(p)=\sqrt{\Big(M+\frac{k}{2\pi}p\Big)^2+4h^2\sin^2\Big(\frac{p}{2}\Big)}\,.
\end{equation}
It is possible to derive the matrix part of the S~matrix~\cite{Lloyd:2014bsa} and express it in terms of the Zhukovsky variables (following the notation of~\cite{Eden:2021xhe})
\begin{equation}
\label{eq:mixedZhukovsky}
\begin{aligned}
x^{\pm}_{\text{\tiny L}}(p) &=
\frac{\big(1+\tfrac{k}{2\pi}p\big)+\sqrt{\big(1+\tfrac{k}{2\pi}p\big)^2+4h^2\sin^2(\tfrac{p}{2})}}{2h\,\sin(\tfrac{p}{2})}\,e^{\pm \frac{i}{2} p},\\
x^{\pm}_{\text{\tiny R}}(p) &=
\frac{\big(1-\tfrac{k}{2\pi}p\big)+\sqrt{\big(1-\tfrac{k}{2\pi}p\big)^2+4h^2\sin^2(\tfrac{p}{2})}}{2h\,\sin(\tfrac{p}{2})}\,e^{\pm \frac{i}{2} p},\\
x^{\pm}_{\text{\tiny 0}}(p) &=
\frac{\big(0+\tfrac{k}{2\pi}p\big)+\sqrt{\big(0+\tfrac{k}{2\pi}p\big)^2+4h^2\sin^2(\tfrac{p}{2})}}{2h\,\sin(\tfrac{p}{2})}\,e^{\pm \frac{i}{2} p}.
\end{aligned}
\end{equation}
Up to taking care of distinguishing ``left'' and ``right'' Zhukovsky variables, the S-matrix elements are essentially the same as in the pure-RR case, so that the same is also true for the crossing equations. In particular, for the massive dressing factors we have~\cite{Lloyd:2014bsa}
\begin{equation}
\begin{aligned}
\sigma^{\bullet\bullet}_{\text{\tiny LL}} (x_{\text{\tiny L}1}^\pm,x_{\text{\tiny L}2}^\pm)^{2}\tilde\sigma^{\bullet\bullet}_{\text{\tiny RL}} (\bar{x}_{\text{\tiny R}1}^\pm,x_{\text{\tiny L}2}^\pm)^{2}&=
\left(\frac{x_{\text{\tiny L}2}^-}{x_{\text{\tiny L}2}^+}\right)^{2}
\frac{(x_{\text{\tiny L}1}^- - x_{\text{\tiny L}2}^+)^{2}}{(x_{\text{\tiny L}1}^- - x_{\text{\tiny L}2}^-)(x_{\text{\tiny L}1}^+ - x_{\text{\tiny L}2}^+)}
\frac{1-\frac{1}{x_{\text{\tiny L}1}^-x_{\text{\tiny L}2}^+}}{1-\frac{1}{x_{\text{\tiny L}1}^+x_{\text{\tiny L}2}^-}},\\
\sigma^{\bullet\bullet}_{\text{\tiny RR}} (\bar{x}_{\text{\tiny R}1}^\pm, x_{\text{\tiny R}2}^\pm)^{2}\tilde\sigma^{\bullet\bullet}_{\text{\tiny LR}} (x_{\text{\tiny L}1}^\pm,x_{\text{\tiny R}2}^\pm)^{2}&=
\left(\frac{x_{\text{\tiny R}2}^-}{x_{\text{\tiny R}2}^+}\right)^{2}
\frac{\big(1-\frac{1}{x^+_{\text{\tiny L}1}x^+_{\text{\tiny R}2}}\big)\big(1-\frac{1}{x^-_{\text{\tiny L}1}x^-_{\text{\tiny R}2}}\big)}{\big(1-\frac{1}{x^+_{\text{\tiny L}1}x^-_{\text{\tiny R}2}}\big)^{2}}
\frac{x_{\text{\tiny L}1}^--x_{\text{\tiny R}2}^+}{x_{\text{\tiny L}1}^+-x_{\text{\tiny R}2}^-}.
\end{aligned}
\end{equation}
despite this apparent simplicity, no solution to these crossing equations is known due to their unusual underlying kinematics. The solution of the mixed-mass and massless crossing equations is also unknown when $k\neq0$.

We will see in a moment that the approach we used above allows us to find a solution, at least formally, in terms of the very same functions used for the pure-RR case. First we introduce the BES factor as a function of the Zhukovsky variables of~\eqref{eq:mixedZhukovsky}. Formally, it satisfies the same crossing equations when these are expressed in terms of the Zhukovsky variables. Like we did above, we can strip out the BES factor to get a simpler set of equations, 
\begin{equation}
\begin{aligned}
    \Big(\sigmabb_{\text{\tiny LL}}(x_{\text{\tiny L}1}^\pm,x_{\text{\tiny L}2}^\pm)\Big)^{-2}\Big(\sigmabbt_{\text{\tiny RL}}(\bar{x}_{\text{\tiny L}1}^\pm,x_{\text{\tiny L}2}^\pm)\Big)^{-2}&=\frac{1-x_{\text{\tiny L}1}^+x_{\text{\tiny L}2}^+}{1-x_{\text{\tiny L}1}^-x_{\text{\tiny L}2}^+}\frac{1-x_{\text{\tiny L}1}^-x_{\text{\tiny L}2}^-}{1-x_{\text{\tiny L}1}^+x_{\text{\tiny L}2}^-}\,,\\
    \Big(\sigmabb_{\text{\tiny RR}}(\bar{x}_{\text{\tiny R}1}^\pm,x_{\text{\tiny R}2}^\pm)\Big)^{-2}\big(\sigmabbt_{\text{\tiny LR}}(x_{\text{\tiny L}1}^\pm,x_{\text{\tiny R}2}^\pm)\Big)^{-2}&=\frac{x_{\text{\tiny L}1}^+ - x_{\text{\tiny R}2}^-}{x_{\text{\tiny L}1}^+ - x_{\text{\tiny R}2}^+}\frac{x_{\text{\tiny L}1}^- - x_{\text{\tiny R}2}^+}{x_{\text{\tiny L}1}^- - x_{\text{\tiny R}2}^-}\,,
\end{aligned}
\end{equation}
which can be straightforwardly expressed in terms of rapidities, giving
\begin{equation}
\begin{aligned}
    \Big(\sigmabb_{\text{\tiny LL}}(x_{\text{\tiny L}1}^\pm,x_{\text{\tiny L}2}^\pm)\Big)^{-2}\Big(\sigmabbt_{\text{\tiny RL}}(\bar{x}_{\text{\tiny L}1}^\pm,x_{\text{\tiny L}2}^\pm)\Big)^{-2}&
    =
    \frac{
    \cosh(\frac{1}{2}\gamma^{++}_{\text{\tiny LL},12})\,\cosh(\frac{1}{2}\gamma^{--}_{\text{\tiny LL},12})
    }{
    \sinh(\frac{1}{2}\gamma^{+-}_{\text{\tiny LL},12})\,\sinh(\frac{1}{2}\gamma^{-+}_{\text{\tiny LL},12})
    }
    \,,\\
    \Big(\sigmabb_{\text{\tiny RR}}(\bar{x}_{\text{\tiny R}1}^\pm,x_{\text{\tiny R}2}^\pm)\Big)^{-2}\big(\sigmabbt_{\text{\tiny LR}}(x_{\text{\tiny L}1}^\pm,x_{\text{\tiny R}2}^\pm)\Big)^{-2}&
    =
    \frac{
    \cosh(\frac{1}{2}\gamma^{+-}_{\text{\tiny LR},12})\,\cosh(\frac{1}{2}\gamma^{-+}_{\text{\tiny LR},12})
    }{
    \sinh(\frac{1}{2}\gamma^{++}_{\text{\tiny LR},12})\,\sinh(\frac{1}{2}\gamma^{--}_{\text{\tiny LR},12})
    }
    \,,
\end{aligned}
\end{equation}
which can be solved in terms of the functions~$\varphi^{\bullet\bullet}$ ans $\tilde{\varphi}^{\bullet\bullet}$ introduced in section~\ref{sec:proposal:massive}, exploiting in particular the crossing relation~\eqref{eq:crossingregularisation}. To this end it is sufficient to set
\begin{equation}
\begin{aligned}
&\left(\sigma_{\text{\tiny LL}}^{\bullet\bullet}(x^\pm_{\text{\tiny L}1},x^\pm_{\text{\tiny L}2})\right)^{-2} =
-
\frac{
\sinh(\frac{1}{2}\gamma^{-+}_{\text{\tiny LL},12})
}{
\sinh(\frac{1}{2}\gamma^{+-}_{\text{\tiny LL},12})
}\,
e^{\varphi^{\bullet\bullet}(\gamma_{\text{L}1}^\pm,\gamma_{\text{L}2}^\pm)}\,\left(\BES(x_{\text{\tiny L}1}^\pm,x_{\text{\tiny L}2}^\pm)\right)^{-2},\\
&\left(\sigma_{\text{\tiny RR}}^{\bullet\bullet}(x^\pm_{\text{\tiny R}1},x^\pm_{\text{\tiny R}2})\right)^{-2} =
-
\frac{
\sinh(\frac{1}{2}\gamma^{-+}_{\text{\tiny RR},12})
}{
\sinh(\frac{1}{2}\gamma^{+-}_{\text{\tiny RR},12})
}\,
e^{\varphi^{\bullet\bullet}(\gamma_{\text{R}1}^\pm,\gamma_{\text{L}2}^\pm)}\,\left(\BES(x_{\text{\tiny R}1}^\pm,x_{\text{\tiny R}2}^\pm)\right)^{-2},
\end{aligned}
\end{equation}
and
\begin{equation}
\begin{aligned}
&\left(\widetilde{\sigma}_{\text{\tiny LR}}^{\bullet\bullet}(x^\pm_{\text{\tiny L}1},x^\pm_{\text{\tiny R}2})\right)^{-2} =
+
\frac{
\cosh(\frac{1}{2}\gamma^{+-}_{\text{\tiny LR},12})
}{
\cosh(\frac{1}{2}\gamma^{-+}_{\text{\tiny LR},12})
}\,
e^{\tilde{\varphi}^{\bullet\bullet}(\gamma_{\text{L}1}^\pm,\gamma_{\text{R}2}^\pm)}\,\left(\BES(x_{\text{\tiny L}1}^\pm,x_{\text{\tiny R}2}^\pm)\right)^{-2},\\
&\left(\widetilde{\sigma}_{\text{\tiny RL}}^{\bullet\bullet}(x^\pm_{\text{\tiny R}1},x^\pm_{\text{\tiny L}2})\right)^{-2} =
+
\frac{
\cosh(\frac{1}{2}\gamma^{+-}_{\text{\tiny RL},12})
}{
\cosh(\frac{1}{2}\gamma^{-+}_{\text{\tiny RL},12})
}\,
e^{\tilde{\varphi}^{\bullet\bullet}(\gamma_{\text{R}1}^\pm,\gamma_{\text{L}2}^\pm)}\,\left(\BES(x_{\text{\tiny R}1}^\pm,x_{\text{\tiny L}2}^\pm)\right)^{-2},
\end{aligned}
\end{equation}
which is consistent with the left-right symmetry of the model~\cite{Lloyd:2014bsa}.
While a detailed analysis of these dressing factors, as well as of those involving massless particles, requires a thorough understanding of the analytic properties of the mixed-flux $x$-, $z$- and $\gamma$-planes, it is very encouraging that the factorisation structure appears to be quite robust. We believe that this approach can be used to resolve the outstanding question of determining the mixed-flux dressing factors, and we plan to present those results elsewhere~\cite{upcoming:mixed}.

\section{Bethe-Yang equations}
\label{sec:BYE}
To conclude, we write the full Bethe-Yang equations using the normalisation as well as the dressing factors constructed above. The auxiliary equations, which were derived in~\cite{Borsato:2016xns}, will be unchanged.

Let us start by summarising the type of excitations. First of all, we will have $N_1$ massive ``left'' particles with $M=+1$; the highest-weight state of their representation is $Y(p)$.  Then, we will have $N_{\bar{1}}$ ``right'' particles with $M=-1$ (highest-weight state $\bar{Z}(p)$. Then, we will have $N_0^{(\dot{\alpha})}$ massless excitations, with $\dot{\alpha}=1,2$ distinguishing whether we are considering the representation with highest-weight state $\chi^{\dot{\alpha}}(p)$ with $\dot{\alpha}=1$ or $\dot{\alpha}=2$. 
Note that the S~matrix (and hence the Bethe equations) are blind to $\dot{\alpha}=1,2$. Nonetheless, the split between $N_0^{(1)}$ and $N_0^{(2)}$ is important to reproduce the correct degeneracy of the states.
From these highest-weight states we can create descendants  by acting with the lowering operators of $psu(1|1)^{\oplus 4}$ centrally extended. There are four such lowering operators, two of which are associated to  the ``left'' part of the algebra (and we associate to them $N_y^{(1)}$ and $N_y^{(2)}$ auxiliary roots) and two of which are associated to the ``right'' part of the algebra ($N_y^{(\bar 1)}$ and $N_y^{(\bar 2)}$). However, due to the central extension relating the left- and right-algebras, these sets are  equivalent (acting with a right charge is tantamount to acting with a left one, at generic values of momentum and coupling). Hence, for regular roots we can treat as a single family $N_y^{( 1)}$ and $N_y^{(\bar 1)}$, as well as $N_y^{( 2)}$ and $N_y^{(\bar 2)}$.%
\footnote{At $h\ll1$ some of the auxiliary will go to infinity, and other will go to zero. This will reproduce the split between left- and right- supercharges, and make it manifest that they can be associated to two copies of $psu(1,1|2)$~\cite{Borsato:2013qpa}.}
We collect in table~\ref{table:roots} the notation for the roots.

\begin{table}[t]
\centering
\begin{tabular}{|l | l|} 
 \hline
 Excitation numbers & Particles \\
 \hline
 $N_1$& Left momentum-carrying mode ($Y(p)$).\\
 $N_{\bar{1}}$& Right momentum-carrying mode ($\bar{Z}(p)$).\\
 $N_0^{(1)}$& Massless momentum-carrying mode, flavour $\dot{\alpha}=1$ ($\chi^1(p)$).\\
 $N_0^{(2)}$& Massless momentum-carrying mode, flavour  $\dot{\alpha}=2$ ($\chi^2(p)$).\\
$N_y^{(1)}$& Auxiliary root with $\alpha=1$ (lowering operator $\gen{Q}^{1}$ or $\widetilde{\gen{S}}^1$).\\
$N_y^{(2)}$& Auxiliary root with $\alpha=2$ (lowering operator $\gen{Q}^{2}$ or $\widetilde{\gen{S}}^2$).\\
 \hline
\end{tabular}
\caption{A summary of the excitations number appearing in the Bethe-Yang equations. We refer to the discussion of the $psu(1|1)^{\oplus4}$ centrally extended as presented for instance in~\cite{upcoming:massless}.}
\label{table:roots}
\end{table}

We begin with the equation for a ``left'' magnon (with $M=+1$), of momentum $p_k$ ($k=1,\dots N_1$) which reads
\begin{equation}
    \begin{aligned}
    &1=
    e^{ip_kL}\prod_{\substack{j=1\\j\neq k}}^{N_1}
    e^{+i p_k}e^{-i p_j}
    \frac{x^-_k-x^+_j}{x^+_k-x^-_j}
    \frac{1-\frac{1}{x^-_kx^+_j}}{1-\frac{1}{x^+_kx^-_j}}\big(\phase^{\bullet\bullet}_{kj}\big)^{-2}
    \prod_{j=1}^{N_{\bar{1}}}e^{-i p_j}
    \frac{1-\frac{1}{x^-_kx^-_j}}{1-\frac{1}{x^+_kx^+_j}}
    \frac{1-\frac{1}{x^-_kx^+_j}}{1-\frac{1}{x^+_kx^-_j}}\big(\tilde{\phase}^{\bullet\bullet}_{kj})^{-2}
    \\
    &\qquad
    \times
    \prod_{\dot{\alpha}=1,2}
    \prod_{j=1}^{N_0^{(\dot{\alpha})}}e^{+\frac{i}{2} p_k}e^{-i p_j}
    \frac{x^-_k-x_j}{1-x^+_kx_j}\big(\phase^{\bullet\circ}_{kj}\big)^{-2}
    \prod_{\alpha=1,2}
    \prod_{j=1}^{N_y^{(\alpha)}}e^{-\tfrac{i}{2}p_k}\frac{x^+_k-y_{j}^{(\alpha)}}{x^-_k-y_{j}^{(\alpha)}}\,,
    \end{aligned}
\end{equation}
To keep the notation light we omitted the index $\dot{\alpha}$ from the rapidity of the massless modes.
For a ``right'' magnon ($M=-1$) with momentum $p_k$ ($k=1,\dots N_{\bar{1}}$) we have
\begin{equation}
    \begin{aligned}
    &1=
    e^{ip_kL}\prod_{\substack{j=1\\j\neq k}}^{N_{\bar{1}}}\frac{x^+_k-x^-_j}{x^-_k-x^+_j}
    \frac{1-\frac{1}{x^-_kx^+_j}}{1-\frac{1}{x^+_kx^-_j}}\big(\phase^{\bullet\bullet}_{kj}\big)^{-2} \prod_{j=1}^{N_1}e^{ip_k}
    \frac{1-\frac{1}{x^+_kx^+_j}}{1-\frac{1}{x^-_kx^-_j}}
    \frac{1-\frac{1}{x^-_kx^+_j}}{1-\frac{1}{x^+_kx^-_j}}\big(\widetilde{\phase}^{\bullet\bullet}_{kj}\big)^{-2}
    \\
    &\qquad\times
    \prod_{\dot{\alpha}=1,2}
    \prod_{j=1}^{N_0^{(\dot{\alpha})}}e^{-\frac{i}{2} p_k}e^{-i p_j}
    \frac{1-x^+_kx_j}{x^-_k-x_j}\big(\phase^{\bullet\circ}_{kj}\big)^{-2}
    \prod_{\alpha=1,2}
    \prod_{j=1}^{N_y^{(\alpha)}}e^{-\tfrac{i}{2}p_k}\frac{1-\frac{1}{x^-_ky_{j}^{(\alpha)}}}{1-\frac{1}{x^+_ky_{j}^{(A)}}}\,.
    \end{aligned}
\end{equation}
A massless magnon of momentum $p_k$ could belong to the representation with highest-weight state $\chi^1(p)$ or to the one with highest-weight state $\chi^2(p)$. The equation for a particle of the former type is
\begin{equation}
    \begin{aligned}
    &1=
    e^{ip_kL}
    \prod_{\substack{j=1\\j\neq k}}^{N_{0}^{(1)}}\big(\phase^{\circ\circ}_{kj}\big)^{-2}
    \prod_{j=1}^{N_{0}^{(2)}}\big(\phase^{\circ\circ}_{kj}\big)^{-2}
     \prod_{j=1}^{N_1}e^{+i p_k}e^{-\frac{i}{2} p_j}
    \frac{x_kx^+_j-1}{x_k-x^-_j}\big(\phase^{\circ\bullet}_{kj}\big)^{-2}
   \\
    &\qquad\times 
    \prod_{j=1}^{N_{\bar{1}}}e^{+i p_k}e^{+\frac{i}{2} p_j}
    \frac{x_k-x^-_j}{x_kx^+_j-1}\big(\phase^{\circ\bullet}_{kj}\big)^{-2}
    \prod_{\alpha=1,2}
    \prod_{j=1}^{N_y^{(\alpha)}}e^{-\tfrac{i}{2}p_k}\frac{x_k-y_{j}^{(\alpha)}}{\frac{1}{x_k}-y_{j}^{(\alpha)}}
   \,.
    \end{aligned}
\end{equation}
Finally, the auxiliary Bethe equations read, for $\alpha =1,2$
\begin{equation}
    1=\prod_{j=1}^{N_1}\frac{y_{k}^{(\alpha)}-x^+_j}{y_{k}^{(\alpha)}-x^-_j}e^{-ip_j/2}
    \prod_{j=1}^{N_{\bar{1}}}\frac{1-\frac{1}{y_{k}^{(\alpha)}x^-_j}}{1-\frac{1}{y_{k}^{(\alpha)}x^+_j}}e^{-ip_j/2}
    \prod_{\dot{\alpha}=1,2}
    \prod_{j=1}^{N_0^{(\dot{\alpha})}}\frac{y_{k}^{(\alpha)}-x_j}{y_{k}^{(\alpha)}-\frac{1}{x_j}}e^{-ip_j/2}\,,
\end{equation}
where $k=1,\dots N_y^{(\alpha)}$.

Let us also summarise the various dressing factors that enter in the Bethe-Yang equations. For massive excitations we have
\begin{equation}
\begin{aligned}
    \big(\phase^{\bullet\bullet}_{12}\big)^{-2}=&-\frac{\sinh\tfrac{\gamma^{-+}_{12}}{2}}{\sinh\tfrac{\gamma^{+-}_{12}}{2}}e^{\varphi^{\bullet\bullet}(\gamma^\pm_1,\gamma^\pm_2)}\ \BES^{-2}(x_1^\pm,x_2^\pm)\,,\\
    \big(\widetilde{\phase}^{\bullet\bullet}_{12}\big)^{-2}=&+\frac{\cosh\tfrac{\gamma^{+-}_{12}}{2}}{\cosh\tfrac{\gamma^{-+}_{12}}{2}}e^{\tilde\varphi^{\bullet\bullet}(\gamma^\pm_1,\gamma^\pm_2)}\ \BES^{-2}(x_1^\pm,x_2^\pm)\,.
\end{aligned}
\end{equation}
For mixed-mass scattering we have
\begin{equation}
\begin{aligned}
    \big(\phase^{\bullet\circ}_{12}\big)^{-2}=&\ i\,\frac{\tanh\tfrac{\gamma^{-\circ}_{12}}{2}}{\tanh\tfrac{\gamma^{+\circ}_{12}}{2}}e^{\tfrac{1}{2}(\varphi^{\bullet\bullet}(\gamma^\pm_1,\gamma_2)+\tilde{\varphi}^{\bullet\bullet}(\gamma^\pm_1,\gamma_2))}\ \BES^{-2}(x_1^\pm,x_2)\,\\
    =&\ i\,\frac{\tanh\tfrac{\gamma^{-\circ}_{12}}{2}}{\tanh\tfrac{\gamma^{+\circ}_{12}}{2}}\varPhi(\gamma_{12}^{+\circ})\varPhi(\gamma_{12}^{-\circ})\,\BES^{-2}(x_1^\pm,x_2)\,.
\end{aligned}
\end{equation}
Finally, for massless scattering we are picking the same solution regardless of the chirality of the scattered particle, and we have
\begin{equation}
    \big(\phase^{\circ\circ}_{12}\big)^{-2}=\ \big(\widetilde{\phase}^{\circ\circ}_{12}\big)^{-2}=a(\gamma_{12})\,\varPhi(\gamma_{12}^{\circ\circ})^2\ \BES^{-2}(x_1,x_2)\,.
\end{equation}

\section{Conclusions}
\label{sec:conclusions}
We have presented a new solution to the crossing equations for $AdS_3\times S^3\times T^4$. The general structure of our solution is such that all of the dressing factors include a BES factor (in the appropriate kinematics) times a piece which depends on the difference of rapidities which we introduced following~\cite{Beisert:2006ib,Fontanella:2019baq}.

In some ways, our solution is similar to that of~\cite{Borsato:2013hoa,Borsato:2016xns}. In fact, for the product of the massive dressing factors, which we called~$\sigmap(x_1^\pm,x_2^\pm)$, we find that our solution coincides with the HL  phase used in~\cite{Borsato:2013hoa}; in fact, as a byproduct of our work, we find a difference representation of the HL phase which is very convenient both for its analytic continuation and  for fusion. 
However, already for the difference of the massive phases~$\sigmam(x_1^\pm,x_2^\pm)$ we find something fundamentally different from~\cite{Borsato:2013hoa}. Our solution is minimal when formulated in the $\gamma^\pm$-plane (see appendix~\ref{app:Fourier}) whereas the one of~\cite{Borsato:2013hoa} appears perhaps  more natural on the $u$-plane. However, as we argue in appendix~\ref{app:monodromy}, the previous proposal is incompatible with parity invariance. This is a strong indication that it needs to be modified.
In any case, it would be nice to carefully compute the one-loop dressing factor $\sigmam(x_1^\pm,x_2^\pm)$, to see whether it agrees with our proposal. It is also intriguing that, as we discussed in section~\ref{sec:proposal:massive}, the difference between our proposal and the existing perturbative computation is quite small, and could be due to a local counterterm. Such counterterms are known to be sometimes necessary in the renormalisation of integrable models, see \textit{e.g.}~\cite{deVega:1981ka,Bonneau:1984pj}.

When considering the dressing factors that involve massless modes, the differences with~\cite{Borsato:2016xns} are rather substantial. The first key difference is that the path for the crossing transformation which we introduce in section~\ref{sec:rapidities:massless} is different from that used in~\cite{Borsato:2016xns}. Our choice is dictated by the compatibility with the mirror transformation, which was not analysed in the literature thus far. The main difference in the functional form of the solutions which we found is that they depend on the BES phase, rather than on its leading and sub-leading order pieces (the AFS and HL phase). One argument made in~\cite{Borsato:2016xns} to justify the appearance of the AFS and HL orders only is that the remaining pieces of the phase would go to zero. We find that, when $h$ is finite, this is not the case --- there is a substantial difference between the AFS and HL orders of the phase, and the whole BES phase. This highlights a difference between asymptotically expanding the BES phase, and going to the massless kinematics (as it was done in~\cite{Borsato:2016xns}), \textit{versus} going to the massless kinematics for the finite-coupling phase. The latter procedure, which we followed here, is most natural when considering  the finite-coupling and finite-volume spectrum of the theory.
Additionally, it is relatively straightforward to analytically continue our proposal to other kinematic regimes (such as the mirror one), while this is harder for the proposal of~\cite{Borsato:2016xns} due to the appearance of the AFS phase.

Indeed, our proposal for the dressing factors and its nice properties in the mirror kinematics give us the necessary tools to study the finite-volume (and finite-coupling) spectrum of the theory by means of the mirror thermodynamic Bethe ansatz. We plan to return to this question in a forthcoming publication~\cite{upcoming:mirror}.

It would also be interesting to see if this approach, based on splitting off a BES factor from a rapidity-difference part of the crossing equations, could lead to solutions for other $AdS_3$ worldsheet S~matrices. A natural candidate is the pure-RR $AdS_3\times S^3\times S^3\times S^1$ background, whose S~matrix and crossing equations were found in~\cite{Borsato:2012ud,Borsato:2015mma}. Another interesting setup is the one where the background is supported both by RR fluxes and Neveu-Schwarz-Neveu-Schwarz ones. In this case we also know the S~matrix and crossing equations~\cite{Hoare:2013pma,Lloyd:2014bsa}, but the kinematics is more intricate~\cite{Hoare:2013lja}. Intriguingly, we have seen that our approach formally works also for this more complicated setup in section~\ref{sec:proposal:mixedflux}, see also~\cite{upcoming:mixed}.
This is particularly exciting in view of the obvious physical significance of the setup: it interpolates between the pure-RR case which we studied here (and which is reminiscent of $AdS_5\times S^5$) and the pure-NSNS case which can be described as a Wess-Zumino-Witten model~\cite{Maldacena:2000hw} and studied in great detail, both by integrability~\cite{Baggio:2018gct, Dei:2018mfl} and by worldsheet CFT techniques~\cite{Giribet:2018ada,Eberhardt:2018ouy,Eberhardt:2021vsx}. 
Finally, it would be interesting to see how this proposal for the dressing factors would amend the current understanding of the hexagon formalism~\cite{Basso:2015zoa,Eden:2016xvg,Fleury:2016ykk} for $AdS_3\times S^3\times T^4$, whose study has been recently initiated~\cite{Eden:2021xhe}.

\section*{Acknowledgements}
AS thanks Juan Maldacena for interesting related discussions. AS gratefully acknowledges support from the IBM Einstein Fellowship.


\appendix

\section{BES phase in the mirror-mirror kinematics}
\label{app:BES}
Here we will study the properties of the BES phase~\cite{Beisert:2006ez} in the mirror-mirror kinematics. For the original BES phase scattering massive particles only (as well as bound states thereof), this was studies in~\cite{Arutyunov:2009kf}. Here we will focus on the case where one or both particles are massless.

\subsection{Mixed-mass BES phase}
\label{app:BES:mixed}
Le us consider the mixed-mass BES phase~\eqref{eq:massless-massive-BES} and let us pick the massless variable to be $x_1$, while $x_2^\pm$ is massive.
In the mirror region, $|x^+_2|<1$ and $|x^-_2|>1$ while  $|x_1|<1$ (in fact, for real particles $-1<x_1<1$). In this region we have
\begin{equation}
\begin{aligned}
\label{eq:mirrormixedBES}
\theta_{\text{BES}}(x_1,x_2^+,x_2^-) &=\Phi(x_1,x_2^+)
-\Phi(x_1,x_2^-)-\Phi(\tfrac{1}{x_1},x_2^+)+\Phi(\tfrac{1}{x_1},x_2^-)
\\
&-\Psi(x_1,x_2^+)+\Psi(x_1,x_2^-)+\Psi(x_2^+,x_1)-\Psi(x_2^+,\tfrac{1}{x_1})
\\
&
+i\,\log\frac{\Gamma\big[1+{i\ov
2}h\big(x_1+\frac{1}{x_1}-x_2^+-{1\ov
x_2^+}\big)\big]}
{\Gamma\big[1-{i\ov 2}h\big(x_1+\frac{1}{x_1}-x_2^+-{1\ov
x_2^+}\big)\big]}
\end{aligned}
\end{equation}
which we can simplify as
\begin{equation}
\begin{aligned}
\theta_{\text{BES}}(x_1,x_2^+,x_2^-) &=2\Phi(x_1,x_2^+)
-2\Phi(x_1,x_2^-)-\Phi(0,x_2^+)+\Phi(0,x_2^-)
\\
&-\Psi(x_1,x_2^+)+\Psi({ x_1},x_2^-) +2\Psi({ x_2^+},x_1)-\Psi({ x_2^+},0)
\\
&
+i\,\log\frac{\Gamma\big[1+{i\ov
2}h\big(x_1+\frac{1}{x_1}-x_2^+-{1\ov
x_2^+}\big)\big]}
{\Gamma\big[1-{i\ov 2}h\big(x_1+\frac{1}{x_1}-x_2^+-{1\ov
x_2^+}\big)\big]}
\, .
\end{aligned}
\end{equation}

Let us now consider how this expression transforms under complex conjugation, which for particles of real mirror momentum gives
\begin{equation}
(x_1)^* = x_1\,,\quad (x_2^\pm)^* = \frac{1}{x_2^\mp}
\end{equation}
and therefore
\begin{equation}
\begin{aligned}
\theta_{\text{BES}} (x_1,x_2^+,x_2^-) - \theta_{\text{BES}}(x_1,x_2^+,x_2^-)^* 
=i\log {h^2\ov 4} (u_1 - u_2 - {i\ov h}) (u_1 - u_2 + {i\ov h})\\
+2\Psi({ x_2^+},x_1)-\Psi({ x_2^+},0) -2\Psi({ x_2^-},x_1)+\Psi({  x_2^-},0)
\, .
\end{aligned}
\end{equation}
By using the formulae (8.8) and (8.9) from~\cite{Arutyunov:2009kf}, we get
\begin{equation}
\begin{aligned}
\theta_{\text{BES}} (x_1,x_2^+,x_2^-) - \theta_{\text{BES}}(x_1,x_2^+,x_2^-)^* 
=-2 i\log x_1  +i\,\log {(x_1 - x^+_2 ) \ov(1-{ x_1  x^+_2} ) }{(1-{ x_1  x^-_2} ) \ov (x_1 - x^-_2 ) }
\, .
\end{aligned}
\end{equation}
Thus, the mixed-mass  BES  factor has the following conjugacy property in the mirror region
\begin{equation}
\begin{aligned}
\BES (x_1,x_2^+,x_2^-)\, \BES(x_1,x_2^+,x_2^-)^* &=
(x_1)^2\,  { (1-{ x_1  x^+_2} ) \ov (x_1  - x^+_2 )}{ (x_1 - x^-_2 ) \ov (1-{ x_1  x^-_2} ) }\,,
\\
\BES (x_1^+,x_1^-,x_2)\,\BES (x_1^+,x_1^-,x_2)^* &=
{1\ov (x_2)^2}\,  {  (x_1^+ - x_2 )\ov(1-{ x_1^+  x_2} ) }{  (1-{ x_1^-  x_2} ) \ov (x_1^- - x_2 ) }\,.
\end{aligned}
\end{equation}

 \paragraph{Improved BES phase.}
In the mirror-mirror kinematics it is useful to introduce an improved dressing factor, which for particles of bound-state number $Q$ and $Q'$ is
\begin{equation}
\label{eq:improvedBES}
\begin{aligned}
&{1\ov i}\log\Sigma^{QQ'}(y_1,y_2)
 =
\Phi(y_1^+,y_2^+)-\Phi(y_1^+,y_2^-)-\Phi(y_1^-,y_2^+)+\Phi(y_1^-,y_2^-)\\
&\qquad\qquad-{1\ov 2}\left(\Psi(y_1^+,y_2^+)+\Psi(y_1^-,y_2^+)-\Psi(y_1^+,y_2^-)-\Psi(y_1^-,y_2^-)\right) \\
&\qquad\qquad+{1\ov
2}\left(\Psi(y_{2}^+,y_1^+)+\Psi(y_{2}^-,y_1^+)-\Psi(y_{2}^+,y_1^-)
-\Psi(y_{2}^-,y_1^-) \right)
 \\
&\qquad\qquad+{1\ov i}\log\frac{ i^{Q}\,\Gamma\big[Q'-{i\ov
2}h\big(y_1^++\frac{1}{y_1^+}-y_2^+-\frac{1}{y_2^+}\big)\big]} {
i^{Q'}\Gamma\big[Q+{i\ov
2}h\big(y_1^++\frac{1}{y_1^+}-y_2^+-\frac{1}{y_2^+}\big)\big]}{1-
{1\ov y_1^+y_2^-}\ov 1-{1\ov
y_1^-y_2^+}}\sqrt{\frac{y_1^+y_2^-}{y_1^-y_2^+}} \,.
\end{aligned}
\end{equation}
We want to compare the above expression, for $Q=0$,  $Q'=1$, and $y_1^\pm=(x_1)^{\pm1}$, $y_2^\pm=x^\pm_2$, with the above expression~\eqref{eq:mirrormixedBES}.
We get
\begin{equation}
\begin{aligned}
&{1\ov i}\log\Sigma^{01}(x_1,x_2^\pm)-\theta_{\text{BES}}(x_1,x_2^\pm) =\\
&\qquad-{1\ov 2}\left(\Psi(x_1,x_2^+)+\Psi(x_1,x_2^+)-\Psi(x_1,x_2^-)-\Psi(x_1,x_2^-)\right) \\
&\qquad+{1\ov
2}\left(\Psi(x_{2}^+,x_1)+\Psi(x_{2}^-,x_1)-\Psi(x_{2}^+,\tfrac{1}{x_1})
-\Psi(x_{2}^-,\tfrac{1}{x_1}) \right)
 \\
&\qquad+{1\ov i}\log\frac{ \Gamma\big[1-{i\ov
2}h\big(x_1+\frac{1}{x_1}-x_2^+-\frac{1}{x_2^+}\big)\big]} {
i\,\Gamma\big[{i\ov
2}h\big(x_1+\frac{1}{x_1}-x_2^+-\frac{1}{x_2^+}\big)\big]}{1-
{1\ov x_1x_2^-}\ov 1-{x_1\ov
x_2^+}}\sqrt{\frac{(x_1)^2x_2^-}{x_2^+}}
\\
&\qquad+\Psi({ x_1},x_2^+)-\Psi({ x_1},x_2^-)-\Psi({ x_2^+},x_1)+\Psi({ x_2^+},\tfrac{1}{x_1})
\\
&\qquad
+{1\ov i}\,\log\frac{\Gamma\big[1+{i\ov
2}h\big(x_1+\frac{1}{x_1}-x_2^+-{1\ov
x_2^+}\big)\big]}
{\Gamma\big[1-{i\ov 2}h\big(x_1+\frac{1}{x_1}-x_2^+-{1\ov
x_2^+}\big)\big]}\,.
\end{aligned}
\end{equation}
Simplifying the above expression we find
\begin{equation}
\begin{aligned}
&{1\ov i}\log\Sigma^{01}(x_1,x_2^\pm)-\theta_{\text{BES}}(x_1,x_2^\pm)\\
&\qquad\quad= 
{1\ov
2}\left(-2\Psi(x_{2}^+,x_1)+2\Psi(x_{2}^-,x_1)+\Psi(x_{2}^+,0)
-\Psi(x_{2}^-,0) \right)
 \\
&\qquad\qquad+{1\ov i}\log{1\ov
2}h\big(x_1+\frac{1}{x_1}-x_2^+-\frac{1}{x_2^+}\big)\,{1-
{1\ov x_1x_2^-}\ov 1-{x_1\ov
x_2^+}}\sqrt{\frac{(x_1)^2x_2^-}{x_2^+}}\,.
\end{aligned}
\end{equation}
By using the identity 
\begin{equation}
\Psi(x_{2}^-,x_1)-\Psi(x_{2}^+,x_1)=
-\frac{2}{i}\log x_1 - {1\ov i} \log
(\frac{1}{x_2^-}-\frac{1}{x_1})(x_2^+ -\frac{1}{x_1})
+\frac{1}{i}\log\frac{x_2^+}{x^-_2}+ i\log{h^2\ov 4},
\end{equation}
and 
\begin{equation}
\Psi(x_{2}^-,0)-\Psi(x_{2}^+,0)=
\frac{1}{i}\log \frac{x_2^+}{x_2^-} +
i\log{h^2\ov 4}\,,
\end{equation}
we finally find
\begin{equation}
\begin{aligned}
&{1\ov i}\log\Sigma^{01}(x_1,x_2^\pm)-\theta_{\text{BES}}(x_1,x_2^\pm)
={1\ov i}\log\frac{1}{x_1^+}\,{1-
{ x_1^+x_2^-}\ov x_1^+-x_2^-}\,.
\end{aligned}
\end{equation}
Thus, we have
\begin{equation}
\left( \BES(x_1,x_2^\pm)\ov \Sigma^{01}(x_1,x_2^\pm) \right)^{-2} =
\frac{1}{(x_1)^2}\tanh^2{\g_{12}^{\circ-}\ov2}\,,
\end{equation}
so that we can write, for the full dressing factor
\begin{equation}
\begin{aligned}
    \big(\sigma^{\circ\bullet}(x_1,x_2^\pm)\big)^{-2}
    =- \frac{i}{(x_1)^2}{\tanh\tfrac{\gamma_{12}^{\circ+}}{2}}{\tanh\tfrac{\gamma_{12}^{\circ-}}{2}}\,
    \PhiSG(\gamma_{12}^{\circ+})\,\PhiSG(\gamma_{12}^{\circ-})\big(\Sigma^{01}(x_1,x_2^\pm)
    \big)^{-2}\,.
\end{aligned}
\end{equation}

\subsection{Massless-massless BES phase}
\label{app:BES:massless}
Let us now consider the massless-massless kinematics. The dressing factor in the string region is given by~\eqref{eq:masslessBES}, while in the mirror kinematics it is
\begin{equation}
\begin{aligned}
\theta^{\circ\circ}_{\text{BES}}(x_1,x_2) &=\Phi(x_1,x_2)-\Phi(x_1,{1\ov x_2})-\Phi({1\ov x_1},x_2)+\Phi({1\ov x_1},{1\ov x_2})
\\
&\quad- \Psi({ x_1},{ x_2})+\Psi({ x_2},{ x_1})+\Psi({ x_1},{ 1\ov x_2})-\Psi({ x_2},{1\ov x_1})
\\
&\quad+ i\,\log\frac{\Gamma\big[1+{i\ov
2}h\big(x_1 +\frac{1}{x_1}-x_2-{1\ov x_2}\big)\big]}
{\Gamma\big[1-{i\ov 2}h\big(x_1 +\frac{1}{x_1}-x_2-{1\ov
x_2^+}\big)\big]}
\\
&=4\Phi(x_1,x_2)-2\Psi({ x_1},{ x_2})
+2\Psi(x_2,x_1)+2\Phi(0,x_1)-2\Phi(0,x_2)
\\
&\quad
-\Psi(x_1,0)+\Psi(x_2,0)
+ i\,\log\frac{\Gamma\big[1+{i\ov
2}h\big(x_1 +\frac{1}{x_1}-x_2-{1\ov x_2}\big)\big]}
{\Gamma\big[1-{i\ov 2}h\big(x_1 +\frac{1}{x_1}-x_2-{1\ov
x_2}\big)\big]}
\, .
\end{aligned}
\end{equation}
Since in the mirror theory $(x)^* = x$, the phase is clearly real.

\paragraph{Improved BES phase.}
Let us now consider the improved BES phase \eqref{eq:improvedBES} for $Q=Q'=0$, and $y_1^\pm = (x_1)^{\pm1}$, $y_2^\pm = (x_2)^{\pm1}$. By repeating the derivation above and using the identity
\begin{equation}
 i\,\log\frac{\Gamma\big[1+{i\ov
2}h\big(x_1^+ +\frac{1}{x_1^+}-x_2^+-{1\ov x_2^+}\big)\big]}
{\Gamma\big[1-{i\ov 2}h\big(x_1^+ +\frac{1}{x_1^+}-x_2^+-{1\ov
x_2^+}\big)\big]}=
{1\ov i}\log-{{\Gamma\big[-{i\ov
2}h\big(x_1^++\frac{1}{x_1^+}-x_2^+-\frac{1}{x_2^+}\big)\big]} \ov {
\Gamma\big[+{i\ov
2}h\big(x_1^++\frac{1}{x_1^+}-y_2^+-\frac{1}{x_2^+}\big)\big]}}\,,
\end{equation}
we find that
\begin{equation}
{1\ov i}\log\Sigma^{00}(x_1^+,x_2^+) = \theta^{\circ\circ}_{\text{BES}}(x_1^+,x_2^+) + {1\ov i}\log \frac{y_2^+}{y_1^+}\sqrt{\frac{(y_1^+)^2}{(y_2^+)^2}}\,.
\end{equation}
Thus, the two phases can only differ by  $\pi$ but since we are interested in $(\BES)^2$, the massless BES factor squared is equal to the massless improved BES factor squared.

\section{Large-tension expansion of the BES phase}
\label{app:BESexpansion}
The BES dressing factor~\cite{Beisert:2006ez} was already introduced in section~\ref{sec:buildingblocks:BES} of the main text. Its asymptotic large-$h$ expansion can be obtained by expanding the integrand in the integral representation of~\cite{Dorey:2007xn}. For the massive-massive kinematics, this is entirely standard, and we report here the expansion for completeness. A little more care is needed for the mixed-mass and massless-massless kinematics, as we discuss below.

\subsection{Massive-massive BES}
Recall that the BES expansion can be written as
\begin{equation}
\label{eq:BESAFSHL}
    \Phi_{\text{BES}}(x_1,x_2) = \Phi_{\text{AFS}}(x_1,x_2) + \Phi_{\text{HL}}(x_1,x_2) +\cdots\,.
\end{equation}
with
\begin{equation}
\label{eq:AFSPhi}
    \Phi_{\text{AFS}}(x_1,x_2) 
    =\frac{h}{2} \left[\frac{1}{x_1}-\frac{1}{x_2}+\big(-x_1+x_2+\frac{1}{x_2}-\frac{1}{x_1}\big) \log \big(1-\frac{1}{x_1 x_2}\big)\right]\,,
\end{equation}
and
\begin{equation}
    \Phi_{\text{HL}}(x_1,x_2) =- \frac{\pi}{2}\oint\frac{{\rm d}w_1}{2\pi i}\oint \frac{{\rm
d}w_2}{2\pi i}\frac{\text{sgn}\big(w_1+\frac{1}{w_1}-w_2-\frac{1}{w_2}\big)}{(w_1-x_1)(w_2-x_2)}\,,
\end{equation}
which can be computed to give
\begin{equation}
\label{eq:HLPhi}
    \begin{aligned} &=\frac{\text{Li}_2\left(1-\frac{1}{x_2}\right)}{2 \pi }-\frac{\text{Li}_2\left(\frac{x_1-1}{x_1-x_2}\right)}{2 \pi
   }+\frac{\text{Li}_2\left(\frac{x_1+1}{x_1-x_2}\right)}{2 \pi }+\frac{\text{Li}_2\left(\frac{x_2}{x_2+1}\right)}{2 \pi
   } -\frac{\text{Li}_2\left(\frac{x_1 \left(x_2-1\right)}{x_1 x_2-1}\right)}{2 \pi }
   \\
   &-\frac{\text{Li}_2\left(\frac{x_1 x_2-1}{x_1
   \left(x_2+1\right)}\right)}{2 \pi }-\frac{\log ^2\left(x_1\right)}{4 \pi }+\frac{\log ^2\left(x_2\right)}{4 \pi }-\frac{\log
   ^2\left(x_1 x_2-1\right)}{4 \pi }-\frac{\log \left(x_2+1\right) \log \left(x_1\right)}{2 \pi }
    \\
   &+\frac{\log \left(x_1
   x_2-1\right) \log \left(x_1\right)}{2 \pi }
  -\frac{\log \left(x_1-1\right) \log \left(x_2-1\right)}{2 \pi }
   -\frac{\log \left(x_2-1\right) \log \left(x_2\right)}{2 \pi }
    \\
   &+\frac{\log \left(x_1+1\right) \log
   \left(x_2+1\right)}{2 \pi }
   +\frac{\log \left(x_1-1\right) \log \left(x_2-x_1\right)}{2
   \pi }-\frac{\log \left(x_1+1\right) \log \left(x_2-x_1\right)}{2 \pi }
    \\
   &+\frac{\log \left(x_2-1\right) \log \left(x_1
   x_2-1\right)}{2 \pi }\,.
    \end{aligned}
\end{equation}
This expression can be further simplified in the physical region for massive particles~\cite{Arutyunov:2006iu}.

\subsection{Mixed-mass dressing factor}

We want to define the BES dressing factor in the case where one particle, \textit{e.g.}\ the first one, is massless, that is $x_1$ is on the upper-half circle.
As discussed in the main text, for finite $h$ we can just use that if one of the integration contours of the BES integral~\eqref{eq:chiBES} is the unit circle, we can slightly deform the other contour, shrinking it a little; for instance, we can deform the contour corresponding to $w_1$ in~\eqref{eq:chiBES}. In this way we can place $x_1$ on the unit circle without encountering any singularity in~\eqref{eq:chiBES}, and get a representation for the mixed-mass BES phase for real momentum. This representation however is not well-suited for the asymptotic large-$h$ expansion. The most natural way to obtain a representation of BES that is well-defined for any $h$ is to analytically continue the phase to complex momentum keeping the relation $x_1^-=1/x_1^+$. An advantage of this way is that for complex $p$ both circles can be unit, and taking the large $h$ limit is the same as for massive particles. The only difference is that if, for example, $|x^+_1|>1$ then $|x^-_1|<1$ and the dressing phase gets two $\Psi$-functions, resulting in~\eqref{eq:massless-massive-BES}.
Now, large $h$ expansion of $\Phi(x_1^+,x_2^+)$ is standard and is given by~\eqref{eq:BESAFSHL}.
The AFS and HL orders are given by~\eqref{eq:AFSPhi} and~\eqref{eq:HLPhi}, respectively.
Next, we need the other terms appearing in \eqref{eq:massless-massive-BES}. We start from
\begin{equation}
\label{eq:AFSPhi0}
\begin{aligned}
\Phi_{\text{AFS}}(0,x_2) &= -{h\ov2}\oint\frac{{\rm d}w_1}{2\pi i}\oint \frac{{\rm
d}w_2}{2\pi i}\frac{1}{w_1(w_2-x_2)}
\\
&\times \big(w_1+\frac{1}{w_1}-w_2-{1\ov w_2}\big)\log{h^2\ov 4e^2}\big(w_1+\frac{1}{w_1}-w_2-{1\ov w_2}\big)^2
\\
&=-\frac{h}{x_2} -\frac{h}{2x_2} \,\log{h^2\ov 4e^2}
\end{aligned}
\end{equation}
and
\begin{equation}
\begin{aligned}
\Phi_{\text{HL}}(0,x_2) &=- {\pi\ov2}\oint\frac{{\rm d}w_1}{2\pi i}\oint \frac{{\rm
d}w_2}{2\pi i}\frac{1}{w_1(w_2-x_2)}
 \text{sign}\big(w_1+\frac{1}{w_1}-w_2-{1\ov w_2}\big)
 \\
 &=- {1\ov2i}\oint\frac{{\rm d}w_1}{2\pi i}\frac{\text{sign}(\Im w_1)
}{w_1}\big(\log(x_2-w_1)-\log(x_2-{1\ov w_1})\big)
\\
&=\frac{1}{2 \pi
   }\left( \text{Li}_2\Big(\frac{1}{x_2^2}\Big)-4 \text{Li}_2\Big(\frac{1}{x_2}\Big)\right)
\end{aligned}
\end{equation}
Finally, the large $h$ expansion of $\Psi({x_1},x_2)$ requires $|x_1|=1$, and it is given by
\begin{equation}
\begin{aligned}
\Psi_{\text{AFS}}&({x_1},x_2)=-{h\ov2}\oint\frac{{\rm d
}w}{2\pi i} \frac{1}{w-x_2}
 \big(x_1+\frac{1}{x_1}-w-{1\ov w}\big)\log{h^2\ov 4e^2}\big(x_1+\frac{1}{x_1}-w-{1\ov w}\big)^2
 \\
&=\frac{h}{2}\Big[
-x_1-\frac{1}{x_1}+\left(x_1+\frac{1}{x_1}-x_2-\frac{1}{x_2}\right)\times 
   \\
   & \qquad\qquad\qquad\qquad\times\left(\log
   \left(1-\frac{1}{x_1 x_2}\right)+\log \left(1-\frac{x_1}{x_2}\right)\right)\Big]
   -\frac{h}{2x_2} \,\log\frac{h^2}{4e^2}\,,
\end{aligned}
\end{equation}
 and
\begin{equation}
\begin{aligned}
\Psi_{\text{HL}}({x_1},x_2)&=-\frac{\pi}{2}\oint\frac{{\rm d
}w}{2\pi i} \frac{1}{w-x_2} \text{sgn}\big(x_1+\frac{1}{x_1}-w-{1\ov w}\big)
\\
&= \frac{i}{2} \Im(x_1)\left(\log
   \left(x_2-\frac{1}{x_1}\right)-\log \left(x_2- x_1\right)\right)\,,
\end{aligned}
\end{equation}

There is another way to obtain a representation of the mixed-mass BES phase which is well-defined for any $h$ which we would like to mention. We write
 \bal
\chi(x_1,x_2)&=i\oint\frac{{\rm d}w_1}{2\pi i}\oint \frac{{\rm
d}w_2}{2\pi i}\frac{1}{(w_1-x_1)(w_2-x_2)}
\\
&\times \log{\Gamma\big[1+{i\ov
2}h\big(w_1+\frac{1}{w_1}-w_2-{1\ov w_2}\big)\big]\ov
\Gamma\big[1-{i\ov 2}h\big(w_1+\frac{1}{w_1}-w_2-{1\ov
w_2}\big)\big]}
{\Gamma\big[1-{i\ov2}h\big(x_1+\frac{1}{x_1}-w_2-{1\ov w_2}\big)\big]\ov
\Gamma\big[1+{i\ov 2}h\big(x_1+\frac{1}{x_1}-w_2-{1\ov w_2}\big)\big]}\, , 
\eal 
The added term vanishes for finite $h$ but its addition makes the integrand to be regular at $w_1=x_1$, and  therefore both circles can be of unit radius. It is not hard to perform this expansion and verify that this results in the same near-BMN expansion as the regularisation introduced above.

\subsection{Massless-massless BES phase}

By the same reasoning as above, we want to define the massless-massless BES phase, when both $x_1$ and $x_2$ are on the unit circle. Again we continue the phase to complex momenta keeping the relation $x_1^-=1/x_1^+$ and $x_2^-=1/x_2^+$. We can simplify the expression of~\eqref{eq:masslessBES} as  follows,
\bal
\theta_{\text{BES}}(x_1,x_2) 
&=4\Phi(x_1,x_2)
-2\Psi(x_2,x_1)+2\Phi(0,x_1)-2\Phi(0,x_2)+\Psi(x_2,0)
\\
&+\Psi({ x_1},x_2)-\Psi({ x_1},{1\ov x_2})+ i\,\log\frac{\Gamma\big[1+{i\ov
2}g\big(x_1 +\frac{1}{x_1}-x_2-{1\ov x_2}\big)\big]}
{\Gamma\big[1-{i\ov 2}g\big(x_1 +\frac{1}{x_1}-x_2-{1\ov
x_2}\big)\big]}
\\
&=4\Phi(x_1,x_2)+2\Psi({ x_1},{ x_2})
-2\Psi(x_2,x_1)+2\Phi(0,x_1)-2\Phi(0,x_2)
\\
&
-\Psi(x_1,0)+\Psi(x_2,0)
+ i\,\log\frac{\Gamma\big[1+{i\ov
2}h\big(x_1 +\frac{1}{x_1}-x_2-{1\ov x_2}\big)\big]}
{\Gamma\big[1-{i\ov 2}h\big(x_1 +\frac{1}{x_1}-x_2-{1\ov
x_2}\big)\big]}
\, .
\eal 
We first consider the terms of the AFS order, where we have~\eqref{eq:AFSPhi} and \eqref{eq:AFSPhi0} as above.
Then, the large $h$ expansion of $\Psi({x_1},x_2)$ requires $|x_1|=1$, and is given by
\bal
&\Psi_{\text{AFS}}({x_1},x_2)=-{h\ov2}\oint\frac{{\rm d
}w}{2\pi i} \frac{1}{w-x_2}
 \big(x_1+\frac{1}{x_1}-w-{1\ov w}\big)\log\frac{h^2\big(x_1+\frac{1}{x_1}-w-{1\ov w}\big)^2}{4e^2}
 \\
&={h\ov2}\left[
-x_1-\frac{1}{x_1}+\big(x_1+\frac{1}{x_1}-x_2-\frac{1}{x_2}\big) \Big[\log
   \big(1-\frac{1}{x_1 x_2}\big)+\log \big(1-{x_1\ov x_2}\big)\Big]\right]
   \\
   & -\frac{h}{2x_2} \,\log{h^2\ov 4e^2}\,,
 \eal
 and
\bal
\Psi_{\text{AFS}}&({x_1},0)=-{h\ov2}\oint\frac{{\rm d
}w}{2\pi i} \frac{1}{w}
 \big(x_1+\frac{1}{x_1}-w-{1\ov w}\big)\log{h^2\ov 4e^2}\big(x_1+\frac{1}{x_1}-w-{1\ov w}\big)^2
 \\
&=-{h}\big(
x_1+\frac{1}{x_1}\big) - \big(x_1+\frac{1}{x_1}\big)\frac{h}{2} \,\log{h^2\ov 4e^2}\,,
 \eal
Finally, we have
\begin{equation}
\begin{aligned}
&i\log\frac{\Gamma\big[1+{i\ov
2}h\big(x_1 +\frac{1}{x_1}-x_2-{1\ov x_2}\big)\big]}
{\Gamma\big[1-{i\ov 2}h\big(x_1 +\frac{1}{x_1}-x_2-{1\ov
x_2}\big)\big]} =\\
&\qquad\qquad\qquad\qquad- {h\ov
2}\big(x_1 +\frac{1}{x_1}-x_2-{1\ov x_2}\big)\log{h^2\ov 4e^2}\big(x_1 +\frac{1}{x_1}-x_2-{1\ov x_2}\big)^2.
\end{aligned}
\end{equation}
By using these formulae one finds that  $\theta_{AFS}(x_1,x_2)$ is given by
\bal
\theta^{\circ\circ}_{AFS}(x_1^+,x_2^+) &={h \left(x_1-x_2\right) \left(1-{1\ov x_1 x_2}\right) \log \left(\frac{x_2-x_1}{x_1 x_2-1}\right)}\,.
\eal 

At the HL order we have
\bal
\Phi_{\text{HL}}(x_1,x_2) &=- {\pi\ov2}\oint\frac{{\rm d}w_1}{2\pi i}\oint \frac{{\rm
d}w_2}{2\pi i}\frac{1}{(w_1-x_1)(w_2-x_2)}
 \text{sign}\big(w_1+\frac{1}{w_1}-w_2-{1\ov w_2}\big)
 \\
 &=- {1\ov2i}\oint\frac{{\rm d}w_1}{2\pi i}\frac{\text{sign}(\Im w_1)
}{w_1-x_1}\big(\log(x_2-w_1)-\log(x_2-{1\ov w_1})\big).
\eal
Next, we need
\bal
\Phi_{\text{HL}}(0,x_2) &=- {\pi\ov2}\oint\frac{{\rm d}w_1}{2\pi i}\oint \frac{{\rm
d}w_2}{2\pi i}\frac{1}{w_1(w_2-x_2)}
 \text{sign}\big(w_1+\frac{1}{w_1}-w_2-{1\ov w_2}\big)
 \\
 &=- {1\ov2i}\oint\frac{{\rm d}w_1}{2\pi i}\frac{\text{sign}(\Im w_1)
}{w_1}\big(\log(x_2-w_1)-\log(x_2-{1\ov w_1})\big)
\\
&=\frac{1}{2 \pi
   }\big( \text{Li}_2\frac{1}{x_2^2}-4 \text{Li}_2\frac{1}{x_2}\big)
\eal
Finally, the large-$h$ expansion of $\Psi({x_1},x_2)$ requires $|x_1|=1$, and is given by
\bal
\Psi_{\text{HL}}({x_1},x_2)&=-{\pi\ov2}\oint\frac{{\rm d
}w}{2\pi i} \frac{1}{w-x_2} \text{sign}\big(x_1+\frac{1}{x_1}-w-{1\ov w}\big)
\\
&= { i\ov2} \Im(x_1)\left(\log
   \left(x_2-\frac{1}{x_1}\right)-\log \left(x_2- x_1\right)\right)\,, 
\eal 
and
\bal 
\Psi_{HL}({x_1},0)&=-{\pi\ov2}\oint\frac{{\rm d
}w}{2\pi i} \frac{1}{w} \text{sign}\big(x_1+\frac{1}{x_1}-w-{1\ov w}\big)
\\
&=-{\pi\ov2}-i\, \Im(x_1)\log x_1\,,
\eal 
By using these formulae one finds numerically that the massless-massless $\theta_{\text{HL}}(x_1,x_2)$ phase is given by
\bal
e^{2i\,\theta_{\text{HL}}(x_1,x_2) }&=e^{2i\theta_{\text{HL}}(\g_1,\g_2)}=- \varPhi(\g_{12}^{\circ\circ})^4.
\eal 

\section{Construction of the monodromy factor by Fourier transform}
\label{app:Fourier}
It is easy and instructive to derive the minimal solution of the monodromy equation~\eqref{eq:varPsimonodromy} and see that it is essentially fixed by minimality (and that it matches~$\widehat{\varPhi}(\gamma)$). We want to solve
\begin{equation}
    \frac{\widehat{\varPhi}(\gamma\pm i\pi)}{\widehat{\varPhi}(\gamma)} = i\big(2\sinh(\gamma)\big)^{\pm1}\,,\qquad\widehat{\varPhi}(\gamma)\,\widehat{\varPhi}(-\gamma)=1\,,\qquad \widehat{\varPhi}(\gamma^*)\,\widehat{\varPhi}(\gamma)^*=1\,,
\end{equation}
under the assumption that the function has no poles or zeros in the physical strip~$|\mathfrak{I}[\gamma]|<i \pi$. 
Let us pick the positive sign in the monodromy equation above. Taking the logarithmic derivative of the equation, we have
\begin{equation}
    \widehat{\varphi}'(\gamma+i\pi)-\widehat{\varphi}'(\gamma)=\coth\gamma\,,
\end{equation}
which is obviously defined up to a constant. Such a constant amounts to a multiplicative factor~$\widehat{\varPhi}(\gamma)\to e^{\text{const.}\, \gamma}\widehat{\varPhi}(\gamma)$, which is actually irrelevant for the whole phase, as it cancels in the ratio of the various $\widehat{\varPhi}(\gamma)$ functions.
Taking the Fourier transform of the whole equation we get
\begin{equation}
    \frac{1}{\sqrt{2\pi}}\int\limits_{-\infty}^{+\infty}\text{d}t \Big(\widehat{\varphi}'(\gamma+i\pi)-\widehat{\varphi}'(\gamma)\Big)
    e^{i\omega \gamma}
    =i\sqrt{\frac{\pi}{2}} \coth \frac{\pi \omega}{2}\,.
\end{equation}
Under the condition that there are no poles or zeros in the strip $0<\mathfrak{I}[\gamma]< \pi$ (the other half of the strip would be needed to solve the equation with the opposite sign) we may deform the integration contour of the first term to find
\begin{equation}
    \frac{1}{\sqrt{2\pi}}\int\limits_{-\infty}^{+\infty}\text{d}t\Big(\widehat{\varphi}'(\gamma)+\text{const.}\Big)\,
    e^{i\omega \gamma}=
    i\sqrt{\frac{\pi}{2}} \frac{e^{\omega\pi}+1}{(e^{\omega\pi}+1)^2}\,,
\end{equation}
where the constant is fixed by demanding that the inverse transform of the right-hand side is regular. This indeed gives
\begin{equation}
    \widehat{\varphi}(\gamma)= \frac{\gamma\,\coth\gamma}{i\pi}\,,
\end{equation}
up to a constant. Of course a posteriori it is also easy to verify that this solution is regular in the physical strip and solves the monodromy equation. Finally, the primitive~$\widehat{\varphi}(z)$ is fixed by requiring that~$\widehat{\varphi}(0)=0$.

\section{Lightcone gauge and S-matrix normalisation}
\label{app:lcgauge}
One subtle point to bear in mind before comparing  our results to the perrturbative computations in the literature is that the S~matrix can be computed in different uniform light-cone gauges~\cite{Arutyunov:2005hd}. This results in a gauge factor which sits in front of the S~matrix and at all loops has the form
\begin{equation}
\label{eq:agauge}
    \Lambda_{12}(a)=e^{-i a (p_1 E(p_2)-p_2E(p_1))}\,,
\end{equation}
where $E(p)$ is the all-loop energy and $a$ is the gauge parameter, $0\leq a \leq 1$. 
As a result of this, the Bethe-Yang equations in the $a$-gauge are independent from~$a$: schematically
\begin{equation}
\begin{aligned}
    1&=e^{i p_k L(a)}\prod_j \Lambda_{kj}(a)\,S(p_k,p_j)\\
    &=e^{i p_k (L(a)- a E_{tot})}e^{i a p_{tot} E_k}\prod_j S(p_k,p_j)=e^{i p_k J}\prod_j S(p_k,p_j)\,,
\end{aligned}
\end{equation}
where we used that in the uniform light-cone gauge~$L(a)=J+a E_{tot}$~\cite{Arutyunov:2009ga} and that  the level-matching constraint sets $p_{tot}=0$ in the absence of winding.%
\footnote{This is not unlike what appears in $T\bar{T}$ deformations, see~\cite{Baggio:2018gct,Frolov:2019nrr,Frolov:2019xzi} for a discussion.}

\section{Near-BMN expansion of Zhukovsky and rapidity variables}
\label{app:Zhukovsky}
The massive Zhukovsky variables are given by
\begin{equation}
    x^\pm(p)=e^{\pm i p/2}\frac{1+\sqrt{1+4h^2\sin^2(p/2)}}{2h\sin(p/2)}\,,
\end{equation}
so that in the near-BMN~\cite{Berenstein:2002jq} limit, where $h\gg1$ and the momentum is rescaled $p\to p/h$ so that it is small, we have
\begin{equation}
    x^\pm(p)=\frac{1+\omega(p)}{p}\left(1\pm\frac{ip}{2h}+\frac{p^2}{24h^2}\frac{1-3\omega(p)}{\omega(p)}\right)+O(h^{-3})\,,\qquad
    \omega(p)=\sqrt{1+p^2}\,,
\end{equation}
where $\omega(p)$ is the tree-level dispersion relation.
The massive rapidity variables~\eqref{eq:gammapm} have the expansion
\begin{equation}
    \gamma^\pm(p)=\log\frac{\omega(p)-p}{\pm i}\pm \frac{ip^2}{2h}+\frac{p^3(4+3p^2)}{24h^2\,\omega(p)}+O(h^{-3})\,.
\end{equation}

The massless Zhukovsky variable is
\begin{equation}
    x^\pm(p)=e^{\pm i p/2}\,\text{sgn}[\sin(p/2)]\,,
\end{equation}
In the near-BMN expansion, we have to distinguish the cases where the momentum is small and positive, or small and negative. We have
\begin{equation}
    x=1+\frac{ip}{2h}-\frac{p^2}{8h^2}+O(h^{-3})\,,\quad(p>0)\,,\qquad
    x=-1-\frac{ip}{2h}+\frac{p^2}{8h^2}+O(h^{-3})\,,\quad(p<0)\,.
\end{equation}
In the $x$-plane we are therefore expanding either around $x=+1$ or $x=-1$, which are branchpoints for $\gamma(x)$, \textit{cf.}~\eqref{eq:gamma}. As a result, the expansion of the rapidity is singular and features terms of the form $\log h$:
\begin{equation}
    \gamma=\log\frac{p}{4h}+\frac{p^2}{16h^2}+O(h^{-3})\,\quad (p>0)\,,\qquad
    \gamma=\log\frac{4h}{-p}-\frac{p^2}{16h^2}+O(h^{-3})\,\quad (p<0)\,.
\end{equation}
This is a first hint of possible divergences in the near-BMN massless kinematics.

\section{Near-BMN expansion of the dressing factors}
\label{app:BMNexpansion}
By combining the formulae of appendices~\ref{app:BES} and~\ref{app:Zhukovsky} we can obtain explicit expressions for the dressing factors.

\subsection{Massive dressing factors}
In the near-BMN kinematics we find, the massive-massive BES phase gives
\begin{equation}
\begin{aligned}
    \log[(\BES)^{-2}]&=
    \frac{1}{h}\frac{p_1^2(\omega_2-1)-p_2^2(\omega_1-1)}{p_1+p_2}\\
    &\quad+
    \frac{1}{h^2}\left(
    \frac{p_1^2p_2^2(\omega_1\omega_2-p_1p_2)}{2\pi(\omega_1p_2-\omega_2 p_1)^2}\log\frac{\omega_2-p_2}{\omega_1-p_1}+\frac{p_1^2p_2^2}{2\pi(\omega_1p_2-\omega_2 p_1)}
    \right)\\
    &\quad+O(h^{-3})\,.
\end{aligned}
\end{equation}
Next we can  expand the part of the dressing factors which we denoted by $\sigmabb(x_1^\pm,x_2^\pm)^{-2}$ and $\sigmabbt(x_1^\pm,x_2^\pm)^{-2}$, which is expressed in terms of the difference of rapidities~$\gamma^\pm$. Using their representation in terms of polylogarithms~\eqref{eq:massivesoldilog} we find
\begin{equation}
\begin{aligned}
    \log(\sigmabb_{12})^{-2}=
    \frac{i}{2\pi h^2}\frac{p_1^2p_2^2(+1+p_1p_2-\omega_1\omega_2)}{\omega_1p_2-\omega_2p_1}\left(+1+\frac{\log\frac{\omega_1-p_1}{\omega_2-p_2}}{\omega_1p_2-\omega_2p_1}\right)+O(h^{-3})\,,\\
    \log(\sigmabbt_{12})^{-2}=
    \frac{i}{2\pi h^2}\frac{p_1^2p_2^2(-1+p_1p_2-\omega_1\omega_2)}{\omega_1p_2-\omega_2p_1}\left(-1+\frac{\log\frac{\omega_1-p_1}{\omega_2-p_2}}{\omega_1p_2-\omega_2p_1}\right)+O(h^{-3})\,,
\end{aligned}
\end{equation}
It is also useful to write the expansions for
\begin{equation}
    \begin{aligned}
    &\log(\sigmap_{12})^{-2}=
    \frac{i}{\pi h^2}\frac{p_1^2p_2^2}{\omega_1p_2-\omega_2p_1}\left(1+(p_1p_2-\omega_1\omega_2)\frac{\log\frac{\omega_1-p_1}{\omega_2-p_2}}{\omega_1p_2-\omega_2p_1}\right)+O(h^{-3})\,,\\
    &\log(\sigmam_{12})^{-2}=
    \frac{i}{\pi h^2}\frac{p_1^2p_2^2}{\omega_1p_2-\omega_2p_1}\left((p_1p_2-\omega_1\omega_2)-\frac{\log\frac{\omega_1-p_1}{\omega_2-p_2}}{\omega_1p_2-\omega_2p_1}\right)+O(h^{-3})\,.
    \end{aligned}
\end{equation}
At this order we can explicitly verify that the sum of the phases is proportional to the HL phase~$\Phi_{12}^{\text{HL}}$,
\begin{equation}
    \log(\sigmap_{12})^{-2}=2\Phi_{12}^{\text{HL}}+O(h^{-3})\,.
\end{equation}

\subsection{Mixed-mass dressing factor}
In this case we will have to discuss the dressing factor along with the whole normalisation of the S~matrix element (we shall choose $\gen{S}|Y_{p_1}\chi_{p_2}\rangle$ for reference), given that our normalisation differs from the one of~\cite{Borsato:2016xns}. To begin with, we consider the Sine-Gordon factor~$\varPhi(z)$ in the mixed-mass kinematics. Using the expressions~\eqref{eq:varphiexplicit} for $\varphi(z)=\log\varPhi(z)$ it is easy to work out its expansions.
In the mixed-mass case we are interested in
\begin{equation}
    \log\sigmabc(x_1^\pm, x_2)\,,\qquad p_2<0\,,
\end{equation}
where it is important to specify the sign of the momentum of the massless particle. In fact, in the near-BMN limit this excitation is relativistic, hence scattering can only be described if the particle moves towards the left. We find
\begin{equation}
\begin{aligned}
    \log(\sigmabc(x_1^\pm, x_2)\big)^{-2}
    &=
    \frac{i}{2h}(\omega_1-p_1)p_2\\
    &\quad+
    \frac{i}{2\pi h^2}(\omega_1-p_1)p_1^2p_2 \log\left(\frac{-(\omega_1-p_1)p_2}{4h}\right)+O(h^{-3})\,.
\end{aligned}
\end{equation}
In a similar way, we expand the mixed-mass BES phase when $p_2<0$. We find
\begin{equation}
\begin{aligned}
    \log(\BES(x_1^\pm, x_2)\big)^{-2}
    &=
    -\frac{i}{h}(\omega_1-1)p_2\\
    &\quad-
    \frac{i}{\pi h^2}(\omega_1-p_1)p_1^2p_2 \log\left(\frac{-(\omega_1-p_1)p_2}{4h}\right)+O(h^{-3})\,.
\end{aligned}
\end{equation}
Once again, this is in good accord with the fact that at order $O(h^{-2})$, \textit{i.e.}\ at one loop, the Sine-Gordon phase gives ``half'' of the BES (or more precisely, HL) phase.

To complete the comparison we have also to take into account the rational prefactor in~\eqref{eq:mixednormalisation} and consider the whole S-matrix element. Expanding the rational prefactor is straightforward. Let us consider the $Y\chi$ scattering element, see~\eqref{eq:mixednormalisation}. We have
\begin{equation}
    \log\left(e^{+\frac{i}{2} p_1}e^{-i p_2}
    \frac{x^-_1-x_2}{1-x^+_1x_2}\right)=-\frac{i}{2h}\Big(2p_2+(\omega_1-p_1)(p_1+p_2)\Big)+O(h^{-3})\,.
\end{equation}

Putting the pieces together we find that at tree level, in the $a=0$ lightcone gauge,
\begin{equation}
\begin{aligned}
    \log\langle Y_1\chi^{\dot{\alpha}}_2|\mathbf{S}|Y_1\chi^{\dot{\alpha}}_2\rangle
    &=
    -\frac{i}{2h}\big(p_1(\omega_1-p_1)+2p_2\omega_1\big)\\
    &\quad-
    \frac{ip_1^2}{2\pi h^2}
    (\omega_1-p_1)p_2\,\log\Big(\tfrac{-(\omega_1-p_1)p_2}{4h}\Big)
    +O(h^{-3})
    \,.
\end{aligned}
\end{equation}

\paragraph{Field-theory result.}
We follow Sundin and Wulff~\cite{Sundin:2016gqe} and rewrite the relevant S-matrix element.
First of all, we note that their result is given in the $a=1/2$ gauge. Rewriting their result in the $a=0$ gauge by means of~\eqref{eq:agauge} we find
\begin{equation}
\begin{aligned}
    \log\langle Y_1\chi^{\dot{\alpha}}_2|\mathbf{S}_{\text{SW}}|Y_1\chi^{\dot{\alpha}}_2\rangle
    &=
    -\frac{i}{2h}\big(p_1(\omega_1-p_1)+2p_2\omega_1\big)\\
    &\quad-
    \frac{ip_1^2}{2\pi h^2}
    (\omega_1-p_1)p_2\,\left[1+\log\Big(\tfrac{\omega_1-p_1}{-2p_2}\Big)
    \right]+O(h^{-3})
    \,.
\end{aligned}
\end{equation}
Notice the different argument of the logarithm.

\subsection{Massless-massless dressing factor}
Like for the mixed-mass dressing factor, we will consider an S-matrix element, to avoid ambiguities with the normalisation. We will pick the highest-weight states scattering, $\gen{S}|\chi^{\dot{\alpha}}_{p_1}\chi^{\dot{\beta}}_{p_2}\rangle$, which is actually independent from the $su(2)_{\circ}$ indices ${\dot{\alpha}},{\dot{\beta}}$ in perturbation theory~\cite{Sundin:2016gqe}. We should compare that perturbative S-matrix element with
\begin{equation}
    \big(\phase^{\circ\circ}(p_1,p_2)\big)^{-2}\qquad
    p_1>0,\quad p_2<0\,,
\end{equation}
see eq.~\eqref{eq:masslessnorm}.
We start by computing
\begin{equation}
    \log\left(-ia(\gamma_{12})\right) =  -i\frac{p_1p_2}{8h^2}+O(h^{-4})
\end{equation}
as well as
\begin{equation}
    \log\left(i\varPhi^2(\gamma_{12})\right)
    =
    +i\frac{p_1p_2}{4\pi h^2} \left(\log\Big(-\frac{p_1p_2}{16h^2}\Big)-1\right)+O(h^{-4})\,,
\end{equation}
where we notice that only the latter piece is proportional to $1/\pi$. Together they give

\begin{equation}
\begin{aligned}
    \log\left(\sigmacc(\gamma_{12})^{-2}\right)&=
    \log\left(a(\gamma_{12})\varPhi^2(\gamma_{12})\right)\\
    &=-i\frac{p_1p_2}{8h^2}
    +i\frac{p_1p_2}{4\pi h^2} \left(\log\Big(-\frac{p_1p_2}{16h^2}\Big)-1\right)+O(h^{-4})\,,
\end{aligned}
\end{equation}
while from the BES phase we get
\begin{equation}
    \log\big(\BES(x_1,x_2)\big)^{-2}=
    -\frac{i}{h}p_1p_2-i\frac{p_1p_2}{2\pi h^2} \left(\log\Big(-\frac{p_1p_2}{16h^2}\Big)-1\right)
    +O(h^{-3})\,,
\end{equation}
where the first term comes from the AFS part of the phase, see appendix~\ref{app:BES:mixed}, while the second term comes from the HL order, which is opposite and twice what we got from the Sine-Gordon part. In conclusion we have
\begin{equation}
    \log\Big(\phase^{\circ\circ}(p_1,p_2)\Big)^{-2}
    =
    -\frac{i}{h}p_1p_2-\frac{i}{h^2}\frac{p_1p_2}{8}
    -i\frac{p_1p_2}{4\pi h^2} \left(\log\frac{-p_1p_2}{16h^2}-1\right)+O(h^{-3})\,.
\end{equation}

\paragraph{Field-theory result.}
We follow once again Sundin and Wulff~\cite{Sundin:2016gqe}, recalling that their results are in the $a=1/2$ gauge. Rewriting them  in the $a=0$ gauge by means of~\eqref{eq:agauge} we~find
\begin{equation}
    \log\langle \chi^{\dot{\alpha}}_1\chi^{\dot{\beta}}_2|\mathbf{S}_{\text{SW}}|\chi^{\dot{\alpha}}_1\chi^{\dot{\beta}}_2\rangle
    =-\frac{i}{h}p_1p_2
    +
    i\frac{p_1p_2}{4\pi h^2}
    \Big(\log(-4p_1p_2)-1\Big)+O(h^{-3})
    \,,
\end{equation}
where the tree-level term matches the expansion above but the one-loop term is rather different from it.

\section{The previous proposal for the monodromy factor}
\label{app:monodromy}

The phase $\theta^-$ of the ratio of massive dressing factors  proposed in~\cite{Borsato:2013hoa} is constructed by using the following function  
\bal
\chi^-(x_1,x_2) = 
\left(\inturl -\intdlr \right)\frac{dw}{8\pi }\frac{1}{x_1-w}\log \left[(x_2-w) \left(1-{1\ov x_2w}\right)\right] - (x_1\leftrightarrow x_2)\,,
\eal
It is easy to show that the phase can be brought to the form
\bal
\chi^-(u_1,u_2) = 
\int_{-2}^2\frac{dv}{8\pi }\left(\frac{\log (u_2-v)}{v-u_1}-\frac{\log (u_1-v)}{v-u_2}\right) \,,
\eal
where
\bal
u_k=x_k+{1\ov x_k}\,,\quad v=w+{1\ov w}\,.
\eal
This representation makes it obvious that the function is not parity odd, which it should be due to parity invariance of the theory~\cite{Arutyunov:2009ga,upcoming:massless}; instead
\begin{equation}
    \chi^-(-u_1,-u_2)\neq -\chi^-(u_1,u_2)\,.
\end{equation}
Even when taking into account that the full phase is given by a linear combination of four $\chi^-(u_1^\pm,u_2^\pm)$ functions, parity invariance is not restored. We conclude that the dressing phase of~\cite{Borsato:2013hoa} breaks parity invariance. This by itself is sufficient to rule out $\chi^-$ as a candidate for a dressing~phase.
It has however another unpleasant property.
In addition to the usual branch points at $u_{1,2}=\pm2$ the function $\chi^-(u_1,u_2)$ has a branch point at $u_{1,2}=\infty$. 
Even though the branch point at infinity disappears in the full phase, its existence for $\chi^-(u_1,u_2)$
 means that the result of the analytic continuation of the phase to the mirror region depends on whether the path used for the analytic continuation goes around the branch point $-2$ or $+2$.
 Obviously, the new phase $\theta^-$ we proposed does not suffer from this ambiguity.

\bibliographystyle{JHEP}
\bibliography{refs}
\end{document}